\documentclass[11pt]{article}%
\usepackage{amsmath}
\usepackage{graphicx}%
\usepackage{amsfonts}%
\usepackage{amssymb}

\newcommand{\eqnb}{\begin{equation}}
\newcommand{\eqne}{\end{equation}}

\newtheorem{The}{Theorem}

\newtheorem{Cor}[The]{Corollary}
\newtheorem{Lem}{Lemma}

\newtheorem{Rem}{Remark}

\begin{document}

\title{A New Theoretical Framework of Pyramid Markov Processes for Blockchain Selfish Mining}
\author{Quan-Lin Li, Yan-Xia Chang\\School of Economics and Management,\\Beijing University of Technology, Beijing 100124, China\\Xiaole Wu\\School of Management\\Fudan University, Shanghai 200433, China\\Guoqing Zhang \\Department of Mechanical, Automotive, and Materials Engineering\\University of Windsor, Windsor, Ontario, Canada}
\maketitle

\begin{abstract}
In this paper, we provide a new theoretical framework of pyramid Markov
processes to solve some open and fundamental problems of blockchain selfish
mining under a rigorous mathematical setting. We first describe a more general
model of blockchain selfish mining with both a two-block leading competitive
criterion and a new economic incentive mechanism. Then we establish a pyramid
Markov process and show that it is irreducible and positive recurrent, and its
stationary probability vector is matrix-geometric with an explicitly
representable rate matrix. Also, we use the stationary probability vector to
study the influence of many orphan blocks on the waste of computing resource.
Next, we set up a pyramid Markov reward process to investigate the long-run
average profits of the honest and dishonest mining pools, respectively. As a
by-product, we build three approximative Markov processes and provide some new
interesting interpretation on the Markov chain and the revenue analysis
reported in the seminal work by Eyal and Sirer (2014). Note that the pyramid
Markov (reward) processes can open up a new avenue in the study of blockchain
selfish mining. Thus we hope that the methodology and results developed in
this paper shed light on the blockchain selfish mining such that a series of
promising research can be developed potentially.

\vskip  0.5cm

\textbf{Keywords:} Bitcoin; blockchain; Proof of Work; selfish mining; forked
structure; orphan block; pyramid Markov process; pyramid Markov reward
process; phase-type distribution; Matrix-geometric solution.

\end{abstract}

\section{Introduction}

Bitcoin has received tremendous attention as the first fully decentralized
distributed cryptocurrency since its advent by Satoshi Nakamoto
\cite{Nak:2008} in 2008. Blockchain is used to securely record a public shared
ledger of Bitcoin payment transactions among Internet users in an open P2P
network. In the past decade, we have witnessed the explosive growth of
blockchain, as exemplified by Bitcoin (2008), Ethereum and smart contract
(2013), Hyperledger Fabric (2015), Libra (Facebook 2019), among others. On the
one hand, the great success of blockchain is based on solving the
cryptographic puzzle by means of the brute force, namely Proof of Work (PoW).
Note that the PoW is a most frequently used consensus protocol (or algorithm).
Readers may refer to Garay et al. \cite{Gar:2015}, Bastiaan \cite{Bas:2015},
Tromp \cite{Tro:2015}, Garay et al. \cite{Gar:2017, Gar:2020}, Kiayias et al.
\cite{Kia:2016, Kia:2018}, Kiayias and Zindros \cite{Kia:2019}, Karakostas and
Kiayias \cite{Kar:2020}, and the references therein. On the other hand, an
economic incentive mechanism in Bitcoin is designed for a lot of miners
according to the shared proportion of their computing powers in solving the
cryptographic puzzle. Based on this, Satoshi Nakamoto \cite{Nak:2008} showed a
key blockchain characteristic: The fairness of PoW, i.e., as long as more than
50\% of the total mining power follows the PoW, the probability that a honest
miner can earn the total revenue (i.e., the block rewards and the transaction
fees) is proportional to his computing power.

In 2014, a seminal work was reported by Eyal and Sirer \cite{Eya:2014} (nearly
at the same time, also Bahack \cite{Bah:2013}), where Eyal and Sirer first
introduced an important concept: \textit{selfish mining}. Further, they set up
a Markov chain in order to study the selfish mining behavior. This leads to a
revenue analysis of the blockchain system. They reported an interesting
finding: The fairness of Bitcoin PoW can be destroyed by the selfish mining
attacks, i.e., the dishonest (or selfish) mining pool secretly withholds
blocks not to broadcast in the open P2P network if the number of blocks mined
by the selfish mining pool remains ahead of the honest mining pool. In this
case, the dishonest mining pool continues to mine on the top of the
selfish-mining block branch such that the dishonest mining pool can earn more
than his fair amount of mining profit. Since then, the Markov chain method of
Eyal and Sirer \cite{Eya:2014} has been widely applied in the literature in
order to discuss the blockchain selfish mining, for example, stubborn mining
by Nayak et al. \cite{Nay:2016}, Wang et al. \cite{WanL:2019} and Liu et al.
\cite{LiuH:2020}; Ethereum (and smart contract) by Niu and Feng
\cite{Niu:2019}; multiple mining pools (or miners) by Liu et al.
\cite{Liu:2018} and Jain \cite{Jai:2019}; multi-stage blockchain by Chang et
al.\cite{Cha:2019}; among others. However, no one has wondered or explained
whether the Markov chain method of Eyal and Sirer \cite{Eya:2014} makes sense
up to now. This motivates us in this paper to set up a pyramid Markov process
to answer such a basic question.

To further understand selfish mining, we introduce two practical factors: The
efficiency-increased ratio, and the mining rate of jumping miners from the
honest mining pool to the dishonest mining pool (the jumping's mining rate, in
short). Note that the efficiency-increased ratio is used to measure the
improved degree of mining efficiency of the dishonest mining pool, while the
jumping's mining rate denotes the social reputation and influence of the
dishonest mining pool. Ulteriorly, to describe the block-pegging process
clearly, we propose a two-block leading competitive criterion. We design a
block-detained probability sequence to express the block-pegging
decision-making of the dishonest mining pool. To our best knowledge, this
paper is the first to add the three practical factors of competitive advantage
into the selfish mining scenario.
In addition, we find a key phenomenon from the interaction effect between the
block-pegging and mining processes: No block is mined during the block-pegging
process, that is, the block-pegging and mining processes must be separated
from two connected time intervals. This is because each block is generated by
means of finding a nonce through solving a cryptographic puzzle of using all
the foregoing information of that blockchain in front of this block. If the
block-pegging and mining processes are not separated, then the cryptographic
puzzle can be possibly chaotic for the cryptographic puzzle design in a block
by block mining environment.

For our more general model of blockchain selfish mining, we set up a pyramid
Markov process to express the mining competition between the dishonest and
honest mining pools, as seen in the two block branches forked at the common
tree root. Note that a key factor of forming the pyramid Markov process is our
important finding that no block is mined during the block-pegging process. The
pyramid Markov process seems fairly complicated. Fortunately, we can easily
show that the pyramid Markov process is irreducible and positive recurrent,
and its stationary probability is matrix-geometric with an explicitly
representable rate matrix. Since our matrix-geometric solution is simpler than
that in Neuts \cite{Neu:1981} for Markov chains of GI/M/1 type, we provide a
unified theoretical framework of pyramid Markov processes, which is applicable
to a wide range of blockchain selfish mining. In addition, from the pyramid
Markov process, we build three approximative Markov processes when the system
states are described as the difference of block numbers on the two block
branches forked at the common tree root. Also, under no network latency, we
provide some new interesting interpretation on the Markov chain (see Figure 1)
and the revenue analysis (see (1) to (3)) given in Eyal and Sirer
\cite{Eya:2014}. Specifically, we show that the Markov chain of Eyal and Sirer
\cite{Eya:2014} is a rough approximation, and there is no valid mathematical
theory behind it.

It is necessary and useful to design the block-leading number (i.e., the
difference of two block numbers in the two forked block branches at the common
tree root). Eyal and Sirer \cite{Eya:2014} used a two-block leading
competitive criterion; while G\H{o}bel et al. \cite{Gob:2016} studied a
one-block leading competitive criterion. In this paper, we focus on the
two-block leading competitive criterion and set up a pyramid Markov process
for our more general model of blockchain selfish mining. From the pyramid
Markov process, it is easy to see that our technique can easily extend and
generalize to a $K$-block leading competitive criterion. This motivates us to
find an optimal number $K^{\ast}$ in time from a revenue management
perspective such that the $K^{\ast}$-block leading competitive criterion can
be adaptively used in the selfish mining scenario.

From the two-dimensional Markov processes, our paper is closely related to
G\H{o}bel et al. \cite{Gob:2016}. However, our pyramid Markov process is
distinguished from that in G\H{o}bel et al. \cite{Gob:2016} in the following
aspects. First, we study the two-block leading competitive criterion, while
G\H{o}bel et al. \cite{Gob:2016} discussed the one-block leading competitive
criterion. This leads to the different starting states of pegging blocks, as
seen in the pyramid Markov process. Second, we assume that the block-pegging
and mining processes must be separated from two connected time intervals, such
that the two-dimensional Markov process is of pyramid type; while in G\H{o}bel
et al. \cite{Gob:2016}, these mining processes can still continue during the
block-pegging period. This leads to that the two-dimensional Markov process of
G\H{o}bel et al. \cite{Gob:2016} is more complicated and difficult than our
pyramid Markov process. At the same time, we show that, if the block-pegging
and mining processes are not separated, then the cryptographic puzzle can be
possibly chaotic for the cryptographic puzzle design in the block by block
mining environment. Thus it is an important finding that the block-pegging and
mining processes can not synchronously go ahead. Third, our pyramid Markov
process is matrix-geometric; while the stationary probability vector of
G\H{o}bel et al. \cite{Gob:2016} is general and complicated, although
G\H{o}bel et al. \cite{Gob:2016} and Javier and Fralix \cite{Jav:2020}
provided two effective methods to deal with the stationary probability vector.
Fourth, we provide three practical and useful factors (e.g., the
efficiency-increased ratio, the jumping's mining rate, and the block-detained
probability sequence) to support the competitive advantages of the dishonest
mining pool; while there is no competitive element in G\H{o}bel et al.
\cite{Gob:2016} such that, when the network latency disappears, all the
selfish miners become honest. Therefore, their simple assumptions would not be
reasonable and enough to support the study of blockchain selfish mining.

More realistically, it is interesting to consider the influence of many orphan
blocks on the waste of computing resource in the blockchain. So far, there
have been some qualitative analysis for the influence of orphan blocks, but
the quantitative research is limited. For the qualitative analysis, readers
may refer to Carlsten et al. \cite{Car:2016}, Velner et al. \cite{Vel:2017},
Stifter et al. \cite{Sti:2019}, Saad et al. \cite{Saa:2019}, Awe et al.
\cite{Awe:2020} and others. To our best knowledge, our pyramid Markov process
is a most effective quantitative technique to analyze the orphan blocks in the
blockchain. To this end, we use the stationary probability vector of the
pyramid Markov process to provide a detailed analysis for the performance
measures of the blockchain selfish mining models with a lot of orphan blocks.

From an economic analysis perspective. we further set up a pyramid Markov
\textit{reward} process to study the revenue analysis of our more general
model of blockchain selfish mining. Based on this, we express the long-run
average profits of the honest and dishonest mining pools, respectively. Also,
we find that the two long-run average profits are multivariate linear in some
key parameters of the system. This enables us to find a simple sufficient
condition under which the blockchain can operate normally. Moreover, we can
measure the mining efficiency of the dishonest mining pool through comparing
to the honest mining pool. Therefore, we develop a unified theoretical
framework of pyramid Markov (reward) processes to open a new avenue, which can
support many potential promising research of blockchain selfish mining.

\vskip                    0.5cm

In summary, the main contributions of this paper are listed as follows:

\begin{itemize}
\item[1.] We describe a more general model of blockchain selfish mining with a
two-block leading competitive criterion and a new economic incentive mechanism
that captures three key factors: The efficiency-increased ratio, the jumping's
mining rate, and a block-detained probability sequence. (Section 3)

\item[2.] We set up a pyramid Markov process to express the dynamics of the
selfish mining, and show that the pyramid Markov process is irreducible and
positive recurrent, and the stationary probability vector is matrix-geometric
with an explicitly representable rate matrix. (Section 4) Also, we use the
stationary probability vector to provide an effective method to analyze the
influence of many orphan blocks on the waste of computing resource of the
blockchain. (Section 5)

\item[3.] We establish a pyramid Markov reward process to investigate the
long-run average profits of the honest and dishonest mining pools,
respectively. Also, we show that the long-run average profits are multivariate
linear in some key parameters. This leads to a sufficient condition under
which the blockchain operates normally. Moreover, we can measure the mining
efficiency of the dishonest mining pool. (Section 6)

\item[4.] We study the transient mining profits in the time interval $[0,t)$
by means of the PH distribution of infinite sizes and the associated PH
renewal process. To our best knowledge, this paper is the first to establish
the transient mining profits in the study of blockchain selfish mining.
(Section 7)

\item[5.] We build three approximative Markov processes when the system states
are taken as the difference of block numbers on the two forked branches at the
common tree root. (Section 8) Further, we discuss a special case without
network latency. Fortunately, by the three approximative Markov processes, we
are able to provide some new interesting interpretation on the Markov chain
and the revenue analysis of Eyal and Sirer \cite{Eya:2014}. We show that the
Markov chain of Eyal and Sirer \cite{Eya:2014} is a rough approximation, and
there is no valid mathematical theory behind it. (Section 9)
\end{itemize}

In addition, we provide the relevant literature in Section 2. In Section 10,
we use some numerical examples to verify the correctness and computability of
our theoretical results. Finally, Section 11 provides some useful concluding remarks.

\section{Relevant Literature}

In this section, we summarize the literature of blockchain selfish mining in
the last six years. At the same time, we relate our paper to three lines of
research by means of a simple classification of these literature.

We first explain the reason why the selfish mining pools can be formed and
developed rapidly. When a block is managed by a miner who can provide the
nonce, the miner receives two parts of profits: The \textit{block reward} and
the \textit{transaction fee}. On the one hand, under such a profit incentive,
each miner is willing to compete in producing and broadcasting a block in an
open P2P network, and hope to peg the block on the blockchain such that he can
receive the profit (the block reward and transaction fee) as much as possible.
If the electricity price is low and the mining profit is high, then more and
more people would like to join the mining process. On the other hand, since
the total number of generating blocks is controlled at an average rate of one
every ten minutes, the probability of individual miner generating and pegging
a block becomes lower and lower, as the number of joined miners increases.
This greatly increases the mining risk of each individual miner. In this
situation, some miners willingly form a \textit{selfish mining pool}. So far,
developing such selfish mining pools has become increasingly widespread, e.g.,
see Lee and Kim \cite{Lee:2019, Lee:2019}. Figure 1 introduces and compares
the best and biggest Bitcoin mining pools in the world in 2020.

\begin{figure}[h]
\centering                    \includegraphics[width=11cm]{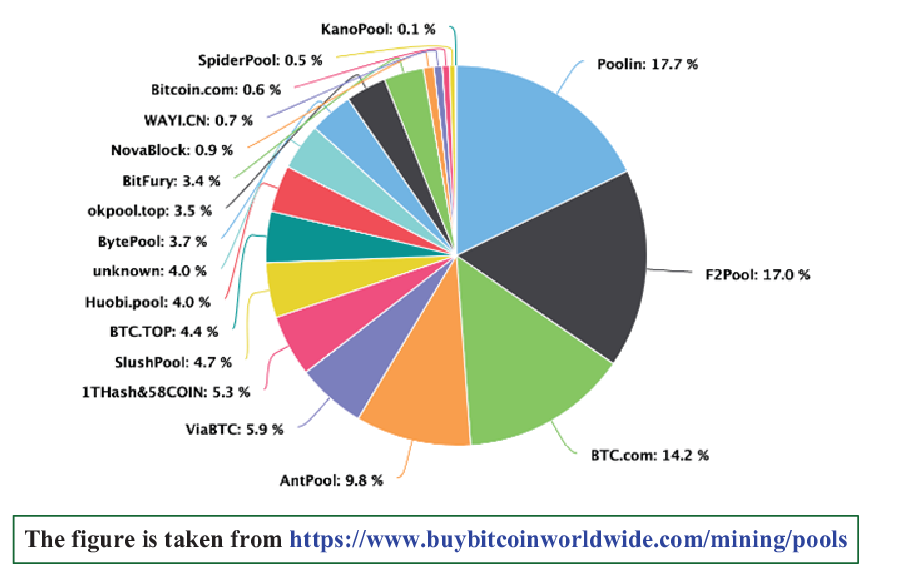}
\caption{The best and biggest Bitcoin mining pools in the world in 2020}%
\label{figure:Fig-0}%
\end{figure}

Now, we discuss that our paper is related to three lines of research on the
blockchain selfish mining, and some available methodologies and results are
also listed in details.

The first research stream started with discussions on the selfish mining
attacks, which can destroy the fairness of Bitcoin PoW. For the blockchain
with two mining pools, Eyal and Sirer \cite{Eya:2014} set up a Markov chain to
express the dynamic of the selfish mining attacks, and also designed a fairly
simple revenue analysis for observing how the selfish mining attacks can
influence on the profit allocation between the honest and dishonest mining
pools. After then, following the Markov chain method of Eyal and Sirer
\cite{Eya:2014}, some researchers extended and generalized such a similar
discussion for some attack strategies of blockchain, among which important
examples include stubborn mining by Nayak et al. \cite{Nay:2016}, Wang et al.
\cite{WanL:2019} and Liu et al. \cite{LiuH:2020}; Ethereum (and smart
contract) by Niu and Feng \cite{Niu:2019}; multiple mining pools (or miners)
by Leelavimolsilp et al. \cite{Leel:2018, Leel:2019}, Liu et al.
\cite{Liu:2018} and Jain \cite{Jai:2019}; multi-stage blockchain by Chang et
al. \cite{Cha:2019}; no block reward by Carlsten et al. \cite{Car:2016}; Power
adjusting by Gao et al. \cite{Gao:2019}; extending Eyal and Sirer's Markov
chain by Gervais et al. \cite{Ger:2015}, Mwale \cite{Mwa:2016}, Moustapha
\cite{Mou:2018}, Mulser \cite{Mul:2018}, Bai et al. \cite{Bai:2019},
Marmolejo-Coss\'{\i}o et al. \cite{Mar:2019}, Dong et al. \cite{Don:2019}, Lee
and Kim \cite{Lee:2018} and Liu et al. \cite{LiuR:2019}. Different from the
previous literature, this paper provides another interesting research
perspective: Does the Markov chain method of Eyal and Sirer \cite{Eya:2014}
make sense? For this purpose, we develop a new theoretical framework of
pyramid Markov processes, and show that the Markov chain of Eyal and Sirer
\cite{Eya:2014} is a rough approximation, and there is no valid mathematical
theory behind it.

G\H{o}bel et al. \cite{Gob:2016} set up a two-dimensional Markov chain to
study the selfish mining attacks. Javier and Fralix \cite{Jav:2020} further
provided a new computational analysis for the stationary probability vector of
the two-dimensional Markov chain given in G\H{o}bel et al. \cite{Gob:2016}.
Our paper is closely related to G\H{o}bel et al. \cite{Gob:2016} from the
two-dimensional structure of Markov chains. Distinctively, we find that the
block-pegging and mining processes must be separated from two connected time
intervals, thus we can establish a pyramid Markov process which is simpler
than that of G\H{o}bel et al. \cite{Gob:2016}, and show that the stationary
probability vector is matrix-geometric with an explicitly representable rate matrix.

The probability theory of blockchain selfish mining is further developed. On
the one hand, Grunspan and P\'{e}rez-Marco \cite{Gru:2018, GruP:2018,
GruPR:2018, Gru:2019} applied martingale to analyze the profitability of
selfish mining, stubborn mining, trailing mining, and Ethereum. Albrecher and
Goffard \cite{Alb:2020} studied the profitability of blockchain selfish mining
by means of the theory of ruin. On the other hand, to dynamically optimize the
selfish mining strategies, the Markov decision processes have been applied
successfully. Important examples include Sapirshtein et al. \cite{Sap:2016},
Sompolinsky and Zohar \cite{Som:2015}, Gervais et al. \cite{Ger:2016},
W\"{u}st \cite{Wus:2016} and Gupta \cite{Gup:2020}.

Our paper is related to the second research stream of blockchain selfish
mining, while their mathematical analysis has still been relatively weak or
more difficult up to now. This greatly motivates us to open up potentially
interesting research on the different directions, including applications of
our pyramid Markov processes. In what follows, we classify the literature into
seven different research directions:

\textbf{(a)} \textit{Generalizations of selfish mining strategies. }Important
examples include the subversive miner strategy by Courtois and Bahack
\cite{Cou:2014}; and the stubborn mining strategy by Nayak et al.
\cite{Nay:2016}. In addition, Lee and Kim \cite{Leel:2019} developed a
detective mining by means of the information of miners.

\textbf{(b)} \textit{The puzzle difficulty. }Note that the Bitcoin miners
competitively solve a PoW puzzle to find the nonce. Once such a nonce is
found, a Bitcoin block can immediately be generated and pegged on the
blockchain. See Nakamoto \cite{Nak:2008}, Narayanan et al. \cite{Nar:2016},
Miller et al. \cite{Mil:2015}, G\H{o}bel et al. \cite{Gob:2016}, Kraft
\cite{Kra:2016} and Davidson and Diamond \cite{Dav:2020}. The difficulty level
used to discover a new block can be constantly adjusted such that, on average,
one block is expected to be discovered every 10 minutes. Readers may refer to,
such as block arrivals by G\H{o}bel et al. \cite{Gob:2016}; zeroblock by Solat
and Potop-Butucaru \cite{SolP:2016, Sol:2016}; smartpool by Luu et al.
\cite{Luu:2017}; difficulty control by Fullmer and Morse \cite{Ful:2018};
pooled mining by Lee and Kim \cite{Lee:2018}; unfairness of blockchain by
Guerraoui and Wang \cite{Gue:2018}; multi-stage blockchain by Chang et al.
\cite{Cha:2019}; bobtail by Bissias and Levine \cite{Bis:2020}; and so on.
Note that, when the difficulty level is changed continuously, the Markov
process of blockchain selfish mining is nonhomogeneous, thus such a transient
analysis is more difficult.

\textbf{(c)} \textit{The forked structure.}\textbf{ }When there exist more
than one mining pools in blockchain (see Leelavimolsilp et al. \cite{Lee:2019,
Lee:2018}, Liu et al. \cite{Liu:2018} and Jain \cite{Jai:2019}), it is
possible that two or more blocks with the same preceding block (parent block)
can be produced and form block branches forked at the preceding block. This
leads to a forked structure starting from the parent block, as the length (or
height) of blockchain increases. Readers may refer to Section 4 of Eyal and
Sirer \cite{Eya:2014} for some explanatory examples. To enhance the
scalability and transaction throughput of blockchain, Sompolinsky and Zohar
\cite{Som:2015} generalized the simple forked tree (i.e., several block
branches forked at a common tree root) to a more general tree with multiple
branching points (called GHOST), and further to a Direct Acyclic Graph
(abbreviated DAG), e.g., see Kiayias and Panagiotakos \cite{Kia:2017}, Lee
\cite{LeeJ:2018} and Choi et al. \cite{Cho:2018}. Typically, it is interesting
and challenging to study the blockchain selfish mining with multiple mining
pools whose forked structure is GHOST or DAG. In this case, our pyramid Markov
processes are related to the fluid and diffusion approximations due to dealing
with a multi-dimensional stochastic system.

\textbf{(d)} \textit{The orphan blocks and uncle blocks.}\textbf{ }In the
blockchain selfish mining, the orphan blocks and uncle blocks can lead to the
waste of computing power resource by means of the longest chain or heaviest
path. Readers may refer to the orphan blocks by Carlsten et al.
\cite{Car:2016}, Velner et al. \cite{Vel:2017}, Stifter et al. \cite{Sti:2019}%
, Saad et al. \cite{Saa:2019} and Awe et al. \cite{Awe:2020}; and the uncle
blocks (in Ethereum) by\textbf{ }Gervais et al. \cite{Ger:2016}, Ritz and
Zugenmaier \cite{Rit:2018}, Niu and Feng \cite{Niu:2019}, Werner et al.
\cite{Wer:2019}, Wang et al. \cite{WanL:2019}, Chang et al. \cite{Chang:2019}
and Liu et al. \cite{LiuH:2020}. It is worthwhile to note that our pyramid
Markov processes are successful and applicable in the study of orphan blocks\textit{.}

\textbf{(e) }\textit{Detection of selfish mining. }In a blockchain, it is
possible to have some different classes of attacks, such as 51\% attack,
selfish mining attack, stalker attack, eclipse attack, physical attack and so
forth. In this case, an interesting topic is how to detect the different
attacks effectively, and to be able to put forward some defensive measures for
the attacks. Important examples include Liu et al. \cite{Liu:2019}, Saad et
al. \cite{SaaN:2019}, Lee and Kim \cite{Lee:2019} and Chicarino et al.
\cite{Chi:2020}.

\textbf{(f) }\textit{Defense against selfish mining.}\textbf{ }The selfish
mining attack is a fundamental challenge in the study of blockchain, since it
breaks the fairness of blockchain (see Guerraoui and Wang \cite{Gue:2018}) and
poses potential threats to the decentralized structure. Thus it is very
important to defense against and prevent the selfish mining, readers may refer
to, such as Zhang \cite{Zha:2015}, Solat and Potop-Butucaru \cite{Sol:2016,
SolP:2016} and Zhang and Preneel \cite{Zhan:2017}.

\textbf{(g) }\textit{The double spending attacks.} The double spending attacks
have a stronger relation with the selfish mining\textit{. }Note that the
block-pegging time can increase due to the network delay (see Decker and
Wattenhofer \cite{Dec:2013}), thus the released blocks will not be confirmed
immediately. In this situation, the selfish mining attacks can make the double
spending attacks, e.g., see Karame et al. \cite{Kar:2012}, Rosenfeld
\cite{Ros:2014} and Javarone and Wright \cite{Jav:2018} for more details.

Other research is also mentioned here, for example, the data and experiment
analysis by Wright \cite{Wri:2017}, Wright and Savanah \cite{WriS:2017},
Eijkel and Fehnker \cite{Eij:2019} and Kedziora et al. \cite{Ked:2020}.

Finally, our work is related to the third research stream in the study of
consensus protocols. So far, the consensus protocols (or algorithms) have been
the core of blockchain development. Normally, the consensus protocol is used
to determine the generation, storage and validation of data, it sets up
operations mode of blockchain and determines performance and security of
blockchain. Now, the frequently used consensus protocols include two different
types: A consensus protocol based on Byzantine fault tolerance algorithm
(BFT), and another consensus protocol based on PoW / Proof of Stake (PoS). The
BFT consensus protocol is usually used in smaller \textit{private chains} or
\textit{alliance chains}; while the PoW/PoS consensus protocol is suitable for
large-scale \textit{public chains}. Furthermore, following BFT, PoW and PoS,
so far over fifty consensus protocols have been developed with advantage of
special needs. Readers may refer to survey papers by, for example, Bissias et
al. \cite{Bis:2016}, Cachin et al. \cite{Cac:2017}, Nguyen and Kim
\cite{Ngu:2018}, Wang et al. \cite{Wan:2019}, Xiao et al. \cite{Xia:2019},
Bano et al. \cite{Ban:2019} and Natoli et al. \cite{Nat:2019}, Huang et al.
\cite{Hua:2020} and fan et al. \cite{Fan:2020}. Recently, it is interesting
and challenging to develop the theory of Markov processes (more generally,
probability theory), including our pyramid Markov processes, in the study of
consensus protocols.

\section{Model Description}

In this section, we describe a more general model of blockchain selfish mining
with two different mining pools: The dishonest mining pool, and the honest
mining pool (i.e., a virtual pool with all the honest miners). Also, the
blockchain selfish mining is controlled by not only a two-block leading
competitive criterion but also an economic incentive mechanism that captures
three types of key parameters: The efficiency-increased ratio $\Re$, the
jumping's mining rate $\gamma$, and the block-detained probability sequence
$\left\{  p_{k}:k=2,3,4,\ldots\right\}  $. In addition, we introduce some
mathematical notation used in our later study.

To avoid the 51\% attack, we assume that the computing power of the honest
mining pool is more than half of the total computing power of the blockchain,
that is, the honest miners are in the majority while the dishonest miners are
in the minority. Note that the honest mining pool follows the two-block
leading competitive criterion, while the dishonest mining pool follows the
selfish mining attack strategy, which is a modification of the two-block
leading competitive criterion through using the block-detained probability
sequence $\left\{  p_{k}:k=2,3,4,\ldots\right\}  $. See Figure 2 for an
intuitive understanding.

\begin{figure}[h]
\centering                    \includegraphics[width=15cm]{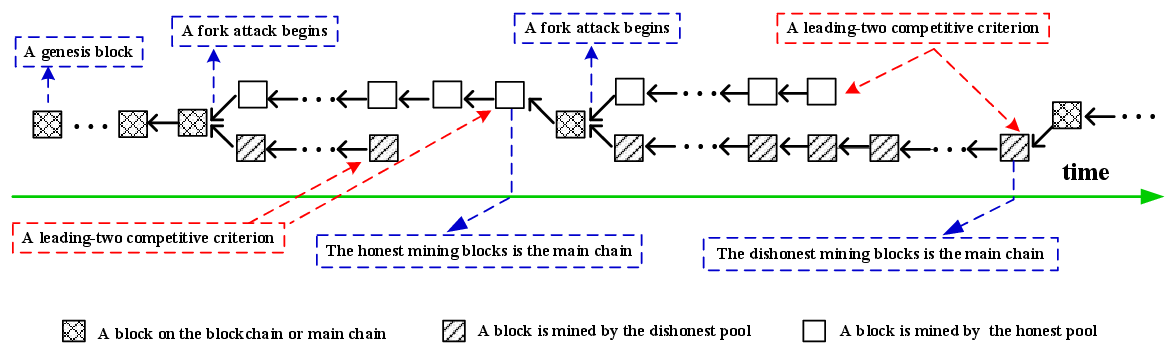}
\caption{The two-block leading competitive criterion}%
\label{figure:Fig-2}%
\end{figure}

In the blockchain selfish mining, the honest and dishonest mining pools
compete fiercely in finding the nonce (i.e., solving the cryptographic puzzle
to give the nonce) to generate the blocks, and they publish the blocks to make
two block branches forked at a common tree root (parent block). For the two
honest and dishonest mining pools under the two-block leading competitive
criterion, by observing the two block branches forked at the common tree root,
we define\textit{ the main chain }as the longer block branch; while another
shorter block branch is called \textit{the chain of orphan blocks}.

In the blockchain, every external transaction first needs to be checked by
paying a certain handling expense (called the transaction fee), and it is sent
to the transaction pool. Then some transactions are randomly taken from the
transaction pool to generate a block, e.g., see Li et al. \cite{Li:2018,
Li:2019} for an intuitive interpretation. On the other hand, the transactions
of an orphan block are returned to the transaction pool but no additional
transaction fee is required again. For convenience of our analysis, we assume
that the transaction pool always has a large enough capacity.

In what follows, we provide some model descriptions for the blockchain selfish
mining as follows.

\textbf{(1) The block-generating processes: }We assume that the blocks mined
by the dishonest and honest mining pools have formed two block branches forked
at a common tree root, and the growths of the two block branches are two
Poisson processes with block-generating rates $\alpha>0$ and $\beta>0$, where
$\alpha=\widetilde{\alpha}\left(  1+\Re\right)  $, $\Re\geq0$ is the
efficiency-increased ratio of the dishonest mining pool; and $\widetilde
{\alpha}>0$ is regarded as a net mining rate when all the miners of the
dishonest mining pool become honest, $\beta$ is a net mining rate of the
honest mining pool.

\textbf{(2) The jumping's mining rate: }The dishonest mining pool has more
mining advantages than the honest mining pool, so it can attract some honest
miners to jump into the dishonest mining pool. Let $\gamma$ be the net mining
rate of such jumping miners. Then after jumping, the block-generating rates of
the dishonest and honest mining pools are given by $\left(  \widetilde{\alpha
}+\gamma\right)  \left(  1+\Re\right)  $ (i.e., $\alpha+$ $\gamma\left(
1+\Re\right)  $) and $\beta-\gamma$, respectively.

To overcome the $51\%$ attacks, it is necessary to limit the jumping's mining
rate $\gamma$. Let $\widetilde{\alpha}+$ $\gamma<\left(  1/2\right)  \left(
\widetilde{\alpha}+\beta\right)  $ and $\beta-\gamma>\left(  1/2\right)
\left(  \widetilde{\alpha}+\beta\right)  $. Then%
\begin{equation}
0\leq\gamma<\frac{1}{2}\left(  \beta-\widetilde{\alpha}\right)  .
\label{Equat-1}%
\end{equation}

From a practical economic perspective, the two key parameters $\Re$ and
$\gamma$ are well related to the dishonest mining pool's operations management
level, social reputation and influence, the reward of each excavated block,
and so on. Based on this, this paper is clearly different from those previous
works in the literature. Figure 3 depicts some relations between the honest
and dishonest mining pools.

\begin{figure}[h]
\centering                    \includegraphics[width=12cm]{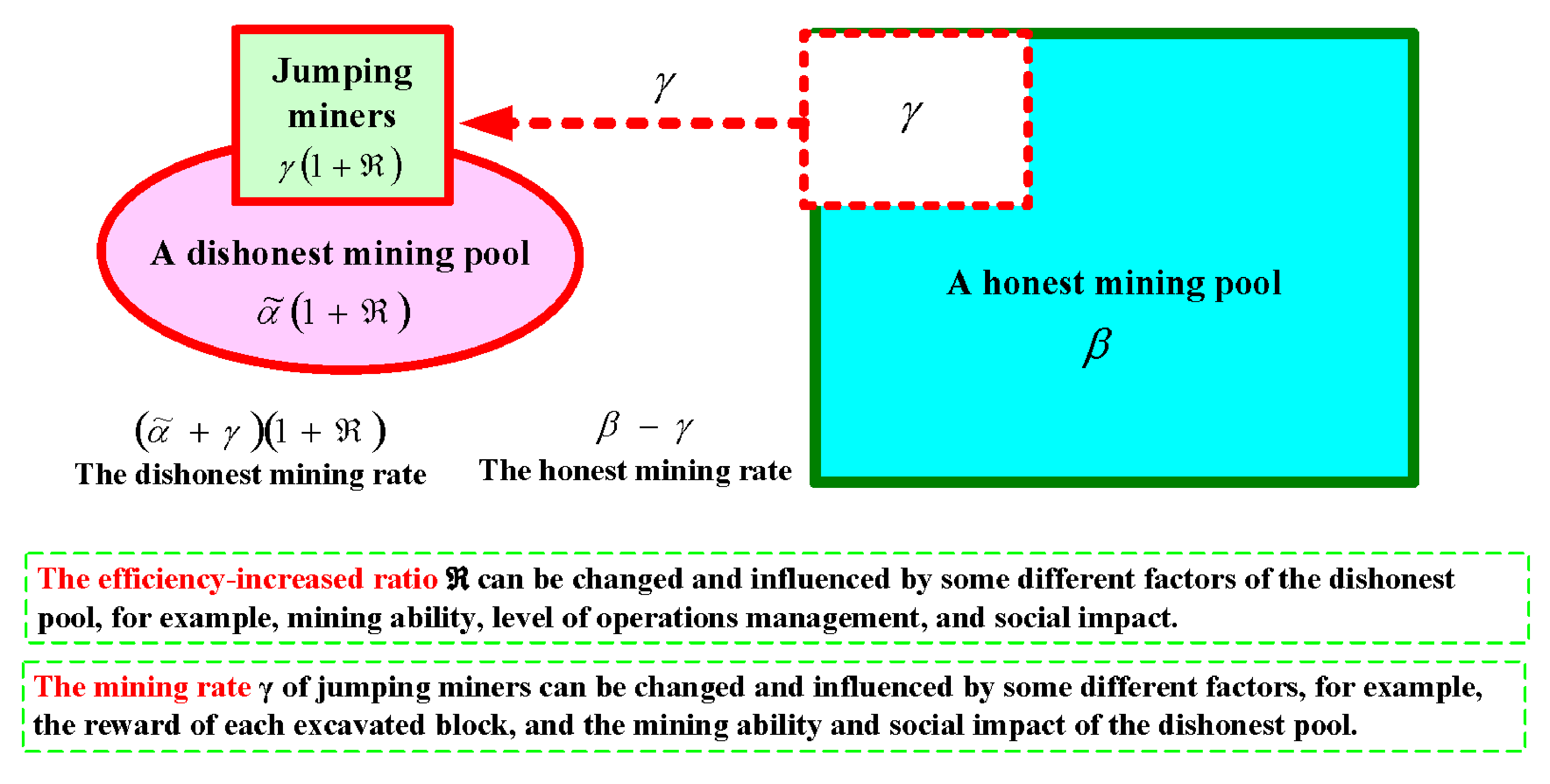}
\caption{Some relations between the honest and dishonest mining pools}%
\label{figure:Fig-1}%
\end{figure}

\textbf{(3) The block-pegging time of the main chain: }The network latency
always results in some delay for each block pegged on the blockchain. Thus the
block-pegging time is used to express the network latency, and it is a time
interval that begins at the broadcast time of a block until the block is
pegged on the blockchain. Currently, the block-pegging time contains two
different parts: The block broadcast time, and the verification time by
various miners. Also, the block broadcast time is always longer than the
verification time, because the verification time is very short.

Once a main chain is formed, then the mining process is terminated
immediately, and \textit{the whole main chain} is pegged on the blockchain. We
assume that the block-pegging time is independent and identically distributed
(i.i.d) and exponential with mean $1/\mu$.

Note that \textit{no new block can be generated during the block-pegging
process of the main chain}. Once the main chain is pegged on the blockchain,
then a new mining competition immediately starts on two new block branches
forked at another common tree root.

\textbf{(4) The return time of the orphan blocks: }Since an orphan block can
not be pegged on the blockchain, it has to return to the transaction pool. We
assume that the return time of each orphan block is exponential with mean
$1/\mu$.

\textbf{(5) A two-block leading competitive criterion:}

\textit{(a) The main chain by the honest mining pool.} Once the honest chain
of blocks is two blocks ahead of the dishonest chain of blocks, then the
honest chain of blocks is the main chain. In this case, the forked process of
two block branches ends immediately, and the whole main chain is pegged on the
blockchain. At the same time, all the blocks of the dishonest chain
immediately become orphan blocks, and all of them need to be returned to the
transaction pool without re-paying any new fee.

\textit{(b) The main chain by the dishonest mining pool.} Once the dishonest
chain of blocks is $k$ blocks ahead of the honest chain of blocks for $k\geq
2$, then the dishonest chain of blocks is taken as the main chain. Based on
this, two different cases are considered as follows:

\textbf{(b-1)} With probability $p_{k}$, the whole main chain is pegged on the
blockchain, and the forked process of two block branches ends immediately.
Also, all the blocks of the honest chain immediately become orphan blocks,
each of which is returned to the transaction pool.

\textbf{(b-2)} With probability $1-p_{k}$, the dishonest mining pool does not
broadcast and peg the main chain such that it continues to mine more blocks to
lengthen the main chain. To peg the main chain on the blockchain finally, we
assume that $\lim_{n\rightarrow\infty}p_{n}=1$.

\textbf{(6) The reward of a block generated and pegged on the blockchain:}
Once a block is generated and pegged on the blockchain, then the mining pool
of generating this block can obtain an appropriate amount of reward (or
compensation) from two different parts:

\textit{(a) A block reward }$r_{B}$\textit{ by the blockchain system}. When a
block is generated and pegged on the blockchain, the mining pool receives a
certain amount of block\textit{ }reward $r_{B}$.

\textit{(b) An average total transaction fee }$r_{F}$\textit{ in the block}.
It is possible that the transactions and their number contained in the blocks
are different from each other. For the sake of simplicity, we assume the
average total transaction fee for a block is $r_{F}$.

Obviously, $r_{B}+r_{F}$ is the total reward received by a mining pool who
generates and pegs the block on the blockchain.

Note that the two mining pools do not receive any block reward and transaction
fee from any orphan block.

\textbf{(7) The mining cost: }The mining cost of the blockchain system
contains two parts:

\textit{(a) The electric charge.} Let $c_{E}$ be the electric price per unit
of net mining rate and per unit time. Then the electric costs per unit time by
the dishonest and honest mining pools are given by $c_{E}\left(
\widetilde{\alpha}+\gamma\right)  $ and $c_{E}\left(  \beta-\gamma\right)  $, respectively.

\textit{(b) The administrative fee.} Let $c_{A}$ be the administrative price
per unit of net mining rate and per unit time. Then the administrative costs
per unit time by the dishonest and honest mining pools are given by
$c_{A}\left(  \widetilde{\alpha}+\gamma\right)  \left(  1+\Re\right)  $ and
$c_{A}\left(  \beta-\gamma\right)  $, respectively.

\textbf{Independence: }We assume that all the random variables defined above
are independent of each other.

\begin{Rem}
To support the selfish mining behavior effectively, the efficiency-increased
ratio $\Re$ measures the improvement of management ability of the dishonest
mining pool, the block-detained probability sequence $\left\{  p_{k}%
:k=2,3,4,\ldots\right\}  $ denotes the unfair competition strategy used by the
dishonest mining pool, and the jumping's mining rate $\gamma$ reflects the
social influence of the dishonest mining pool in the world.
\end{Rem}

\begin{Rem}
If $\lim_{n\rightarrow\infty}p_{n}<1$, then with probability $1-\lim
_{n\rightarrow\infty}p_{n}$, the dishonest mining pool does not broadcast the
mined blocks so that some main chains will become the chain of orphan blocks.
This case is clearly impractical from the mining purpose. Thus this paper will
not consider the case with $\lim_{n\rightarrow\infty}p_{n}<1$.
\end{Rem}

\begin{Rem}
(a) Note that such a block reward is halved after every $210,000$ blocks are
mined. The block reward was reduced from its initial level of $50$ Bitcoins to
$25$ Bitcoins in November $2012$, further reduced to $12.5$ Bitcoins in July
$2016$, and then again in July $2020$ to its current level of $6.25$ Bitcoins.
Finally, the block reward will be reduced $32$ more times before eventually
reaching zero sometime around $2140$. (b) As $r_{B}$ decreases after every
period of four years, designing a sufficient large transaction fee $r_{F}$
will be the main economic source to maintain the operations of a blockchain system.
\end{Rem}

\section{ A Pyramid Markov Process}

In this section, we set up a new pyramid Markov process to analyze the
blockchain selfish mining with a two-block leading competitive criterion and a
new economic incentive mechanism. Note that all the key factors or parameters
designed in the blockchain selfish mining are well related to the physical and
dynamic structure of the pyramid Markov process. In particular, we show that
the pyramid Markov process is irreducible and positive recurrent, and that the
stationary probability vector is matrix-geometric with an explicitly
representable rate matrix.

In the two block branches forked at a common tree root, we denote by $I(t)$
and $J(t)$ the numbers of blocks mined by the honest and dishonest mining
pools at time $t$, respectively. It is clear that $\left\{  \left(
I(t),J(t)\right)  :t\geq0\right\}  $ is a continuous-time Markov process whose
state space is given by%
\[
\Omega=\Omega_{\widetilde{\mathbf{0}}}\cup\left(  \bigcup\limits_{k=0}%
^{\infty}\Omega_{k}\right)  ,
\]
where%
\begin{align*}
\text{Level }\widetilde{\mathbf{0}}  &  :\Omega_{\widetilde{\mathbf{0}}%
}=\left\{  \left(  0,0\right)  \right\}  ,\\
\text{Level }0  &  :\Omega_{\mathbf{0}}=\left\{  \left(  0,1\right)  ,\left(
0,2\right)  ,\left(  0,3\right)  ,\ldots\right\}  ,\\
\text{Level }1  &  :\Omega_{\mathbf{1}}=\left\{  \left(  1,0\right)  ,\left(
1,1\right)  ,\left(  1,2\right)  ,(1,3),\ldots\right\}  ,\\
\text{Level }k  &  :\Omega_{k}=\left\{  \left(  k,k-2\right)  ,\left(
k,k-1\right)  ,\left(  k,k\right)  ,(k,k+1),\left(  k,k+2\right)
,\ldots\right\}  ,\text{ }k\geq2.
\end{align*}
Based on this, the state transition relation of the Markov process $\left\{
\left(  I(t),J(t)\right)  :t\geq0\right\}  $ is depicted in Figure 4. Note
that the Markov process $\left\{  \left(  I(t),J(t)\right)  :t\geq0\right\}  $
is of pyramid type, thus it is also called a pyramid Markov process (note that
such a pyramid form can become clearer in multiple mining pools).

\begin{figure}[h]
\centering        \includegraphics[width=16cm]{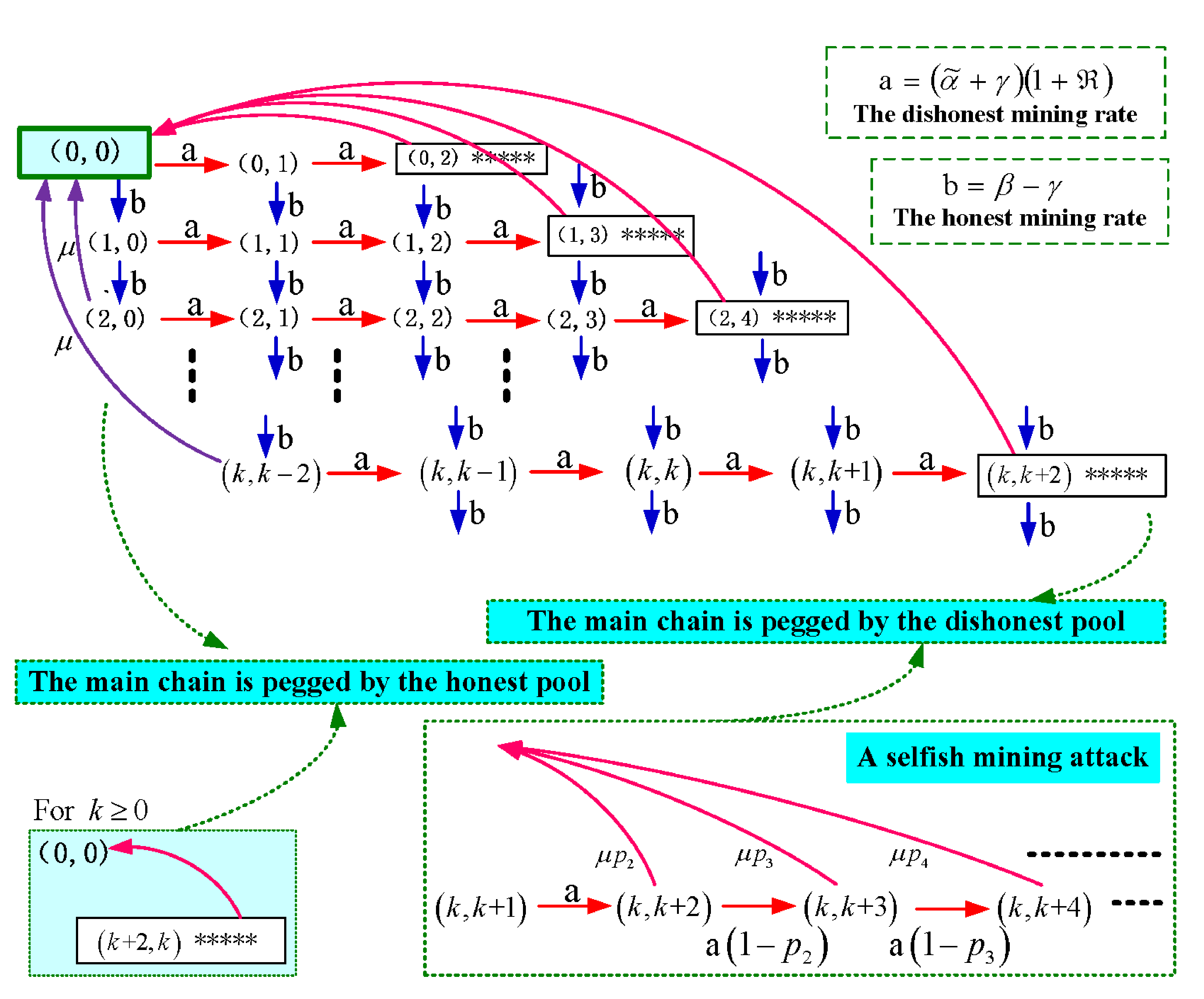}  \caption{The state
transition relation of the pyramid Markov process}%
\label{figure:Fig-3}%
\end{figure}

\begin{Rem}
State $\left(  0,0\right)  $ plays a key role in setting up the pyramid Markov
process. In fact, State $\left(  0,0\right)  $ describes the tree root as the
starting point of a fork attack, see Figure 4 for more details. If the pyramid
Markov process enters State $\left(  0,0\right)  $, then the fork attack ends
immediately, and the main chain is pegged on the blockchain.
\end{Rem}

By using Figure 4, the infinitesimal generator of the Markov process $\left\{
\left(  I(t),J(t)\right)  :t\geq0\right\}  $ is given by
\begin{equation}
Q=\left(
\begin{array}
[c]{ccccccc}%
Q_{\widetilde{\mathbf{0}},\widetilde{\mathbf{0}}} & Q_{\widetilde{\mathbf{0}%
},0} & Q_{\widetilde{\mathbf{0}},1} &  &  &  & \\
Q_{0,\widetilde{\mathbf{0}}} & Q_{0,0} & Q_{0,1} &  &  &  & \\
Q_{1,\widetilde{\mathbf{0}}} &  & Q_{1,1} & Q_{1,2} &  &  & \\
B &  &  & A & C &  & \\
B &  &  &  & A & C & \\
\vdots &  &  &  &  & \ddots & \ddots
\end{array}
\right)  , \label{Equat-2}%
\end{equation}
where $a=\left(  \widetilde{\alpha}+\gamma\right)  \left(  1+\Re\right)  $,
$b=\beta-\gamma$, $\xi_{k}=a\left(  1-p_{k}\right)  +b+\mu p_{k}$ for $k\geq
2$,%
\[
Q_{\widetilde{\mathbf{0}},\widetilde{\mathbf{0}}}=-\text{ }\left(  a+b\right)
,Q_{\widetilde{\mathbf{0}},0}=\left(  a,0,0,\ldots\right)  ,Q_{\widetilde
{\mathbf{0}},1}=\left(  b,0,0,\ldots\right)  ;
\]%
\[
Q_{0,\widetilde{\mathbf{0}}}=\left(
\begin{array}
[c]{c}%
0\\
\mu p_{2}\\
\mu p_{3}\\
\mu p_{4}\\
\vdots
\end{array}
\right)  ,Q_{0,1}=\left(
\begin{array}
[c]{cccccc}%
0 & b &  &  &  & \\
& 0 & b &  &  & \\
&  & 0 & b &  & \\
&  &  & 0 & b & \\
&  &  &  & \ddots & \ddots
\end{array}
\right)  ,
\]%
\[
Q_{0,0}=\left(
\begin{array}
[c]{ccccc}%
-\text{ }\left(  a+b\right)  & a &  &  & \\
& -\xi_{2}\text{ } & a\left(  1-p_{2}\right)  &  & \\
&  & -\text{ }\xi_{3} & a\left(  1-p_{3}\right)  & \\
&  &  & -\text{ }\xi_{4} & \ddots\\
&  &  &  & \ddots
\end{array}
\right)  ;
\]%
\[
Q_{1,\widetilde{\mathbf{0}}}=\left(
\begin{array}
[c]{c}%
0\\
0\\
0\\
\mu p_{2}\\
\mu p_{3}\\
\mu p_{4}\\
\vdots
\end{array}
\right)  ,Q_{1,2}=\left(
\begin{array}
[c]{ccccccc}%
b &  &  &  &  &  & \\
& b &  &  &  &  & \\
&  & b &  &  &  & \\
&  &  & b &  &  & \\
&  &  &  & b &  & \\
&  &  &  &  & b & \\
&  &  &  &  &  & \ddots
\end{array}
\right)  ,
\]%
\[
Q_{1,1}=\left(
\begin{array}
[c]{ccccccc}%
-\text{ }\left(  a+b\right)  & a &  &  &  &  & \\
& -\text{ }\left(  a+b\right)  & a &  &  &  & \\
&  & -\text{ }\left(  a+b\right)  & a &  &  & \\
&  &  & -\xi_{2} & a\left(  1-p_{2}\right)  &  & \\
&  &  &  & -\xi_{3} & a\left(  1-p_{3}\right)  & \\
&  &  &  &  & \ddots & \ddots
\end{array}
\right)  ;
\]%
\[
B=\left(
\begin{array}
[c]{c}%
\mu\\
0\\
0\\
0\\
\mu p_{2}\\
\mu p_{3}\\
\vdots
\end{array}
\right)  ,C=\left(
\begin{array}
[c]{ccccccc}%
0 &  &  &  &  &  & \\
b & 0 &  &  &  &  & \\
& b & 0 &  &  &  & \\
&  & b & 0 &  &  & \\
&  &  & b & 0 &  & \\
&  &  &  & b & 0 & \\
&  &  &  &  & \ddots & \ddots
\end{array}
\right)  ,
\]%
\[
A=\left(
\begin{array}
[c]{cccccccc}%
-\text{ }\left(  a+\mu\right)  & a &  &  &  &  &  & \\
& -\text{ }\left(  a+b\right)  & a &  &  &  &  & \\
&  & -\text{ }\left(  a+b\right)  & a &  &  &  & \\
&  &  & -\text{ }\left(  a+b\right)  & a &  &  & \\
&  &  &  & -\xi_{2} & a\left(  1-p_{2}\right)  &  & \\
&  &  &  &  & -\xi_{3} & a\left(  1-p_{3}\right)  & \\
&  &  &  &  &  & \ddots & \ddots
\end{array}
\right)  .
\]

In the above matrices, we only give the non-zero elements, while all the zero
elements are empty with an easy understanding from the context of the state
space $\Omega$.

\begin{The}
The pyramid Markov process $Q$ must be irreducible and positive recurrent.
\end{The}

\textbf{Proof.} It is easy to see from Figure 4 that the pyramid Markov
process $Q$ is irreducible, since for any two states $\left(  n_{1}%
,n_{2}\right)  $ and $\left(  m_{1},m_{2}\right)  $, there must exist a state
transition path such that the pyramid Markov process $Q$ can arrive at state
$\left(  m_{1},m_{2}\right)  $ from state $\left(  n_{1},n_{2}\right)  $.

To prove the stability, it is a little bit complicated. Here, we use the
\textit{double} mean drift method for dealing with not only each level but
also the whole process., e.g., see Chapter 3 of Li \cite{Li:2010}.

\begin{figure}[h]
\centering                    \includegraphics[width=14cm]{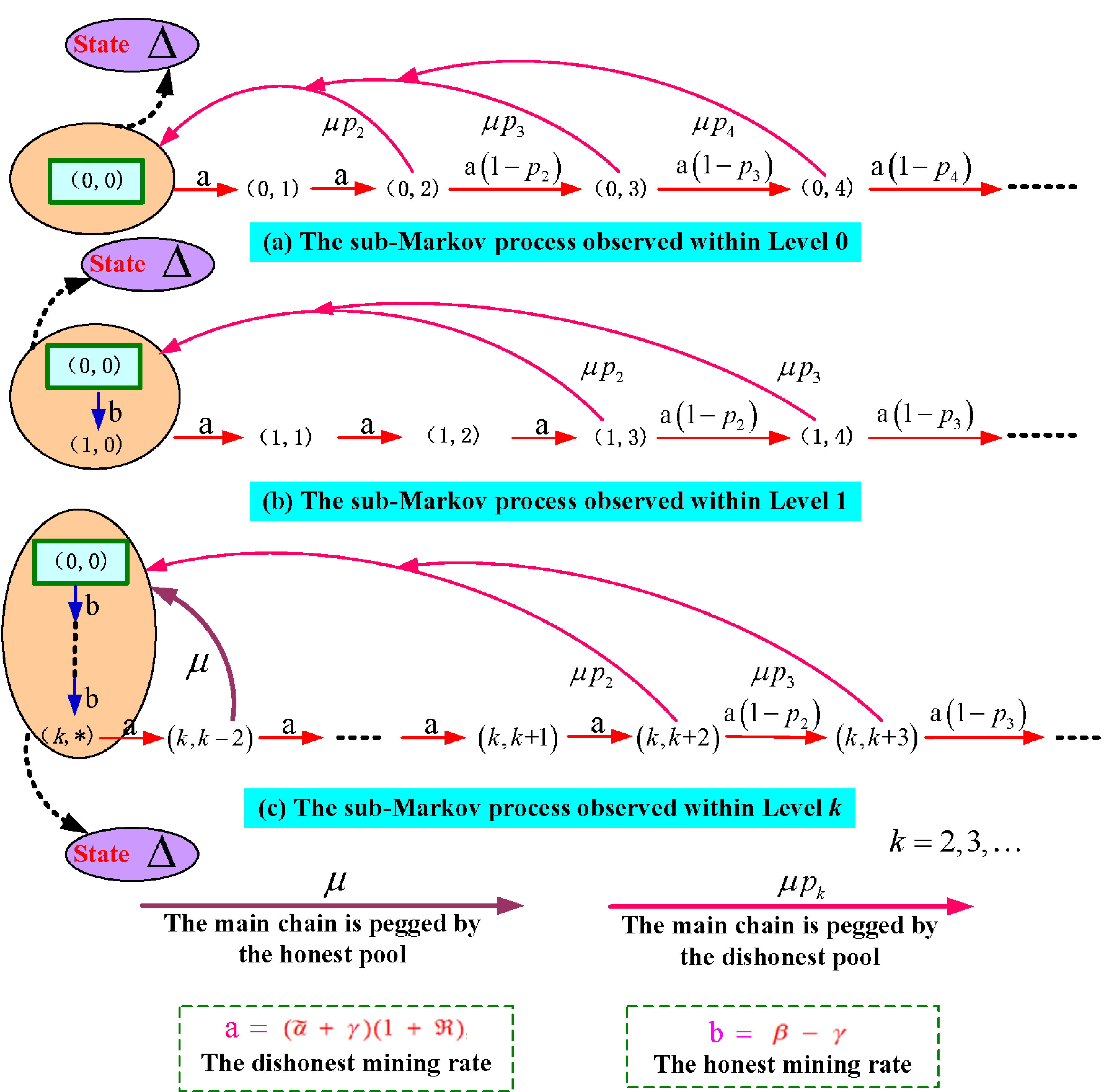}
\caption{Three different sub-Markov processes }%
\label{figure:Fig-4}%
\end{figure}

We first prove the sub-Markov process only observed within Level $k$ is
positive recurrent for $k=0,1,2,\ldots$. From Figure 5, our analysis is to
consider the following two different cases:

\textbf{Case one:} The sub-Markov process $\left\{  J\left(  t\right)
,I\left(  t\right)  =0\text{ or }1:t\geq0\right\}  $ observed within Level $0$
or Level $1$. Here, we only discuss the case with Level $0$, while another
case with Level $1$ can be dealt with similarly.

It is easy to see from (a) of Figure 5 that the sub-Markov process $\left\{
J\left(  t\right)  ,I\left(  t\right)  =0:t\geq0\right\}  $ observed within
Level $0$ has the state space
$\Omega$%
$_{\text{Level }0}=\left\{  \Delta,1,2,3,\ldots\right\}  $, and its
infinitesimal generator is given by%
\[
\mathbf{Q}_{\text{Level }0}=\left(
\begin{array}
[c]{cccccc}%
-a & a &  &  &  & \\
0 & -a & a &  &  & \\
\mu p_{2} &  & -\left[  \mu p_{2}+a\left(  1-p_{2}\right)  \right]  & a\left(
1-p_{2}\right)  &  & \\
\mu p_{3} &  &  & -\left[  \mu p_{3}+a\left(  1-p_{3}\right)  \right]  &
a\left(  1-p_{3}\right)  & \\
\vdots &  &  &  & \ddots & \ddots
\end{array}
\right)  .
\]
Here, we apply the mean drift method to consider the stability of the
sub-Markov process $\mathbf{Q}_{\text{Level }0}$. To this end, it is observed
on state $k$ that the mean drift rate of moving left to state $\Delta$ is
given by $k\cdot\mu p_{k}$, where $k$ denotes the $k$-step state transitions;
while the mean drift rate of moving right to state $k+1$ is given by $1\cdot
a\left(  1-p_{k}\right)  $, where $1$ denotes the $1$-step state transitions.
Note that $\lim_{k\rightarrow\infty}p_{k}=1$, it is clear that $\lim
_{k\rightarrow\infty}k\cdot\mu p_{k}=\lim_{k\rightarrow\infty}k\mu=\infty$ and
$\lim_{k\rightarrow\infty}1\cdot a\left(  1-p_{k}\right)  =0$. This shows that
there exists a sufficient large positive integer $K$ such that for any $n>K$,%
\[
n\cdot\mu p_{n}>1\cdot a\left(  1-p_{n}\right)  ,
\]
that is, for any $n>K$, the mean drift rate of moving left to state $\Delta$
is bigger than the mean drift rate of moving right to state $n+1$. Therefore,
the sub-Markov process $\mathbf{Q}_{\text{Level }0}$ is stable.

Similarly, the sub-Markov process $\mathbf{Q}_{\text{Level }1}$ is stable by
means of (b) of Figure 5.

\textbf{Case two:} The sub-Markov process $\left\{  J\left(  t\right)
,I\left(  t\right)  =k:t\geq0\right\}  $ observed within Level $k$ for
$k=2,3,4,\ldots$. Here, we only discuss a case with Level $k$. It is easy to
see from (c) of Figure 5 that the sub-Markov process $\left\{  J\left(
t\right)  ,I\left(  t\right)  =k:t\geq0\right\}  $ observed within Level $k$
has the state space $\Omega_{\text{Level }k}=\left\{  \Delta
,k-2,k-1,k,k+1,k+2,k+3,\ldots\right\}  $, and its infinitesimal generator is
given by%
\[
\mathbf{Q}_{\text{Level }k}=\left(
\begin{array}
[c]{ccccccccccc}%
-a & a &  &  &  &  &  &  &  &  & \\
\mu & -\left(  a+\mu\right)  & a &  &  &  &  &  &  &  & \\
0 &  & -a & a &  &  &  &  &  &  & \\
0 &  &  & -a & a &  &  &  &  &  & \\
0 &  &  &  & -a & a &  &  &  &  & \\
\mu p_{2} &  &  &  &  & -\eta_{2} & \delta_{2} &  &  &  & \\
\mu p_{3} &  &  &  &  &  & -\eta_{3} & \delta_{3} &  &  & \\
\mu p_{4} &  &  &  &  &  &  & -\eta_{4} & \delta_{4} &  & \\
\mu p_{5} &  &  &  &  &  &  &  & -\eta_{5} & \delta_{5} & \\
\vdots &  &  &  &  &  &  &  &  & \ddots & \ddots
\end{array}
\right)  ,
\]
where $\eta_{k}=\mu p_{k}+a\left(  1-p_{k}\right)  $ and $\delta_{k}=a\left(
1-p_{k}\right)  $ for $k=3,4,5,\ldots$.

By using a similar analysis of mean drift to that in the sub-Markov process
$\mathbf{Q}_{\text{Level }0}$, we indicate that the sub-Markov process
$\mathbf{Q}_{\text{Level }k}$ is positive recurrent for $k=2,3,4,\ldots$.

In what follows, we compute the mean drift rates of the pyramid Markov process
$Q$ on Level $k$ for a large positive integer $k$.

Corresponding to the state space $\Omega_{\text{Level }k}=\left\{
\Delta,k-2,k-1,k,k+1,k+2,k+3,\ldots\right\}  $, we write the stationary
probability vector of the sub-Markov process $\mathbf{Q}_{\text{Level }k}$ as%
\[
\theta=\left(  \theta_{\Delta},\theta_{-2},\theta_{-1},\theta_{0},\theta
_{1},\theta_{2},\theta_{3},\ldots\right)  .
\]
Through solving the system of linear equations $\theta\mathbf{Q}_{\text{Level
}k}=0$ and $\theta e=1$, we obtain that for $k=2,3,4,\ldots$,%
\[
\theta_{k}=\frac{\delta_{k-1}\delta_{k-2}\delta_{k-3}\cdots\delta_{2}a}%
{\eta_{k}\eta_{k-1}\eta_{k-2}\cdots\eta_{3}\eta_{2}}\frac{a}{a+\mu}%
\theta_{\Delta},
\]%
\[
\theta_{1}=\theta_{0}=\theta_{-1}=\theta_{-2}=\frac{a}{a+\mu}\theta_{\Delta},
\]
and%
\[
\theta_{\Delta}=\frac{1}{1+\left(  4+\frac{\delta_{k-1}\delta_{k-2}%
\delta_{k-3}\cdots\delta_{2}a}{\eta_{k}\eta_{k-1}\eta_{k-2}\cdots\eta_{3}%
\eta_{2}}\right)  \frac{a}{a+\mu}}.
\]

Observing state $\Delta$ of the sub-Markov process $\mathbf{Q}_{\text{Level
}k}$, let $\widetilde{\theta}=\left(  \theta_{-2},\theta_{-1},\theta
_{0},\theta_{1},\theta_{2},\theta_{3},\ldots\right)  $. Then it is easy to see
from Figure 4 that for the pyramid Markov process $Q$ on Level $k$, the mean
drift rate of moving left to state $\Delta$ (see Level $\widetilde{0}$) is
given by%
\[
\left(  k+1\right)  \cdot\widetilde{\theta}Be=\left(  k+1\right)  \left(
\mu\theta_{-2}+\mu\sum_{k=2}^{\infty}\theta_{k}p_{k}\right)  >\mu\theta
_{-2}\left(  k+1\right)  ,
\]
where $e$ is a column vector of ones with a suitable size, and $k+1$ denotes
the $\left(  k+1\right)  $-level state transitions; while the mean drift rate
of moving right to Level $k+1$ is given by%
\[
1\cdot\widetilde{\theta}Ce=b\sum_{k=-1}^{\infty}\theta_{k}<b,
\]
where $1$ denotes the $1$-level state transitions. Since%
\[
\lim_{k\rightarrow\infty}\left(  k+1\right)  \cdot\widetilde{\theta}%
Be>\lim_{k\rightarrow\infty}\mu\theta_{-2}\left(  k+1\right)  =\infty
>b>1\cdot\widetilde{\theta}Ce,
\]
there exists a sufficient large positive integer $\mathbf{K}$ such that for
any $n>\mathbf{K}$,%
\[
\left(  n+1\right)  \cdot\widetilde{\theta}Be>1\cdot\widetilde{\theta}Ce.
\]
This shows that in the pyramid Markov process $Q$ on Level $n$ for any
$n>\mathbf{K}$, the mean drift rate of moving left to state $\Delta$ (see
Level $\widetilde{0}$) is bigger than the mean drift rate of moving right to
Level $n+1$.

Based on the above discussion, for the pyramid Markov process $Q$, we obtain
two basic results: (i) The sub-Markov process $\mathbf{Q}_{\text{Level }k}$ is
stable for $k=0,1,2,3,4,\ldots$. (ii) For the pyramid Markov process $Q$ on
Level $n$ with any $n>\mathbf{K}$, the mean drift rate of moving left to state
$\Delta$ (see Level $\widetilde{0}$) is bigger than the mean drift rate of
moving right to Level $n+1$. Therefore, it follows from Chapter 3 of Li
\cite{Li:2010} that the pyramid Markov process $Q$ must be positive recurrent.
This completes the proof. $\square$

Note that the pyramid Markov process $Q$ is irreducible and positive
recurrent. For $i,j=0,1,2,\ldots$, we define the probabilities%
\[
p_{i,j}\left(  t\right)  =P\left\{  I(t)=i,J(t)=j\right\}  ,
\]
and%
\[
\pi_{i,j}=\lim_{t\rightarrow+\infty}p_{i,j}\left(  t\right)  .
\]
We write%
\[
\pi=\left(  \pi_{\widetilde{0}},\pi_{0},\pi_{1},\pi_{2},\pi_{3},\pi_{4}%
,\pi_{5},\ldots\right)  ,
\]
where by corresponding to the state space $\Omega$,%
\begin{align*}
\pi_{\widetilde{0}}  &  =\pi_{0,0},\\
\pi_{0}  &  =\left(  \pi_{0,1},\pi_{0,2},\pi_{0,3},\pi_{0,4},\ldots\right)
,\\
\pi_{1}  &  =\left(  \pi_{1,0},\pi_{1,1},\pi_{1,2},\pi_{1,3},\pi_{1,4}%
,\ldots\right)  ,\\
\pi_{k}  &  =\left(  \pi_{k,k-2},\pi_{k,k-1},\pi_{k,k},\pi_{k,k+1},\pi
_{k,k+2},\ldots\right)  ,\ \ \ k\geq2.
\end{align*}

To express the stationary probability vector $\pi=\left(  \pi_{\widetilde{0}%
},\pi_{0},\pi_{1},\pi_{2},\pi_{3},\pi_{4},\pi_{5},\ldots\right)  $, we need to
compute some inverse matrices for a class of upper triangular matrices of
infinite size, for example, the matrices $Q_{0,0}$, $Q_{1,1}$ and $A$ given in
(\ref{Equat-2}).

The following lemma provides the inverse of an upper triangular matrix of
infinite size, which is useful in our later study.

\begin{Lem}
\label{Lem:Inverse}Let $d_{k}\neq0$ for $k=0,1,2,\ldots$. Then the upper
triangular matrix%
\[
D=\left(
\begin{array}
[c]{cccccc}%
d_{0} & f_{0} &  &  &  & \\
& d_{1} & f_{1} &  &  & \\
&  & d_{2} & f_{2} &  & \\
&  &  & d_{3} & f_{3} & \\
&  &  &  & \ddots & \ddots
\end{array}
\right)
\]
is invertible, and there exists a unique inverse matrix as follows:%
\[
D^{-1}=\left(
\begin{array}
[c]{cccccc}%
\frac{1}{d_{0}} & -\frac{f_{0}}{d_{0}d_{1}} & \frac{f_{0}f_{1}}{d_{0}%
d_{1}d_{2}} & -\frac{f_{0}f_{1}f_{2}}{d_{0}d_{1}d_{2}d_{3}} & \frac{f_{0}%
f_{1}f_{2}f_{3}}{d_{0}d_{1}d_{2}d_{3}d_{4}} & \cdots\\
& \frac{1}{d_{1}} & -\frac{f_{1}}{d_{1}d_{2}} & \frac{f_{1}f_{2}}{d_{1}%
d_{2}d_{3}} & -\frac{f_{1}f_{2}f_{3}}{d_{1}d_{2}d_{3}d_{4}} & \cdots\\
&  & \frac{1}{d_{2}} & -\frac{f_{2}}{d_{2}d_{3}} & \frac{f_{2}f_{3}}%
{d_{2}d_{3}d_{4}} & \cdots\\
&  &  & \frac{1}{d_{3}} & -\frac{f_{3}}{d_{3}d_{4}} & \cdots\\
&  &  &  & \frac{1}{d_{4}} & \cdots\\
&  &  &  &  & \ddots
\end{array}
\right)
\]
whose $\left(  k,k+l\right)  st$ element is given by%
\[
D_{k,k+l}^{-1}=\left(  -1\right)  ^{l}\frac{f_{k}f_{k+1}f_{k+2}\cdots
f_{k+l-1}}{d_{k}d_{k+1}d_{k+2}\cdots d_{k+l-1}d_{k+l}},
\]
for $k=0,1,2,3,\ldots$, and $l=0,1,2,3,\ldots$.
\end{Lem}

\textbf{Proof.} Note that $d_{k}\neq0$ for $k=0,1,2,\ldots$, the upper
triangular matrix $D$ is invertible. Let%
\[
X=\left(
\begin{array}
[c]{ccccc}%
x_{0,0} & x_{0,1} & x_{0,2} & x_{0,3} & \cdots\\
& x_{1,1} & x_{1,2} & x_{1,3} & \cdots\\
&  & x_{2,2} & x_{2,3} & \cdots\\
&  &  & x_{3,3} & \cdots\\
&  &  &  & \ddots
\end{array}
\right)  .
\]
Then the invertible upper triangular matrix $D$ must have the unique upper
triangular inverse matrix. Then it follows from $XD=I$ that for
$k=0,1,2,3,\ldots$, and $l=0,1,2,3,\ldots$,%
\[
x_{k,k+l}=\left(  -1\right)  ^{l}\frac{f_{k}f_{k+1}f_{k+2}\cdots f_{k+l-1}%
}{d_{k}d_{k+1}d_{k+2}\cdots d_{k+l-1}d_{k+l}}.
\]
This proof is completed. $\square$

By using Lemma \ref{Lem:Inverse}, it is easy to explicitly express the inverse
matrices of the matrices $Q_{0,0}$, $Q_{1,1}$ and $A$. Here, we omit their details.

The following lemma indicates that each element of the matrix $\left(
I-\mathbf{R}\right)  ^{-1}$ of infinite size is finite. This result is
necessary and useful in our later analysis.

\begin{Lem}
Let $\mathbf{R}=C\left(  -A\right)  ^{-1}$ and $\left(  I-\mathbf{R}\right)
^{-1}=\sum_{k=0}^{\infty}\mathbf{R}^{k}$. Then each element of the matrix
$\left(  I-\mathbf{R}\right)  ^{-1}$ of infinite size is finite.
\end{Lem}

\textbf{Proof.} Note that%
\[
\left(  I-\mathbf{R}\right)  ^{-1}=\left[  I-C\left(  -A\right)  ^{-1}\right]
^{-1}=\left[  -\left(  A+C\right)  \right]  ^{-1}\left(  -A\right)  ,
\]
where%
\[
A+C=\left(
\begin{array}
[c]{cccccccc}%
-\text{ }\left(  a+\mu\right)  & a &  &  &  &  &  & \\
b & -\text{ }\left(  a+b\right)  & a &  &  &  &  & \\
& b & -\text{ }\left(  a+b\right)  & a &  &  &  & \\
&  & b & -\text{ }\left(  a+b\right)  & a &  &  & \\
&  &  & b & -\xi_{2} & a\left(  1-p_{2}\right)  &  & \\
&  &  &  & b & -\xi_{3} & a\left(  1-p_{3}\right)  & \\
&  &  &  &  & \ddots & \ddots & \ddots
\end{array}
\right)  ,
\]
which is the infinitesimal generator of an irreducible birth-death process,
having%
\[
-\left(  A+C\right)  e=\left(  \mu,0,0,0,\mu p_{2},\mu p_{3},\mu p_{4}%
,\ldots\right)  ^{T}.
\]
By using Section 3 in Chapter 1 of Li \cite{Li:2010}, the LU-type
RG-factorization is given by%
\[
A+C=\left(  I-R_{U}\right)  U\left(  I-G_{L}\right)  ,
\]
where%
\[
U=\text{diag}\left(  U_{0},U_{1},U_{2},U_{3},\ldots\right)  ,
\]%
\[
R_{U}=\left(
\begin{array}
[c]{cccccc}%
0 & R_{0} &  &  &  & \\
& 0 & R_{1} &  &  & \\
&  & 0 & R_{2} &  & \\
&  &  & 0 & R_{3} & \\
&  &  &  & \ddots & \ddots
\end{array}
\right)  ,\text{ \ }G_{L}=\left(
\begin{array}
[c]{ccccc}%
0 &  &  &  & \\
G_{1} & 0 &  &  & \\
& G_{2} & 0 &  & \\
&  & G_{3} & 0 & \\
&  &  & \ddots & \ddots
\end{array}
\right)  ,
\]
the real number sequence $\left\{  R_{k}:k\geq0\right\}  $ is the minimal
positive solution to the system of nonlinear equations%
\begin{align*}
R_{1}R_{0}b-R_{0}\left(  a+b\right)  +a  &  =0,\\
R_{2}R_{1}b-R_{1}\left(  a+b\right)  +a  &  =0,\\
R_{3}R_{2}b-R_{2}\left(  a+b\right)  +a  &  =0,\\
R_{k+1}R_{k}b-R_{k}\xi_{k-1}+a  &  =0,\text{ \ }k\geq3;
\end{align*}
while the real number sequence $\left\{  G_{l}:l\geq1\right\}  $ is the
minimal positive solution to the system of nonlinear equations%
\begin{align*}
b-G_{1}\left(  a+b\right)  +G_{2}G_{1}a  &  =0,\\
b-G_{2}\left(  a+b\right)  +G_{3}G_{2}a  &  =0,\\
b-G_{3}\left(  a+b\right)  +G_{4}G_{3}a  &  =0,\\
b-G_{k}\xi_{k-2}+G_{k+1}G_{k}a  &  =0,\text{ \ }k\geq4.
\end{align*}
Furthermore, the real number sequence $\left\{  U_{k}:k\geq0\right\}  $ is
given by%
\begin{align*}
U_{0}  &  =-\text{ }\left(  a+\mu\right)  +R_{0}b=-\text{ }\left(
a+\mu\right)  +aG_{1},\\
U_{1}  &  =-\text{ }\left(  a+b\right)  +R_{1}b=-\text{ }\left(  a+b\right)
+aG_{2},\\
U_{2}  &  =-\text{ }\left(  a+b\right)  +R_{2}b=-\text{ }\left(  a+b\right)
+aG_{3},\\
U_{3}  &  =-\text{ }\left(  a+b\right)  +R_{3}b=-\text{ }\left(  a+b\right)
+aG_{4},\\
U_{k}  &  =-\xi_{k-2}+R_{k}b=-\xi_{k-2}+aG_{k+1},\text{ \ }k\geq4.
\end{align*}
Thus, we obtain%
\begin{align*}
\left(  I-\mathbf{R}\right)  ^{-1}  &  =\left[  -\left(  A+C\right)  \right]
^{-1}\left(  -A\right) \\
&  =\left(  I-G_{L}\right)  ^{-1}\text{diag}\left(  -U_{0}^{-1},-U_{1}%
^{-1},-U_{2}^{-1},-U_{3}^{-1},\ldots\right)  \left(  I-R_{U}\right)
^{-1}\left(  -A\right)  .
\end{align*}
Following Theorem 1.2 in Chapter 1 of Li \cite{Li:2010}, some matrix
computation indicates that each element of the matrix $\left(  I-\mathbf{R}%
\right)  ^{-1}$ is finite. This proof is completed. $\square$

The following theorem provides an explicit expression for the stationary
probability vector $\pi=\left(  \pi_{\widetilde{0}},\pi_{0},\pi_{1},\pi
_{2},\pi_{3},\pi_{4},\pi_{5},\ldots\right)  $ by means of a new
matrix-geometric solution with the rate matrices%
\begin{equation}
\mathbf{R}=C\left(  -A\right)  ^{-1} \label{Equat-3}%
\end{equation}
and%
\begin{equation}
\widetilde{\mathbf{R}}=Q_{1,2}\left(  -A\right)  ^{-1}. \label{Equat-4}%
\end{equation}

\begin{The}
\label{The:StatP}The stationary probability vector $\pi=\left(  \pi
_{\widetilde{0}},\pi_{0},\pi_{1},\pi_{2},\pi_{3},\pi_{4},\pi_{5}%
,\ldots\right)  $ of the pyramid Markov process $Q$ is matrix-geometric, given
by%
\begin{equation}
\pi_{k}=\pi_{1}\widetilde{\mathbf{R}}\mathbf{R}^{k-2},\text{ \ }%
k=2,3,4,\ldots, \label{Equat-5}%
\end{equation}%
\[
\pi_{0}=\pi_{\widetilde{0}}Q_{\widetilde{0},0}\left(  -Q_{0,0}\right)  ^{-1},
\]
and $\pi_{\widetilde{0}}$ and $\pi_{1}$ are determined by means of solving the
following system of linear equations%
\begin{equation}
\left(  \pi_{\widetilde{0}},\pi_{1}\right)  \left(
\begin{array}
[c]{cc}%
Q_{\widetilde{0},\widetilde{0}}+Q_{\widetilde{0},0}\left(  -Q_{0,0}\right)
^{-1}Q_{0,\widetilde{0}} & Q_{\widetilde{0},1}+Q_{\widetilde{0},0}\left(
-Q_{0,0}\right)  ^{-1}Q_{0,1}\\
Q_{1,\widetilde{0}}+\widetilde{\mathbf{R}}\left(  I-\mathbf{R}\right)
^{-1}B & Q_{1,1}%
\end{array}
\right)  =0 \label{Equ-0}%
\end{equation}
with the normalized condition%
\begin{equation}
\pi_{\widetilde{0}}\left[  1+Q_{\widetilde{0},0}\left(  -Q_{0,0}\right)
^{-1}e\right]  +\pi_{1}\left[  I+\widetilde{\mathbf{R}}\left(  I-R\right)
^{-1}\right]  e=1. \label{Equ-0-1}%
\end{equation}
\end{The}

\textbf{Proof.} From $\pi Q=0$ and $\pi e=1$, it is easy to see that%
\begin{equation}
\left\{
\begin{array}
[c]{l}%
\pi_{\widetilde{0}}Q_{\widetilde{0},\widetilde{0}}+\pi_{0}Q_{0,\widetilde{0}%
}+\pi_{1}Q_{1,\widetilde{0}}+\sum\limits_{k=2}^{\infty}\pi_{k}B=0,\\
\pi_{\widetilde{0}}Q_{\widetilde{0},0}+\pi_{0}Q_{0,0}=0,\\
\pi_{\widetilde{0}}Q_{\widetilde{0},1}+\pi_{0}Q_{0,1}+\pi_{1}Q_{1,1}=0,\\
\pi_{1}Q_{1,2}+\pi_{2}A=0,\\
\pi_{k}C+\pi_{k+1}A=0,\text{ \ \ \ }k\geq2.
\end{array}
\right.  \label{Equ-1}%
\end{equation}
From the second equation of (\ref{Equ-1}), it is easy to see that%
\[
\pi_{0}=\pi_{\widetilde{0}}Q_{\widetilde{0},0}\left(  -Q_{0,0}\right)  ^{-1},
\]
and from the fourth and fifth equations of (\ref{Equ-1}), it is easy to check
that for $k=2,3,4,5,\ldots,$
\[
\pi_{k}=\pi_{1}\widetilde{\mathbf{R}}\mathbf{R}^{k-2}.
\]
Also, by using the first and third equations of (\ref{Equ-1}), the boundary
equation (\ref{Equ-0}) is obtained. Finally, the normalized condition
(\ref{Equ-0-1}) is given easily by means of $\pi_{\widetilde{0}}+\sum
_{k=0}^{\infty}\pi_{k}e=1$. This proof is completed. $\square$

\begin{Rem}
Although the pyramid Markov process of blockchain selfish mining is very
complicated, we obtain that the stationary probability vector is
matrix-geometric, in which the two rate matrices can be explicitly expressed
as $\mathbf{R}=C\left(  -A\right)  ^{-1}$ and\ $\widetilde{\mathbf{R}}%
=Q_{1,2}\left(  -A\right)  ^{-1}$. Obviously, our matrix-geometric solution is
different from that in Markov chains of GI/M/1 type given in Neuts
\cite{Neu:1981} whose rate matrix $\mathbf{R}$ only has one numerical solution
from the nonlinear equation: $\sum_{k=0}^{\infty}\mathbf{R}^{k}A_{k}=0$.
\end{Rem}

In the remainder of this section, we simply explain three useful special cases.

\textbf{Case one: }$p_{2}=1$. In this case, the main chain with two blocks by
the dishonest mining pool is pegged on the blockchain with probability one.
The dishonest mining pool can not continue to mine more blocks. Thus there do
not exist all the states $\left(  k,k+l\right)  $ for $k=0,1,2,\ldots$ and
$l=3,4,5,\ldots$, and the infinitesimal generator of the pyramid Markov
process is simplified greatly.

\textbf{Case two: }$0\leq p_{i}<1$ for $2\leq i\leq n_{0}-1$ and $p_{n_{0}}%
=1$. In this case, the pyramid Markov process does not have all the states
$\left(  k,k+l\right)  $ for $k=0,1,2,\ldots$ and $l=n_{0}+1,n_{0}%
+2,n_{0}+3,\ldots$. Also, the sizes of the two rate matrices: $\mathbf{R}%
=C\left(  -A\right)  ^{-1}\ $and $\widetilde{\mathbf{R}}=Q_{1,2}\left(
-A\right)  ^{-1}$ are finite, so the computation of the matrices involved is ordinary.

\textbf{Case three: }$p_{k}=1$ for $k\geq1$, and the two mining pools follow
the one-block leading competitive criterion.

In this case, this becomes the model given by G\H{o}bel et al. \cite{Gob:2016}%
, which is modified by means of our block-pegging rule: No new block can be
generated during the block-pegging process of the main chain.

\section{Analysis of Orphan Blocks}

In this section, by using the stationary probability vector of the pyramid
Markov process, we analyze some interesting performance measures of the main
chain and the chain of orphan blocks. To our best knowledge, this is the first
one to develop an effective quantitative method in the study of orphan blocks.

Once the honest and dishonest mining pools simultaneously begin a mining
process, it is seen from Figure 2 that the selfish mining can result in two
block branches forked at a common tree root, in which one chain comes from the
honest mining pool, while the other from the dishonest mining pool. Also, the
two block branches forked at the common tree root are terminated immediately
once the main chain begins to peg on the blockchain.

If the main chain is taken by the honest mining pool, then all the blocks on
the other chain by the dishonest mining pool become orphan blocks. On the
contrary, if the main chain comes from the dishonest mining pool, then all the
blocks on the other chain by the honest mining pool are orphan blocks.
Obviously, the main chain and the chain of orphan blocks end at the same
moment that the main chain is confirmed.

Note that the main chain is taken by the honest mining pool at state $\left(
k,k-2\right)  $ with probability $\pi_{k,k-2}$ for $k=2,3,4,\ldots$; while the
main chain is taken by the dishonest mining pool at state $\left(
k,k+l\right)  $ with probability $\pi_{k,k+l}$ for $k=0,1,2,\ldots
,l=2,3,4,\ldots$. In addition, all the other states can not result in such a
main chain. It is easy to see that $\sum_{k=2}^{\infty}\pi_{k,k-2}$ and
$\sum_{k=0}^{\infty}\sum_{l=2}^{\infty}\pi_{k,k+l}$ are the probabilities that
the main chain is taken by the honest and dishonest mining pools, respectively.

Let $P_{\text{H}}$ and $P_{\text{D}}$ be the probabilities that the main chain
is taken by the honest and dishonest mining pools, respectively. Then
\[
P_{\text{H}}=\frac{\sum\limits_{k=2}^{\infty}\pi_{k,k-2}}{\sum\limits_{k=2}%
^{\infty}\pi_{k,k-2}+\sum\limits_{k=0}^{\infty}\sum\limits_{l=2}^{\infty}%
\pi_{k,k+l}},\text{ \ \ }P_{\text{D}}=\frac{\sum\limits_{k=0}^{\infty}%
\sum\limits_{l=2}^{\infty}\pi_{k,k+l}}{\sum\limits_{k=2}^{\infty}\pi
_{k,k-2}+\sum\limits_{k=0}^{\infty}\sum\limits_{l=2}^{\infty}\pi_{k,k+l}}.
\]

\textbf{(a) The average stationary lengths of the two chains}

Between the two block branches forked at the common tree root, one is taken as
the main chain, while the other is regarded as the chain of orphan blocks.

Let $L_{\text{M}}$ and $L_{\text{O}}$ be the average stationary lengths of the
main chain and the chain of orphan blocks, respectively.

The following theorem provides expressions for $L_{\text{M}}$ and
$L_{\text{O}}$ by using the stationary probability vector of the pyramid
Markov process.

\begin{The}
\label{The:Ave}(a) The average stationary length of the main chain is given by%
\begin{equation}
L_{\text{M}}=P_{\text{H}}\sum\limits_{k=2}^{\infty}k\pi_{k,k-2}+P_{\text{D}%
}\sum\limits_{k=0}^{\infty}\sum\limits_{l=2}^{\infty}\left(  k+l\right)
\pi_{k,k+l}. \label{Equat-6}%
\end{equation}
(b)The average stationary length of the chain of orphan blocks is given by%
\begin{equation}
L_{\text{O}}=P_{\text{H}}\sum\limits_{k=2}^{\infty}\left(  k-2\right)
\pi_{k,k-2}+P_{\text{D}}\sum\limits_{k=0}^{\infty}\sum\limits_{l=2}^{\infty
}k\pi_{k,k+l}. \label{Equat-7}%
\end{equation}
\end{The}

\textbf{Proof.} We only prove (a), while (b) can be proved similarly.

Let $N_{\text{M}}$ be the average stationary length of the main chain. Also,
we introduce two events as follows:%
\[
E_{\text{H}}=\left\{  \text{The main chain is taken by the honest mining
pool}\right\}
\]
and%
\[
E_{\text{D}}=\left\{  \text{The main chain is taken by the dishonest mining
pool}\right\}  .
\]
Then $P_{\text{H}}=P\left\{  E_{\text{H}}\right\}  $ and $P_{\text{D}%
}=P\left\{  E_{\text{D}}\right\}  $. By applying the law of total probability,
we have%
\begin{equation}
L_{\text{M}}=E\left[  N_{\text{M}}\right]  =E\left[  N_{\text{M}}\text{
$\vert$
}E_{\text{H}}\right]  P\left\{  E_{\text{H}}\right\}  +E\left[  N_{\text{M}%
}\text{
$\vert$
}E_{\text{D}}\right]  P\left\{  E_{\text{D}}\right\}  . \label{Equat-8}%
\end{equation}
Note that $E\left[  N_{\text{M}}\text{
$\vert$
}E_{\text{H}}\right]  =\sum_{k=2}^{\infty}k\pi_{k,k-2}$, since the length of
the honest main chain is $k$ with probability $\pi_{k,k-2}$; $E\left[
N_{\text{M}}\text{
$\vert$
}E_{\text{D}}\right]  =\sum_{k=0}^{\infty}\sum_{l=2}^{\infty}\left(
k+l\right)  \pi_{k,k+l}$, because the length of the dishonest main chain is
$k+l$ with probability $\pi_{k,k+l}$. Thus it follows from (\ref{Equat-8})
that%
\[
L_{\text{M}}=P_{\text{H}}\sum\limits_{k=2}^{\infty}k\pi_{k,k-2}+P_{\text{D}%
}\sum\limits_{k=0}^{\infty}\sum\limits_{l=2}^{\infty}\left(  k+l\right)
\pi_{k,k+l}.
\]
This completes the proof. $\square$

\textbf{(b) The blockchain pegging rate and the orphan block removal rate}

Let $\Upsilon_{\text{M}}$ and $\Upsilon_{\text{O}}$ be the stationary
block-pegging rate of the main chain, and the stationary orphan block removal
rate of the chain of orphan blocks, respectively. Then the following theorem
provides an expressions for $\Upsilon_{\text{M}}$ and $\Upsilon_{\text{O}}$.

\begin{The}
(a) The stationary block-pegging rate of the main chain is given by%
\begin{equation}
\Upsilon_{\text{M}}=\mu\left[  P_{\text{H}}\sum\limits_{k=2}^{\infty}%
k\pi_{k,k-2}+P_{\text{D}}\sum\limits_{k=0}^{\infty}\sum\limits_{l=2}^{\infty
}\left(  k+l\right)  p_{l}\pi_{k,k+l}\right]  . \label{Equat-9}%
\end{equation}
(b) The stationary orphan block removal rate of the chain of orphan blocks is
given by%
\begin{equation}
\Upsilon_{\text{O}}=\mu\left[  P_{\text{H}}\sum\limits_{k=2}^{\infty}\left(
k-2\right)  \pi_{k,k-2}+P_{\text{D}}\sum\limits_{k=0}^{\infty}\sum
\limits_{l=2}^{\infty}k\pi_{k,k+l}\right]  =\mu L_{\text{O}}. \label{Equat-10}%
\end{equation}
\end{The}

\textbf{Proof.} The proof is similar to that in Theorem \ref{The:Ave}, while
the only difference between both of them comes from the following two points:

\textit{Point one:} In the main chain (resp. the chain of orphan blocks), each
of the blocks that is submitted to the blockchain (resp. the transaction pool)
has an exponential network delay time with parameter $\mu$. Thus the $k$
blocks can have the exponential network delay time with parameter $k\mu$.

\textit{Point two:} When the main chain is taken by the dishonest mining pool,
it is key to observe at state $\left(  k,k+l\right)  $ that with probability
$p_{l}$, the main chain is published on the blockchain; while with probability
$1-p_{l}$, the main chain is detained to continue mining more blocks such that
it is not broadcasted in the blockchain network. In this case, the dishonest
mining pool hopes to obtain more mining profit through winning on mining more
blocks. On the contrary, the stationary orphan block removal probability of
the chain of orphan blocks at state $\left(  k,k+l\right)  $ is $\pi_{k,k+l}$
because the orphan blocks are removed to the transaction pool with probability
one. This completes the proof. $\square$

Now, we introduce two useful ratios of the blockchain selfish mining, which
are necessary and useful in design of blockchain.

We define the ratio of two average stationary lengths as%
\begin{equation}
\phi=\frac{L_{\text{O}}}{L_{\text{M}}}=\frac{P_{\text{H}}\sum\limits_{k=2}%
^{\infty}\left(  k-2\right)  \pi_{k,k-2}+P_{\text{D}}\sum\limits_{k=0}%
^{\infty}\sum\limits_{l=2}^{\infty}k\pi_{k,k+l}}{P_{\text{H}}\sum
\limits_{k=2}^{\infty}k\pi_{k,k-2}+P_{\text{D}}\sum\limits_{k=0}^{\infty}%
\sum\limits_{l=2}^{\infty}\left(  k+l\right)  \pi_{k,k+l}}, \label{Equat-11}%
\end{equation}
and the stationary ratio of block removal and pegging rates as%
\begin{equation}
\psi=\frac{\Upsilon_{\text{O}}}{\Upsilon_{\text{M}}}=\frac{P_{\text{H}}%
\sum\limits_{k=2}^{\infty}\left(  k-2\right)  \pi_{k,k-2}+P_{\text{D}}%
\sum\limits_{k=0}^{\infty}\sum\limits_{l=2}^{\infty}k\pi_{k,k+l}}{P_{\text{H}%
}\sum\limits_{k=2}^{\infty}k\pi_{k,k-2}+P_{\text{D}}\sum\limits_{k=0}^{\infty
}\sum\limits_{l=2}^{\infty}\left(  k+l\right)  p_{l}\pi_{k,k+l}}.
\label{Equat-12}%
\end{equation}
It is easy to see from $0\leq p_{l}\leq1$ that $0<\phi<\psi<1$. This shows
that the main chain may not be pegged on the blockchain due to the blockchain
selfish mining.

In the remainder of this section, we develop some local performance measures
of the blockchain, and discuss their monotonous properties for the jumping's
mining rate $\gamma$ and the efficiency-increased ratio $\Re$. Note that the
monotonous properties are useful in our later study.

Let $L_{\text{M}}^{\left(  \text{H}\right)  }$ and $L_{\text{M}}^{\left(
\text{D}\right)  }$ be the average stationary lengths of the main chain by the
honest and dishonest mining pools, respectively; and $L_{\text{O}}^{\left(
\text{H}\right)  }$ and $L_{\text{O}}^{\left(  \text{D}\right)  }$ the average
stationary lengths of the chain of orphan blocks by the honest and dishonest
mining pools, respectively. We write%
\[
\Lambda=\sum\limits_{k=0}^{\infty}\sum\limits_{l=2}^{\infty}\left(
k+l\right)  \pi_{k,k+l}.
\]

The following corollary shows that the blockchain selfish mining leads to a
larger waste of mining resources.

\begin{Cor}
\label{Cor:QueM}For the blockchain selfish mining, we have

(a) $0<L_{\text{M}}^{\left(  \text{H}\right)  }-L_{\text{O}}^{\left(
\text{H}\right)  }<2$,

(b) $0<L_{\text{M}}^{\left(  \text{D}\right)  }-L_{\text{O}}^{\left(
\text{D}\right)  }<\Lambda$, and

(c) $0<L_{M}-L_{O}<2P_{H}+\Lambda P_{D}.$
\end{Cor}

\textbf{Proof.} We only prove (b) and (c), while (a) can be proved similarly.

It is easy to check that%
\[
L_{\text{M}}^{\left(  \text{D}\right)  }=\sum\limits_{k=0}^{\infty}%
\sum\limits_{l=2}^{\infty}\left(  k+l\right)  \pi_{k,k+l}%
\]
and%
\[
L_{\text{O}}^{\left(  \text{D}\right)  }=\sum\limits_{k=0}^{\infty}%
\sum\limits_{l=2}^{\infty}k\pi_{k,k+l}.
\]
Thus we obtain%
\[
L_{\text{M}}^{\left(  \text{D}\right)  }-L_{\text{O}}^{\left(  \text{D}%
\right)  }=\sum\limits_{k=0}^{\infty}\sum\limits_{l=2}^{\infty}l\pi_{k,k+l}>0
\]
and%
\[
L_{\text{M}}^{\left(  \text{D}\right)  }-L_{\text{O}}^{\left(  \text{D}%
\right)  }=\sum\limits_{k=0}^{\infty}\sum\limits_{l=2}^{\infty}l\pi
_{k,k+l}<\sum\limits_{k=0}^{\infty}\sum\limits_{l=2}^{\infty}\left(
k+l\right)  \pi_{k,k+l}=\Lambda.
\]

Now, we prove (c). Since%
\[
L_{M}=P_{H}L_{M}^{\left(  H\right)  }+P_{D}L_{M}^{\left(  D\right)  }%
\]
and%
\[
L_{O}=P_{H}L_{O}^{\left(  H\right)  }+P_{D}L_{O}^{\left(  D\right)  },
\]
we obtain%
\begin{align*}
0  &  <L_{M}-L_{O}=P_{H}\left(  L_{M}^{\left(  H\right)  }-L_{O}^{\left(
H\right)  }\right)  +P_{D}\left(  L_{M}^{\left(  D\right)  }-L_{O}^{\left(
D\right)  }\right) \\
&  <2P_{H}+\Lambda P_{D}.
\end{align*}
This completes the proof. $\square$

\begin{Cor}
\label{Cor:Mon-1}In the blockchain selfish mining, we have

(a) the two average stationary lengths $L_{\text{M}}^{\left(  \text{H}\right)
}$ and $L_{\text{O}}^{\left(  \text{H}\right)  }$ decrease as the jumping's
mining rate $\gamma$ increases;

(b) the two average stationary lengths $L_{\text{M}}^{\left(  \text{D}\right)
}$ and $L_{\text{O}}^{\left(  \text{D}\right)  }$ increase as the jumping's
mining rate $\gamma$ increases.
\end{Cor}

\textbf{Proof.} We only prove (a), while (b) can be proved similarly.

If decreasing the mining rate of the honest mining pool and simultaneously
increasing the mining rate of the dishonest mining pool, then the probability
$\pi_{k,k-2}$ decreases for $k\geq2$.

As the jumping's mining rate $\gamma$ increases, the mining rate $\beta
-\gamma$ of the honest mining pool decreases, and simultaneously the mining
rate $\left(  \widetilde{\alpha}+\gamma\right)  \left(  1+\Re\right)  $ of the
dishonest mining pool increases. Since
\[
L_{\text{M}}^{\left(  \text{H}\right)  }=\sum\limits_{k=2}^{\infty}%
k\pi_{k,k-2}%
\]
and%
\[
L_{\text{O}}^{\left(  \text{H}\right)  }=\sum\limits_{k=2}^{\infty}\left(
k-2\right)  \pi_{k,k-2},
\]
it is easy to see that the two average stationary lengths $L_{\text{M}%
}^{\left(  \text{H}\right)  }$ and $L_{\text{O}}^{\left(  \text{H}\right)  }$
decrease as the jumping's mining rate $\gamma$ increases. This completes the
proof. $\square$

\begin{Cor}
\label{Cor:Mon-2}In the blockchain selfish mining, we have

(a) the two average stationary lengths $L_{\text{M}}^{\left(  \text{H}\right)
}$ and $L_{\text{O}}^{\left(  \text{H}\right)  }$ decrease as the
efficiency-increased ratio $\Re$ increases; and

(b) the two average stationary lengths $L_{\text{M}}^{\left(  \text{D}\right)
}$ and $L_{\text{O}}^{\left(  \text{D}\right)  }$ increase as the
efficiency-increased ratio $\Re$ increases.
\end{Cor}

\textbf{Proof.} We only prove (a), while (b) can be proved similarly.

We write%
\[
h\left(  \Re\right)  =\frac{\beta-\gamma}{\beta-\gamma+\left(  \widetilde
{\alpha}+\gamma\right)  \left(  1+\Re\right)  }%
\]
and%
\[
d\left(  \Re\right)  =\frac{\left(  \widetilde{\alpha}+\gamma\right)  \left(
1+\Re\right)  }{\beta-\gamma+\left(  \widetilde{\alpha}+\gamma\right)  \left(
1+\Re\right)  }=\frac{\widetilde{\alpha}+\gamma}{\widetilde{\alpha}%
+\gamma+\frac{\beta-\gamma}{1+\Re}}.
\]
It is easy to check that $h\left(  \Re\right)  $ decreases and $d\left(
\Re\right)  $ increases as the efficiency-increased ratio $\Re$ increases.

When decreasing the relative mining rate $h\left(  \Re\right)  $ of the honest
mining pool and simultaneously increasing the relative mining rate $d\left(
\Re\right)  $ of the dishonest mining pool, the probability $\pi_{k,k-2}$ can
decrease. Thus, the two average stationary lengths $L_{\text{M}}^{\left(
\text{H}\right)  }$ and $L_{\text{O}}^{\left(  \text{H}\right)  }$ decrease as
the efficiency-increased ratio $\Re$ increases. This completes the proof.
$\square$

Although Corollaries \ref{Cor:Mon-1} and \ref{Cor:Mon-2} provide some
monotonous properties for the local performance measures (e.g., the average
stationary lengths $L_{\text{M}}^{\left(  \text{H}\right)  }$ and
$L_{\text{O}}^{\left(  \text{H}\right)  }$, and $L_{\text{M}}^{\left(
\text{D}\right)  }$ and $L_{\text{O}}^{\left(  \text{D}\right)  }$), there do
not exist such a monotonicity for the total performance measures (e.g.,
$L_{\text{M}}$ and $L_{\text{O}}$; $\Upsilon_{\text{M}}$ and $\Upsilon
_{\text{O}}$; and $\phi$ and $\psi$). This can be observed in Figure 11 of
Section 10.

\section{Markov Reward Processes}

In this section, we set up a pyramid Markov reward process to evaluate the
long-run average profits of the honest and dishonest mining pools,
respectively. Note that our results apply the Markov reward process to provide
a more complete and practical mining profit than that in Eyal and Sirer
\cite{Eya:2014}.

For the pyramid Markov process $\left\{  \left(  I\left(  t\right)  ,J\left(
t\right)  \right)  :t\geq0\right\}  $, let $f_{\text{H}}\left(  I\left(
t\right)  ,J\left(  t\right)  \right)  $ and $f_{\text{D}}\left(  I\left(
t\right)  ,J\left(  t\right)  \right)  $ be the reward functions received by
the honest and dishonest mining pools at time $t$, respectively. Then the
average profits in the time interval $[0,t)$ of the honest and dishonest
mining pools are respectively given by%
\[
R_{\text{H}}\left(  t\right)  =E\left[  \frac{1}{t}\int_{0}^{t}f_{\text{H}%
}\left(  I\left(  t\right)  ,J\left(  t\right)  \right)  \text{d}t\right]
\]
and%
\[
R_{\text{D}}\left(  t\right)  =E\left[  \frac{1}{t}\int_{0}^{t}f_{\text{D}%
}\left(  I\left(  t\right)  ,J\left(  t\right)  \right)  \text{d}t\right]  .
\]

Since the pyramid Markov process $\left\{  \left(  I\left(  t\right)
,J\left(  t\right)  \right)  :t\geq0\right\}  $ is irreducible and positive
recurrent, by using Chapter 10 of Li \cite{Li:2010}, the long-run average
profits of the honest and dishonest mining pools are respectively given by%
\begin{equation}
\mathbf{R}_{\text{H}}=\lim_{t\rightarrow+\infty}R_{\text{H}}\left(  t\right)
=\sum_{i=0}^{1}\sum_{j=0}^{\infty}\pi_{i,j}f_{\text{H}}\left(  i,j\right)
+\sum_{i=2}^{\infty}\sum_{j=i-2}^{\infty}\pi_{i,j}f_{\text{H}}\left(
i,j\right)  \label{Equ-3}%
\end{equation}
and%
\begin{equation}
\mathbf{R}_{\text{D}}=\lim_{t\rightarrow+\infty}R_{\text{D}}\left(  t\right)
==\sum_{i=0}^{1}\sum_{j=0}^{\infty}\pi_{i,j}f_{\text{D}}\left(  i,j\right)
+\sum_{i=2}^{\infty}\sum_{j=i-2}^{\infty}\pi_{i,j}f_{\text{D}}\left(
i,j\right)  . \label{Equ-4}%
\end{equation}

Note that the stationary probability vector $\pi=\left(  \pi_{\widetilde{0}%
},\pi_{0},\pi_{1},\pi_{2},\pi_{3},\pi_{4},\pi_{5},\ldots\right)  $ is given in
Theorem \ref{The:StatP}, it is easy to see from (\ref{Equ-3}) and
(\ref{Equ-4}) that we first need to express the two reward functions
$f_{\text{H}}\left(  i,j\right)  $ and $f_{\text{D}}\left(  i,j\right)  $ for
$i,j=0,1,2,3,\ldots$. From the model description of Section 2, we can obtain%
\[
f_{\text{H}}\left(  i,j\right)  =\left\{
\begin{array}
[c]{ll}%
k\mu\left(  r_{B}+r_{F}\right)  -\left(  c_{E}+c_{A}\right)  \left(
\beta-\gamma\right)  , & \text{for }i=k,j=k-2,\text{ }k\geq2,\\
-\left(  c_{E}+c_{A}\right)  \left(  \beta-\gamma\right)  , & \text{for all
the other states}.
\end{array}
\right.
\]
and%
\[
f_{\text{D}}\left(  i,j\right)  =\left\{
\begin{array}
[c]{ll}%
\begin{array}
[c]{l}%
\left(  k+l\right)  \mu p_{l}\left(  r_{B}+r_{F}\right) \\
-\left(  \widetilde{\alpha}+\gamma\right)  \left[  c_{E}+c_{A}\left(
1+\Re\right)  \right]  ,
\end{array}
&
\begin{array}
[c]{l}%
\text{for }i=k,j=k+l,\text{ }\\
k\geq0,l\geq2;
\end{array}
\\
-\left(  \widetilde{\alpha}+\gamma\right)  \left[  c_{E}+c_{A}\left(
1+\Re\right)  \right]  , & \text{for all the other states}.
\end{array}
\right.
\]

Now, we introduce the Hadamard product of two vectors $W=\left(  w_{1}%
,w_{2},w_{3},\ldots\right)  $ and $V=\left(  v_{1},v_{2},v_{3},\ldots\right)
$ as follows:%
\[
W\odot V=\left(  w_{1}v_{1},w_{2}v_{2},w_{3}v_{3},\ldots\right)  .
\]
It is easy to check that%
\[
W\odot V_{1}+W\odot V_{2}=W\odot\left(  V_{1}+V_{2}\right)  .
\]
Let $\mathbf{e}_{1}=\left(  1,0,0,0,\ldots\right)  $.

The following theorem provides an explicit expression for the long-run average
profit of the honest mining pool.

\begin{The}
\label{The:ProH}The long-run average profit of the honest mining pool is given
by%
\begin{equation}
\mathbf{R}_{\text{H}}=\mu\left(  r_{B}+r_{F}\right)  \left\{  \mathbf{e}%
_{1}\odot\left\{  \pi_{1}\widetilde{\mathbf{R}}\left[  \left(  I-\mathbf{R}%
\right)  ^{-1}+\left(  I-\mathbf{R}\right)  ^{-2}\right]  \right\}  \right\}
e-\left(  c_{E}+c_{A}\right)  \left(  \beta-\gamma\right)  . \label{Equat-13}%
\end{equation}
\end{The}

\textbf{Proof.} Note that%
\begin{align*}
\mathbf{R}_{\text{H}}  &  =\sum_{i=0}^{1}\sum_{j=0}^{\infty}\pi_{i,j}%
f_{\text{H}}\left(  i,j\right)  +\sum_{i=2}^{\infty}\sum_{j=i-2}^{\infty}%
\pi_{i,j}f_{\text{H}}\left(  i,j\right) \\
&  =\mu\left(  r_{B}+r_{F}\right)  \sum_{k=2}^{\infty}k\pi_{k,k-2}-\left(
c_{E}+c_{A}\right)  \left(  \beta-\gamma\right)  \left(  \sum_{i=0}^{1}%
\sum_{j=0}^{\infty}+\sum_{i=2}^{\infty}\sum_{j=i-2}^{\infty}\right)  \pi
_{i,j};
\end{align*}
since $\left(  \sum_{i=0}^{1}\sum_{j=0}^{\infty}+\sum_{i=2}^{\infty}%
\sum_{j=i-2}^{\infty}\right)  \pi_{i,j}=1$ and%
\begin{align*}
\sum_{k=2}^{\infty}k\pi_{k,k-2}  &  =\sum_{k=2}^{\infty}\left(  \mathbf{e}%
_{1}\odot k\pi_{k}\right)  e=\left\{  \mathbf{e}_{1}\odot\left(  \pi
_{1}\widetilde{\mathbf{R}}\sum_{k=2}^{\infty}k\mathbf{R}^{k-2}\right)
\right\}  e\\
&  =\left\{  \mathbf{e}_{1}\odot\left\{  \pi_{1}\widetilde{\mathbf{R}}\left[
\left(  I-\mathbf{R}\right)  ^{-1}+\left(  I-\mathbf{R}\right)  ^{-2}\right]
\right\}  \right\}  e,
\end{align*}
we obtain%
\[
\mathbf{R}_{\text{H}}=\mu\left(  r_{B}+r_{F}\right)  \left\{  \mathbf{e}%
_{1}\odot\left\{  \pi_{1}\widetilde{\mathbf{R}}\left[  \left(  I-\mathbf{R}%
\right)  ^{-1}+\left(  I-\mathbf{R}\right)  ^{-2}\right]  \right\}  \right\}
e-\left(  c_{E}+c_{A}\right)  \left(  \beta-\gamma\right)  .
\]
This completes the proof. $\square$

\begin{Cor}
\label{CoroH}In the blockchain selfish mining, we have

(a) the long-run average profit $\mathbf{R}_{\text{H}}$ of the honest mining
pool decreases as the jumping's mining rate $\gamma$ increases;

(b) the long-run average profit $\mathbf{R}_{\text{H}}$ of the honest mining
pool decreases as the efficiency-increased ratio $\Re$ increases.
\end{Cor}

\textbf{Proof.} Note that%
\begin{equation}
\mathbf{R}_{\text{H}}=\mu\left(  r_{B}+r_{F}\right)  \sum_{k=2}^{\infty}%
k\pi_{k,k-2}-\left(  c_{E}+c_{A}\right)  \left(  \beta-\gamma\right)  ,
\label{Equa-1}%
\end{equation}
as the jumping's mining rate $\gamma$ (or the efficiency-increased ratio $\Re
$) increases, it is easy to see from Corollaries \ref{Cor:Mon-1} and
\ref{Cor:Mon-2} that the long-run average profit $\mathbf{R}_{\text{H}}$ of
the honest mining pool decreases. This completes the proof. $\square$

Now, we consider the long-run average profit of the dishonest mining pool,
which is a bit more complicated than that of the long-run average profit of
the honest mining pool.

Let%
\[
\overline{\mathbf{p}}_{0}=\left(  0,2p_{2},3p_{3},4p_{4},5p_{5},\ldots\right)
,
\]%
\[
\overline{\mathbf{p}}_{1}=\left(  0,0,0,3p_{2},4p_{3},5p_{4},6p_{5}%
,\ldots\right)
\]
and%
\[
\overline{\mathbf{p}}_{k}=\left(  0,0,0,0,\left(  k+2\right)  p_{2},\left(
k+3\right)  p_{3},\left(  k+4\right)  p_{4},\left(  k+5\right)  p_{5}%
,\ldots\right)  ,\text{ }k\geq2.
\]

The following theorem provides an explicit expression for the long-run average
profit of the dishonest mining pool.

\begin{The}
\label{The:ProD}The long-run average profit of the dishonest mining pool is
given by%
\begin{align}
\mathbf{R}_{\text{D}}  &  =\mu\left(  r_{B}+r_{F}\right)  \left\{
\overline{\mathbf{p}}_{0}\odot\left[  \pi_{\widetilde{0}}Q_{\widetilde{0}%
,0}\left(  -Q_{0,0}\right)  ^{-1}\right]  +\overline{\mathbf{p}}_{1}\odot
\pi_{1}+\sum\limits_{k=2}^{\infty}\overline{\mathbf{p}}_{k}\odot\left(
\pi_{1}\widetilde{\mathbf{R}}\mathbf{R}^{k-2}\right)  \right\}  e\nonumber\\
&  -\left(  \widetilde{\alpha}+\gamma\right)  \left[  c_{E}+c_{A}\left(
1+\Re\right)  \right]  . \label{Equat-14}%
\end{align}
\end{The}

\textbf{Proof.} Note that%
\begin{align*}
\mathbf{R}_{\text{D}}  &  =\sum_{i=0}^{1}\sum_{j=0}^{\infty}\pi_{i,j}%
f_{\text{D}}\left(  i,j\right)  +\sum_{i=2}^{\infty}\sum_{j=i-2}^{\infty}%
\pi_{i,j}f_{\text{D}}\left(  i,j\right) \\
&  =\mu\left(  r_{B}+r_{F}\right)  \sum_{k=0}^{\infty}\sum_{l=2}^{\infty}%
p_{l}\left(  k+l\right)  \pi_{k,k+l}\\
&  -\left(  \widetilde{\alpha}+\gamma\right)  \left[  c_{E}+c_{A}\left(
1+\Re\right)  \right]  \left(  \sum_{i=0}^{1}\sum_{j=0}^{\infty}+\sum
_{i=2}^{\infty}\sum_{j=i-2}^{\infty}\right)  \pi_{i,j};
\end{align*}
since $\left(  \sum_{i=0}^{1}\sum_{j=0}^{\infty}+\sum_{i=2}^{\infty}%
\sum_{j=i-2}^{\infty}\right)  \pi_{i,j}=1$, and%
\[
\sum_{k=0}^{\infty}\sum_{l=2}^{\infty}p_{l}\left(  k+l\right)  \pi
_{k,k+l}=\left\{  \overline{\mathbf{p}}_{0}\odot\left[  \pi_{\widetilde{0}%
}Q_{\widetilde{0},0}\left(  -Q_{0,0}\right)  ^{-1}\right]  +\overline
{\mathbf{p}}_{1}\odot\pi_{1}+\sum\limits_{k=2}^{\infty}\overline{\mathbf{p}%
}_{k}\odot\left(  \pi_{1}\widetilde{\mathbf{R}}\mathbf{R}^{k-2}\right)
\right\}  e,
\]
this gives the desired result. We complete the proof. $\square$

\begin{Cor}
\label{CoroD}In the blockchain selfish mining, we have

(a) the long-run average profit $\mathbf{R}_{\text{D}}$ of the dishonest
mining pool increases as the jumping's mining rate $\gamma$ increases;

(b) the long-run average profit $\mathbf{R}_{\text{D}}$ of the dishonest
mining pool increases as the efficiency-increased ratio $\Re$ increases.
\end{Cor}

\textbf{Proof.} We only need to note that%
\begin{equation}
\mathbf{R}_{\text{D}}=\mu\left(  r_{B}+r_{F}\right)  \sum_{k=0}^{\infty}%
\sum_{l=2}^{\infty}p_{l}\left(  k+l\right)  \pi_{k,k+l}-\left(  \widetilde
{\alpha}+\gamma\right)  \left[  c_{E}+c_{A}\left(  1+\Re\right)  \right]  .
\label{Equa-2}%
\end{equation}
By using Corollaries \ref{Cor:Mon-1} and \ref{Cor:Mon-2}, it is easy to give
our desired result. We complete the proof. $\square$

The following theorem provides a multivariate linear structure for the two
long-run average profits $\mathbf{R}_{\text{H}}$ and $\mathbf{R}_{\text{D}}$.
This result is necessary and useful for discussing economics of blockchain
from both design perspective and operations management. Its proof is
straightforward from (\ref{Equa-1}) and (\ref{Equa-2}) and is omitted here.

\begin{The}
In the blockchain selfish mining, the long-run average profits $\mathbf{R}%
_{\text{H}}$ and $\mathbf{R}_{\text{D}}$ are multivariate\ linear in the three
key parameters: $r_{B}+r_{F}$, $c_{E}$ and $c_{A}$.
\end{The}

The following theorem provides a sufficient condition under which the
blockchain can operate normally.

\begin{The}
In the blockchain selfish mining, there exists a minimal positive number
$\mathbf{V}$ such that for $r_{B}+r_{F}>\mathbf{V}$, the blockchain can
operate normally.
\end{The}

\textbf{Proof.} For the blockchain selfish mining, to guarantee the normal
operations of blockchain, we need to satisfy two basic conditions:
$\mathbf{R}_{\text{H}}>0$ and $\mathbf{R}_{\text{D}}>0$. It follows from
(\ref{Equa-1}) and (\ref{Equa-2}) that $\mathbf{R}_{\text{H}}>0$ and
$\mathbf{R}_{\text{D}}>0$ if%
\[
\left(  r_{B}+r_{F}\right)  >\frac{\left(  c_{E}+c_{A}\right)  \left(
\beta-\gamma\right)  }{\mu\sum_{k=2}^{\infty}k\pi_{k,k-2}}%
\]
and%
\[
\left(  r_{B}+r_{F}\right)  >\frac{\left(  \widetilde{\alpha}+\gamma\right)
\left[  c_{E}+c_{A}\left(  1+\Re\right)  \right]  }{\mu\sum_{k=0}^{\infty}%
\sum_{l=2}^{\infty}p_{l}\left(  k+l\right)  \pi_{k,k+l}}.
\]
Let%
\[
\mathbf{V}=\max\left\{  \frac{\left(  c_{E}+c_{A}\right)  \left(  \beta
-\gamma\right)  }{\mu\sum_{k=2}^{\infty}k\pi_{k,k-2}},\frac{\left(
\widetilde{\alpha}+\gamma\right)  \left[  c_{E}+c_{A}\left(  1+\Re\right)
\right]  }{\mu\sum_{k=0}^{\infty}\sum_{l=2}^{\infty}p_{l}\left(  k+l\right)
\pi_{k,k+l}}\right\}  .
\]
Thus it is clear that if $r_{B}+r_{F}>\mathbf{V}$, then $\mathbf{R}_{\text{H}%
}>0$ and $\mathbf{R}_{\text{D}}>0$. This completes the proof. $\square$

In the remainder of this section, we further provide a long-run economic ratio
of the dishonest mining pool over the honest mining pool. We focus on the
economic ratio using the per unit net mining rate. Define%
\begin{equation}
\Im=\frac{\frac{1}{\widetilde{\alpha}+\gamma}\mathbf{R}_{\text{D}}}%
{\frac{1}{\beta-\gamma}\mathbf{R}_{\text{H}}}=\frac{\beta-\gamma}%
{\widetilde{\alpha}+\gamma}\frac{\mathbf{R}_{\text{D}}}{\mathbf{R}_{\text{H}}%
}. \label{Equa-3}%
\end{equation}
Obviously, the economic ratio $\Im$ measures the mining advantage of the
dishonest mining pool.

The following theorem provides a useful approximate evaluation for the
long-run economic ratio if $r_{B}+r_{F}$ is significantly larger than the costs.

\begin{The}
In the blockchain selfish mining, if max$\left\{  c_{E},c_{A}\right\}
/\left(  r_{B}+r_{F}\right)  \approx0$, then there exists a positive constant
$\mathbf{C}$ such that $\Im\approx\mathbf{C}$.
\end{The}

\textbf{Proof.} It follows from (\ref{Equa-3}), (\ref{Equa-1}) and
(\ref{Equa-2}) that for max$\left\{  c_{E},c_{A}\right\}  /\left(  r_{B}%
+r_{F}\right)  \approx0$,
\begin{align*}
\Im &  =\frac{\beta-\gamma}{\widetilde{\alpha}+\gamma}\frac{\mathbf{R}%
_{\text{D}}}{\mathbf{R}_{\text{H}}}\\
&  =\frac{\beta-\gamma}{\widetilde{\alpha}+\gamma}\frac{\mu\left(  r_{B}%
+r_{F}\right)  \sum\limits_{k=0}^{\infty}\sum\limits_{l=2}^{\infty}%
p_{l}\left(  k+l\right)  \pi_{k,k+l}-\left(  \widetilde{\alpha}+\gamma\right)
\left[  c_{E}+c_{A}\left(  1+\Re\right)  \right]  }{\mu\left(  r_{B}%
+r_{F}\right)  \sum\limits_{k=2}^{\infty}k\pi_{k,k-2}-\left(  c_{E}%
+c_{A}\right)  \left(  \beta-\gamma\right)  }\\
&  \approx\frac{\beta-\gamma}{\widetilde{\alpha}+\gamma}\frac{\sum
\limits_{k=0}^{\infty}\sum\limits_{l=2}^{\infty}p_{l}\left(  k+l\right)
\pi_{k,k+l}}{\sum\limits_{k=2}^{\infty}k\pi_{k,k-2}}.
\end{align*}
Let%
\[
\mathbf{C}=\frac{\beta-\gamma}{\widetilde{\alpha}+\gamma}\frac{\sum
\limits_{k=0}^{\infty}\sum\limits_{l=2}^{\infty}p_{l}\left(  k+l\right)
\pi_{k,k+l}}{\sum\limits_{k=2}^{\infty}k\pi_{k,k-2}}.
\]
This completes the proof. $\square$

In addition, we consider the long-run block-pegging rate ratio of the two
mining pools. We define%
\[
\tau=\frac{\frac{1}{\widetilde{\alpha}+\gamma}\sum\limits_{k=0}^{\infty}%
\sum\limits_{l=2}^{\infty}\left(  k+l\right)  \mu p_{l}\pi_{k,k+l}}%
{\frac{1}{\beta-\gamma}\sum\limits_{k=2}^{\infty}k\mu\pi_{k,k-2}}%
=\frac{\beta-\gamma}{\widetilde{\alpha}+\gamma}\frac{\sum\limits_{k=0}%
^{\infty}\sum\limits_{l=2}^{\infty}\left(  k+l\right)  p_{l}\pi_{k,k+l}}%
{\sum\limits_{k=2}^{\infty}k\pi_{k,k-2}}.
\]

The following corollary provides an interesting monotonicity for both the
long-run economic ratio $\Im$ and the long-run block-pegging rate ratio $\tau$
with respect to the efficiency-increased ratio $\Re$. The proof is easy by
means of (b) of Corollaries \ref{CoroH} and \ref{CoroD} and is omitted here.

\begin{Cor}
\label{Cor:Mono}In the blockchain selfish mining, each of the two long-run
ratios $\Im$ and $\tau$ increases, as the efficiency-increased ratio $\Re$ increases.
\end{Cor}

\begin{Rem}
By establishing the pyramid Markov reward process, this paper investigates the
long-run average profits of the honest and dishonest mining pools,
respectively. Further, we show that the long-run average profits are
multivariate linear in the three key parameters: $r_{B}+r_{F}$, $c_{E}$ and
$c_{A}$. Based on this, we can measure the mining efficiency of the dishonest
mining pool. Thus, our method of pyramid Markov reward processes can
effectively improve the revenue analysis given in (3) of Eyal and Sirer
\cite{Eya:2014}.
\end{Rem}

\section{The Mining Profits in the Time Interval $[0,t)$}

In this section, we compute the mining profits in the time interval $[0,t)$ of
the honest and dishonest mining pools, respectively. This is an interesting
but difficult topic due to a high computational complexity and fewer available
results of the transient solution of Markov processes, e.g., see Li
\cite{Li:2010}. To this end, we develop a new effective method that applies
the PH distribution of infinite size and the associated PH renewal process to
determine the expected numbers that the main chains by the honest mining pool
and the dishonest mining pool are pegged on the blockchain, respectively.

For simplicity of calculation, we only consider the stable case of the
blockchain system, that is, the initial probability vector of the Markov
process $\left\{  \left(  I\left(  t\right)  ,J\left(  t\right)  \right)
:t\geq0\right\}  $ is observed at its states following the stationary state
probability distribution $\pi=\left(  \pi_{\widetilde{0}},\pi_{0},\pi_{1}%
,\pi_{2},\pi_{3},\pi_{4},\pi_{5},\ldots\right)  $.

To compute the expected number of generating the main chains by the honest
(resp. dishonest) mining pool, it is a key that state $\left(  0,0\right)  $
is generated by either the honest mining pool or the dishonest mining pool. In
this case, we write state $\left(  0,0\right)  $ by the honest mining pool
(resp. the dishonest mining pool) as an absorbing state $\Delta$ of the Markov
process $\left\{  \left(  I\left(  t\right)  ,J\left(  t\right)  \right)
:t\geq0\right\}  $. The case that state $\left(  0,0\right)  $ by the honest
mining pool is regarded as an absorbing state is illustrated in Figure 6;
while the case that state $\left(  0,0\right)  $ by the dishonest mining pool
is regarded as an absorbing state is illustrated in Figure 7.

In what follows we consider the honest and dishonest mining processes, both of
which have a common initial probability vector $\pi=\left(  \pi_{\widetilde
{0}},\pi_{0},\pi_{1},\pi_{2},\pi_{3},\pi_{4},\pi_{5},\ldots\right)  $ at time
$0$, and show that each of their first passage times to the absorbing state
$\Delta$ follows a PH distribution of infinite size, e,g., see Chapter 8 of Li
\cite{Li:2010} and Chapter 2 of Neuts \cite{Neu:1981}.

\textbf{(a) The honest mining pool}

Now, we compute the expected number of generating the main chains by the
honest mining pool. It is key to compute the first passage time that the
honest mining process enters the absorbing state $\Delta$ for the first time.
Clearly, the honest mining process is a continuous-time Markov process whose
state transition relation of such a process is depicted in Figure 6.

\begin{figure}[h]
\centering                      \includegraphics[width=14cm]{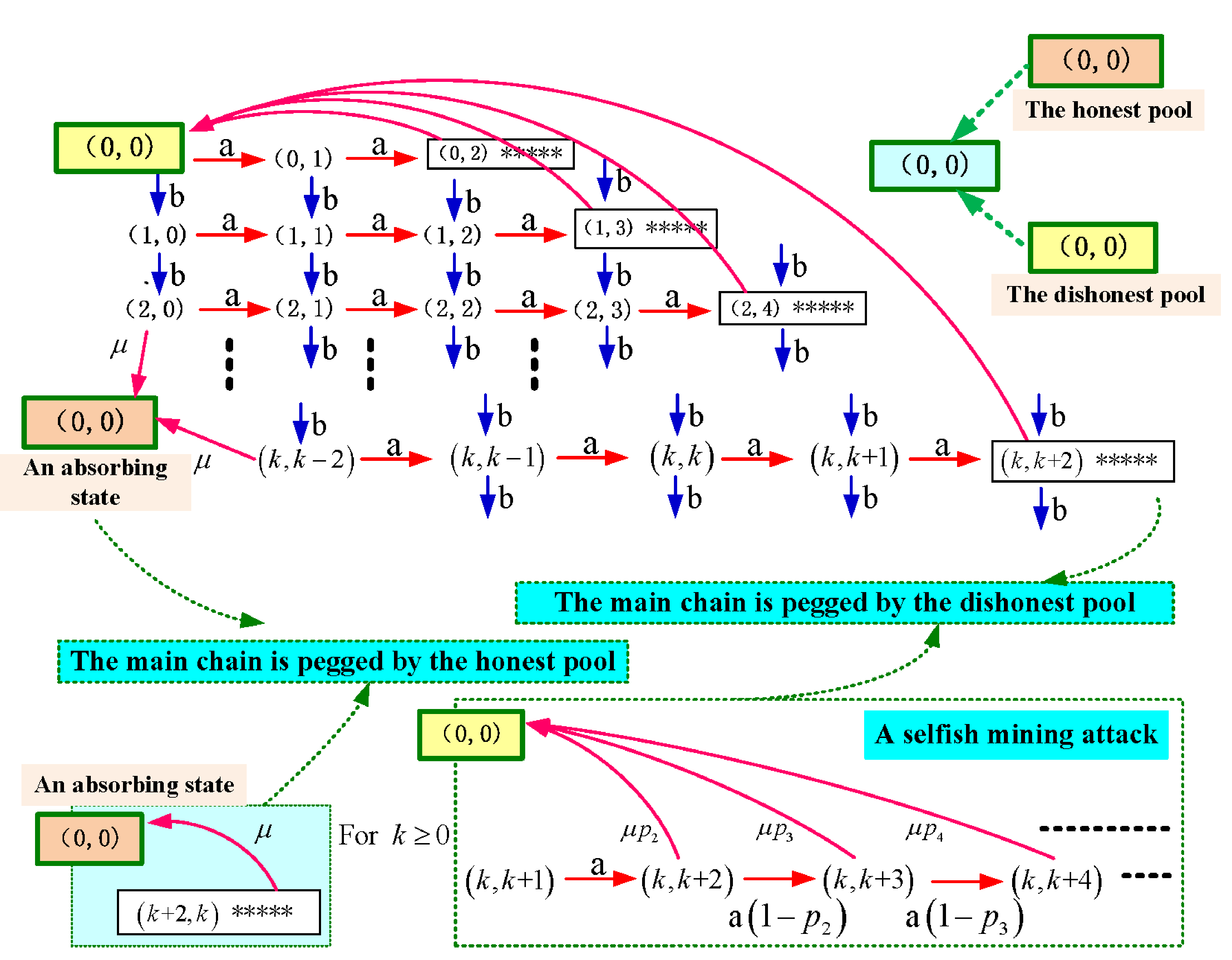}
\caption{The state transition relation of the pyramid Markov process
$\mathbf{Q}_{\text{H}}$ }%
\label{figure:Fig-5}%
\end{figure}

From Figure 6, we write the infinitesimal generator of the honest mining
process with the absorbing state $\Delta$ as%
\[
\mathbf{Q}_{\text{H}}=\left(
\begin{array}
[c]{cc}%
0 & 0\\
T^{0} & T
\end{array}
\right)  ,
\]
where%
\begin{equation}
T=\left(
\begin{array}
[c]{ccccccc}%
Q_{\widetilde{\mathbf{0}},\widetilde{\mathbf{0}}} & Q_{\widetilde{\mathbf{0}%
},0} & Q_{\widetilde{\mathbf{0}},1} &  &  &  & \\
Q_{0,\widetilde{\mathbf{0}}} & Q_{0,0} & Q_{0,1} &  &  &  & \\
\widehat{Q}_{1,\widetilde{\mathbf{0}}} &  & Q_{1,1} & Q_{1,2} &  &  & \\
\widehat{B} &  &  & A & C &  & \\
\widehat{B} &  &  &  & A & C & \\
\vdots &  &  &  &  & \ddots & \ddots
\end{array}
\right)  ,\text{ \ }T^{0}=\left(
\begin{array}
[c]{c}%
0\\
\mathbf{0}\\
\mathbf{0}\\
\mathcal{T}^{0}\\
\mathcal{T}^{0}\\
\vdots
\end{array}
\right)  , \label{Equat-15}%
\end{equation}%
\[
\widehat{B}=\left(
\begin{array}
[c]{c}%
0\\
0\\
0\\
0\\
\mu p_{2}\\
\mu p_{3}\\
\vdots
\end{array}
\right)  ,\text{ \ \ }\mathcal{T}^{0}=\left(
\begin{array}
[c]{c}%
\mu\\
0\\
0\\
0\\
0\\
0\\
\vdots
\end{array}
\right)  ,\text{\ }%
\]
and all the other block-elements are the same as those in (\ref{Equat-2}).

When the main chain by the honest mining pool is pegged on the blockchain, we
denote by $\chi_{\text{H}}$ the time duration starting from the initial mining
moment $0$ until the main chain by the honest mining pool is pegged on the
blockchain for the first time. We call $\chi_{\text{H}}$ the first
generated-pegged time of the main chain by the honest mining pool.

\begin{The}
The first generated-pegged time $\chi_{\text{H}}$ of the main chain by the
honest mining pool is of phase type of infinite size with irreducible
representation $\left(  \pi,T\right)  $. At the same time, we have%
\[
E\left[  \chi_{\text{H}}\right]  =-\pi T_{\max}^{-1}\mathbf{e},
\]
where $T_{\max}^{-1}$ is the maximal non-positive inverse matrix of the matrix
$T$ of infinite sizes.
\end{The}

\textbf{Proof.} To check whether the first generated-pegged time
$\chi_{\text{H}}$ of the main chain by the honest mining pool is of phase
type, it is key to prove that the matrix $T$ is invertible.

To do this, it is easy to see from Figure 6 that the Markov process $T$ is
irreducible. Since $Te\lvertneqq0$, the matrix $T$ is invertible and has a
maximal non-positive inverse matrix $T_{\max}^{-1}$.

To compute the maximal non-positive inverse matrix $T_{\max}^{-1}$, we rewrite
the matrix $T$ as%
\[
T=\left(
\begin{array}
[c]{cccccc}%
\mathcal{B}_{1} & \mathcal{B}_{0} &  &  &  & \\
\mathcal{B}_{2} & A & C &  &  & \\
\mathcal{B}_{2} &  & A & C &  & \\
\mathcal{B}_{2} &  &  & A & C & \\
\vdots &  &  &  & \ddots & \ddots
\end{array}
\right)  ,
\]
wher%
\[
\mathcal{B}_{1}=\left(
\begin{array}
[c]{ccc}%
Q_{\widetilde{\mathbf{0}},\widetilde{\mathbf{0}}} & Q_{\widetilde{\mathbf{0}%
},0} & Q_{\widetilde{\mathbf{0}},1}\\
Q_{0,\widetilde{\mathbf{0}}} & Q_{0,0} & Q_{0,1}\\
\widehat{Q}_{1,\widetilde{\mathbf{0}}} &  & Q_{1,1}%
\end{array}
\right)  ,\text{ }\mathcal{B}_{0}=\left(
\begin{array}
[c]{c}%
0\\
\mathbf{0}\\
Q_{1,2}%
\end{array}
\right)  ,
\]%
\[
\mathcal{B}_{2}=\left(  \widehat{B},\mathbf{0},\mathbf{0}\right)  .
\]
By using Chapter 2 of Li \cite{Li:2010} (e.g., Pages 111 and 112), it is easy
to check that the matrix $T$ exists the UL-type RG-factorization%
\[
T=\left(  I-R_{U}\right)  U_{D}\left(  I-G_{L}\right)  ,
\]
where%
\[
R_{U}=\left(
\begin{array}
[c]{cccccc}%
\mathbf{0} & R_{0} &  &  &  & \\
& \mathbf{0} & R &  &  & \\
&  & \mathbf{0} & R &  & \\
&  &  & \mathbf{0} & R & \\
&  &  &  & \ddots & \ddots
\end{array}
\right)  ,
\]%
\[
U_{D}=\text{diag}\left(  U_{0},U,U,U,\ldots\right)  ,
\]%
\[
G_{L}=\left(
\begin{array}
[c]{ccccc}%
\mathbf{0} &  &  &  & \\
G_{1} & \mathbf{0} &  &  & \\
G_{2} &  & \mathbf{0} &  & \\
G_{3} &  &  & \mathbf{0} & \\
\vdots &  &  &  & \ddots
\end{array}
\right)  .
\]
we obtain%
\[
\left(  I-R_{U}\right)  ^{-1}=\left(
\begin{array}
[c]{cccccc}%
I & R_{0} & R_{0}R & R_{0}R^{2} & R_{0}R^{3} & \cdots\\
& I & R & R^{2} & R^{3} & \cdots\\
&  & I & R & R^{2} & \cdots\\
&  &  & I & R & \cdots\\
&  &  &  & \ddots & \ddots
\end{array}
\right)  ,
\]%
\[
U_{D}^{-1}=\text{diag}\left(  U_{0}^{-1},U^{-1},U^{-1},U^{-1},\ldots\right)
,
\]%
\[
\left(  I-G_{L}\right)  ^{-1}=\left(
\begin{array}
[c]{ccccc}%
I &  &  &  & \\
G_{1} & I &  &  & \\
G_{2} &  & I &  & \\
G_{3} &  &  & I & \\
\vdots &  &  &  & \ddots
\end{array}
\right)  .
\]
This gives%
\begin{align*}
T_{\max}^{-1}  &  =\left(
\begin{array}
[c]{ccccc}%
I &  &  &  & \\
G_{1} & I &  &  & \\
G_{2} &  & I &  & \\
G_{3} &  &  & I & \\
\vdots &  &  &  & \ddots
\end{array}
\right)  \left(
\begin{array}
[c]{ccccc}%
U_{0}^{-1} &  &  &  & \\
& U^{-1} &  &  & \\
&  & U^{-1} &  & \\
&  &  & U^{-1} & \\
&  &  &  & \ddots
\end{array}
\right) \\
&  \left(
\begin{array}
[c]{cccccc}%
I & R_{0} & R_{0}R & R_{0}R^{2} & R_{0}R^{3} & \cdots\\
& I & R & R^{2} & R^{3} & \cdots\\
&  & I & R & R^{2} & \cdots\\
&  &  & I & R & \cdots\\
&  &  &  & \ddots & \ddots
\end{array}
\right)  ,
\end{align*}
which can further be computed easily. Therefore, the first generated-pegged
time $\chi_{\text{H}}$ of the main chain by the honest mining pool is of phase
type of infinite size with the irreducible representation $\left(
\pi,T\right)  $, because the Markov process $T+T^{0}\pi$ is irreducible. This
completes the proof. $\square$

When the mining difficulty level is not adjusted, it is easy to see that the
number $N_{\text{H}}(t)$ of generating the main chain by the honest mining
pool in the time interval $[0,t)$ is a renewal process whose interarrival
times are i.i.d. and the first generated-pegged time is $\chi_{\text{H}}$.
Since the first generated-pegged time $\chi_{\text{H}}$ is of phase type of
infinite size with irreducible representation $\left(  \pi,T\right)  $, its
associated renewal process is a PH renewal process of infinite size.
Therefore, $N_{\text{H}}(t)$ is also the number of renewals of the PH renewal
process in the time interval $(0,t]$. Let $J(t)$ be the phase of the PH
renewal process at time $t$. It is clear that $J(t)\in\left\{  0,1,2,\ldots
\right\}  $. We write%
\[
P_{i,j}(n,t)=P\left\{  N_{\text{H}}(t)=n,J(t)=j\quad|\text{ }N_{\text{H}%
}(0)=0,J(0)=i\right\}
\]
for $n\geq0$ and $i,j=0,1,2,\ldots$. We denote by $P(n,t)$ a matrix of
infinite size with the $\left(  i,j\right)  $th element $P_{i,j}(n,t)$. It is
clear from Neuts \cite{Neu:1981} and Li \cite{Li:2010} that the matrix
sequence $\left\{  P(n,t)\right\}  $ satisfies the Chapman-Kolmogorov
differential equations as follows:%
\begin{align}
\frac{\text{d}}{\text{d}t}P(0,t)  &  =TP(0,t),\nonumber\\
\frac{\text{d}}{\text{d}t}P(n,t)  &  =TP(n,t)+T^{0}\pi P(n-1,t),\text{ }%
n\geq1,\nonumber
\end{align}
with $P(n,0)=\delta_{n0}I$. Let $P^{\ast}(z,t)=\overset{\infty}{\underset
{n=0}{\sum}}z^{n}P(n,t)$. Then%
\[
P^{\ast}(z,t)=\exp\left\{  (T+zT^{0}\pi)t\right\}  ,\text{ }t\geq0.
\]
Let $Q_{\text{H}}^{\ast}=T+$ $T^{0}\pi$. Note that the Markov process
$Q_{\text{H}}^{\ast}$ is irreducible and positive recurrent, and thus there
must exist the stationary probability vector $\upsilon$ such that $\upsilon
Q_{\text{H}}^{\ast}=0$ and $\upsilon\mathbf{e}=1$. In this case, we have%
\begin{align*}
E\left[  N_{\text{H}}(t)\right]   &  =\frac{\text{d}}{\text{d}z}\upsilon
P^{\ast}(z,t)e_{|z=1}e=\upsilon T^{0}\pi t\exp(Q_{\text{H}}^{\ast}%
t)\mathbf{e}\\
&  =\frac{t}{E\left[  \left(  \chi_{\text{H}}\right)  \right]  }\pi
\exp(Q_{\text{H}}^{\ast}t)\mathbf{e},
\end{align*}
since $\upsilon T^{0}=1/E\left[  \left(  \chi_{\text{H}}\right)  \right]  $.

Now, we compute the stationary average number $\Psi_{\text{H}}$ of blocks in
the main chain by the honest mining pool. It follows from Figure 6 that%
\begin{equation}
\Psi_{\text{H}}=\sum_{k=2}^{\infty}k\pi_{k,k-2}. \label{BlockN-1}%
\end{equation}

Based on the above analysis, the mining profit in the time interval $[0,t)$ of
the honest mining pool is given by%
\begin{align*}
\mathcal{R}_{\text{H}}\left(  t\right)   &  =E\left[  N_{\text{H}}(t)\right]
\left(  r_{B}+r_{F}\right)  \Psi_{\text{H}}-t\left(  c_{E}+c_{A}\right)
\left(  \beta-\gamma\right) \\
&  =\frac{t\left(  r_{B}+r_{F}\right)  \Psi_{\text{H}}}{E\left[  \left(
\chi_{\text{H}}\right)  \right]  }\pi\exp(Q_{\text{H}}^{\ast}t)\mathbf{e}%
-t\left(  c_{E}+c_{A}\right)  \left(  \beta-\gamma\right)  .
\end{align*}

\textbf{(b) The dishonest mining pool}

By a similar analysis to that in (a), we can discuss the mining profit in the
time interval $[0,t)$ of the dishonest mining pool. Here, we provide a simple
outline of such an analysis.

Our aim is to compute the expected number of generating the main chains by the
dishonest mining pool. To this end, it is key to compute the first passage
time that the honest mining process enters the absorbing state $\Delta$ for
the first time. The dishonest mining process is a continuous-time pyramid
Markov process whose state transition relation of such a process is depicted
in Figure 7.

\begin{figure}[h]
\centering                      \includegraphics[width=14cm]{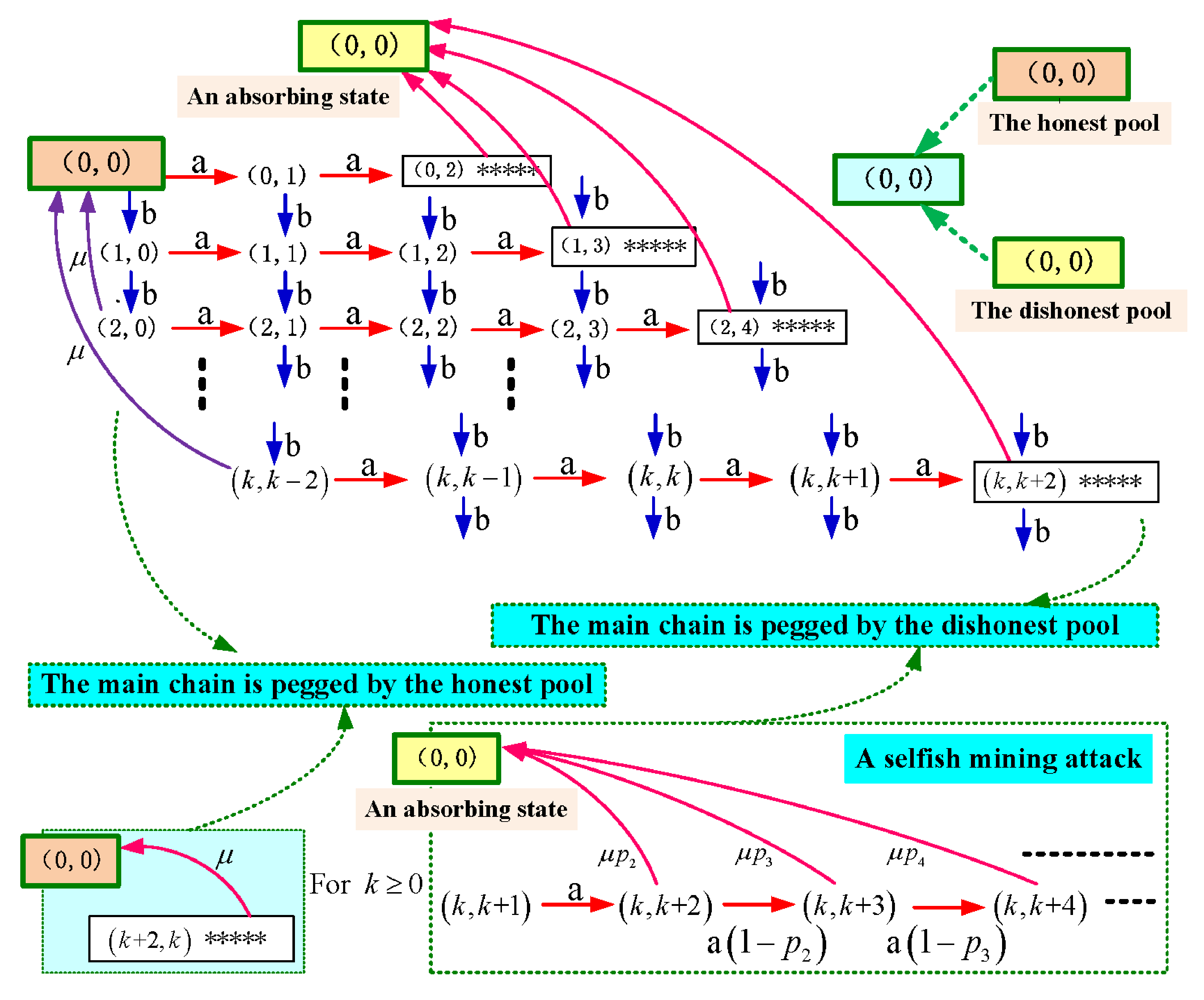}
\caption{The state transition relation of the pyramid Markov process
$\mathbf{Q}_{\text{D}}$ }%
\label{figure:Fig-6}%
\end{figure}

From Figure 7, the infinitesimal generator of the pyramid Markov process with
an absorbing state $\left(  0,0\right)  $ is%
\[
\mathbf{Q}_{\text{D}}=\left(
\begin{array}
[c]{cc}%
0 & 0\\
S^{0} & S
\end{array}
\right)  ,
\]
where%
\begin{equation}
S=\left(
\begin{array}
[c]{ccccccc}%
Q_{\widetilde{\mathbf{0}},\widetilde{\mathbf{0}}} & Q_{\widetilde{\mathbf{0}%
},0} & Q_{\widetilde{\mathbf{0}},1} &  &  &  & \\
\mathbf{0} & Q_{0,0} & Q_{0,1} &  &  &  & \\
\mathbf{0} &  & Q_{1,1} & Q_{1,2} &  &  & \\
\widetilde{B} &  &  & A & C &  & \\
\widetilde{B} &  &  &  & A & C & \\
\vdots &  &  &  &  & \ddots & \ddots
\end{array}
\right)  ,\text{ \ }S^{0}=\left(
\begin{array}
[c]{c}%
0\\
S_{0}^{0}\\
S_{1}^{0}\\
\mathbb{S}^{0}\\
\mathbb{S}^{0}\\
\vdots
\end{array}
\right)  , \label{Equat-16}%
\end{equation}
where%
\[
\widetilde{B}=\left(
\begin{array}
[c]{c}%
\mu\\
0\\
0\\
0\\
0\\
0\\
\vdots
\end{array}
\right)  ,S_{0}^{0}=\left(
\begin{array}
[c]{c}%
0\\
\mu p_{2}\\
\mu p_{3}\\
\mu p_{4}\\
\mu p_{5}\\
\mu p_{6}\\
\vdots
\end{array}
\right)  ,S_{1}^{0}=\left(
\begin{array}
[c]{c}%
0\\
0\\
0\\
\mu p_{2}\\
\mu p_{3}\\
\mu p_{4}\\
\vdots
\end{array}
\right)  ,\mathbb{S}^{0}=\left(
\begin{array}
[c]{c}%
0\\
0\\
0\\
0\\
\mu p_{3}\\
\mu p_{4}\\
\vdots
\end{array}
\right)  .
\]

When the main chain by the dishonest mining pool is pegged on the blockchain,
we denote by $\chi_{\text{D}}$ the time duration starting from the initial
mining moment $0$ until the main chain by the dishonest mining pool is pegged
on the blockchain for the first time. We call $\chi_{\text{D}}$ the first
generated-pegged time of the main chain by the dishonest mining pool.

\begin{The}
The first generated-pegged time $\chi_{\text{H}}$ of the main chain by the
dishonest mining pool is of phase type of infinite size with irreducible
representation $\left(  \pi,S\right)  $. At the same time, we have%
\[
E\left[  \chi_{\text{D}}\right]  =-\pi S_{\max}^{-1}\mathbf{e},
\]
where $S_{\max}^{-1}$ is the maximal non-positive inverse matrix of the matrix
$S$ of infinite sizes.
\end{The}

Now, we compute the stationary average number $\Psi_{\text{D}}$ of blocks in
the main chain by the honest mining pool. It follows from Figure 6 that%
\begin{equation}
\Psi_{\text{D}}=\sum_{k=0}^{\infty}\sum_{l=2}^{\infty}\left(  k+l\right)
p_{l}\pi_{k,k+l}. \label{BlockN-2}%
\end{equation}
Therefore, the mining profit in the time interval $[0,t)$ of the dishonest
mining pool is given by%
\[
\mathcal{R}_{\text{H}}\left(  t\right)  =\frac{t\left(  r_{B}+r_{F}\right)
\Psi_{\text{D}}}{E\left[  \left(  \chi_{\text{D}}\right)  \right]  }\pi
\exp(Q_{\text{D}}^{\ast}t)e-t\left(  \widetilde{\alpha}+\gamma\right)  \left[
c_{E}+c_{A}\left(  1+\Re\right)  \right]  ,
\]
where $Q_{\text{D}}^{\ast}=S+S^{0}\pi$.

\section{A One-Dimensional Markov Model}

In this section, we provide a new one-dimensional Markov model to further deal
with the mining processes of the two honest and dishonest mining pools. To
this end, we establish a new continuous-time Markov process to express the
blockchain selfish mining.

Let $N_{\text{H}}(t)$ and $N_{\text{D}}(t)$ denote the numbers of blocks mined
by the honest and dishonest mining pools in the time interval $[0,t)$,
respectively. We write $\mathcal{N}\left(  t\right)  =N_{\text{D}%
}(t)-N_{\text{H}}(t)$. It is easy to see that $\left\{  \mathcal{N}\left(
t\right)  :t\geq0\right\}  $ is a one-dimensional Markov process whose state
space is $\mathbf{E}=\left\{  \left(  0,0\right)  ,-2,-1,0,1,2,\ldots\right\}
$, and its state transition relations are depicted in Figure 8.

\begin{figure}[h]
\centering                   \includegraphics[width=14cm]{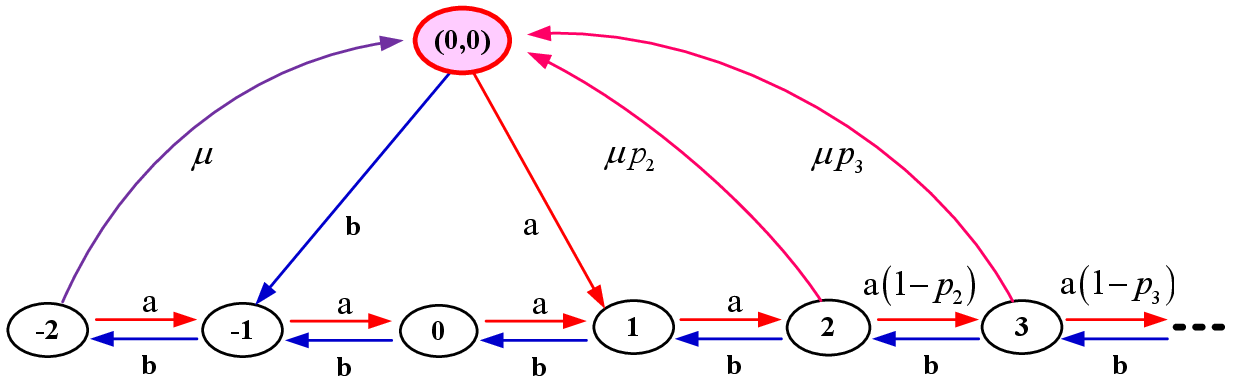}
\caption{The state transition relations of one-dimensional Markov processes }%
\label{figure:Fig-7}%
\end{figure}

By using Figure 8, the infinitesimal generator of the Markov process $\left\{
\mathcal{N}\left(  t\right)  :t\geq0\right\}  $ is given by%
\[
Q_{E}=\left(
\begin{array}
[c]{ccccccccc}%
-\left(  a+b\right)  & 0 & b & 0 & a &  &  &  & \\
\mu & -\left(  a+\mu\right)  & a & 0 & 0 &  &  &  & \\
0 & b & -\left(  a+b\right)  & a & 0 &  &  &  & \\
0 &  & b & -\left(  a+b\right)  & a &  &  &  & \\
0 &  &  & b & -\left(  a+b\right)  & a &  &  & \\
\mu p_{2} &  &  &  & b & -\xi_{2} & a\left(  1-p_{2}\right)  &  & \\
\mu p_{3} &  &  &  &  & b & -\xi_{3} & a\left(  1-p_{3}\right)  & \\
\vdots &  &  &  &  &  & \ddots & \ddots & \ddots
\end{array}
\right)  .
\]

Now, we use the mean-drift method to study the stability of the Markov process
$Q_{E}$. For $k\geq5$, it is easy to see from the infinitesimal generator
$Q_{E}$ that the downward rate from level $k$ to levels $k-1$ and $0$ is given
by $r_{\text{High }\rightarrow\text{ Low}}^{\left(  k\right)  }=b+\mu p_{k-3}%
$; while the upward rate from level $k$ to levels $k+1$ is given by
$r_{\text{Low }\rightarrow\text{ High }}^{\left(  k\right)  }=a\left(
1-p_{k-3}\right)  $. Note that $\lim_{k\rightarrow\infty}p_{k}=1$, and we
obtain that $\lim_{k\rightarrow\infty}r_{\text{High }\rightarrow\text{ Low}%
}^{\left(  k\right)  }=b+\mu>0$ and $\lim_{k\rightarrow\infty}r_{\text{Low
}\rightarrow\text{ High }}^{\left(  k\right)  }=0$. This gives%
\[
\lim_{k\rightarrow\infty}r_{\text{High }\rightarrow\text{ Low}}^{\left(
k\right)  }>\lim_{k\rightarrow\infty}r_{\text{Low }\rightarrow\text{ High }%
}^{\left(  k\right)  }.
\]
Based on this, there exists a big enough positive integer $n_{0}$ such that
$r_{\text{High }\rightarrow\text{ Low}}^{\left(  k\right)  }>r_{\text{Low
}\rightarrow\text{ High }}^{\left(  k\right)  }$ for each $k>n_{0}$.
Therefore, by using the mean drift method, it is easy to check that the Markov
process $Q_{E}$ is irreducible and positive recurrent.

If the Markov process $Q_{E}$ is irreducible and positive recurrent, then
there must exist one unique stationary probability vector $\Psi=\left(
\psi_{0,0},\psi_{-2},\psi_{-1},\psi_{0},\psi_{1},\psi_{2},\psi_{3}%
,\ldots\right)  $, which satisfies the system of linear equations: $\Psi
Q_{\mathbf{E}}=0$ and $\Psi\mathbf{e}=1$.

Now, we provide a matrix-analytic method to compute the stationary probability
vector $\Psi$ from the system of linear equations: $\Psi Q_{\mathbf{E}}=0$ and
$\Psi\mathbf{e}=1$. To do this, we write the infinitesimal generator
$Q_{\mathbf{E}}$ as the standard structured form of the matrix-analytic method
as follows:%
\[
Q_{\mathbf{E}}=\left(
\begin{array}
[c]{cccccc}%
B_{1} & B_{0} &  &  &  & \\
B_{2} & A_{1}^{\left(  1\right)  } & A_{0}^{\left(  1\right)  } &  &  & \\
\mu p_{4} & A_{2}^{\left(  2\right)  } & A_{1}^{\left(  2\right)  } &
A_{0}^{\left(  2\right)  } &  & \\
\mu p_{5} &  & A_{2}^{\left(  3\right)  } & A_{1}^{\left(  3\right)  } &
A_{0}^{\left(  3\right)  } & \\
\vdots &  &  & \ddots & \ddots & \ddots
\end{array}
\right)  ,
\]
where%
\[
B_{1}=\left(
\begin{array}
[c]{cccccc}%
-\left(  a+b\right)  & 0 & b & 0 & a & \\
\mu & -\left(  a+\mu\right)  & a & 0 & 0 & \\
0 & b & -\left(  a+b\right)  & a & 0 & \\
0 &  & b & -\left(  a+b\right)  & a & \\
0 &  &  & b & -\left(  a+b\right)  & a\\
\mu p_{2} &  &  &  & b & -\xi_{2}%
\end{array}
\right)  ,
\]%
\[
B_{0}=\left(  0,0,0,0,0,a\left(  1-p_{2}\right)  \right)  ^{T},
\]%
\[
B_{2}=\left(  \mu p_{3},0,0,0,0,b\right)  ;
\]%
\[
A_{1}^{\left(  1\right)  }=-\xi_{3},\text{ \ }A_{0}^{\left(  1\right)
}=a\left(  1-p_{3}\right)  ;
\]
for $k\geq2$%
\[
A_{1}^{\left(  k\right)  }=-\xi_{k+2},\text{ \ }A_{0}^{\left(  k\right)
}=a\left(  1-p_{k+2}\right)  ,\text{ \ }A_{2}^{\left(  k\right)  }=b.
\]

Let the sequence of numbers $\left\{  R_{k}:k\geq2\right\}  $ be the minimal
positive solution to the system of quadratic equations%
\[
A_{0}^{\left(  k-1\right)  }+R_{k}A_{1}^{\left(  k\right)  }+R_{k}R_{k+1}%
A_{2}^{\left(  k+1\right)  }=0,\text{ \ }k\geq2,
\]
or%
\begin{equation}
a\left(  1-p_{k+1}\right)  -\xi_{k+2}R_{k}+bR_{k}R_{k+1}=0,\text{ \ }k\geq2.
\label{Equ-7}%
\end{equation}

\begin{Rem}
It is a key to solve the system of nonlinear equations (\ref{Equ-7}). To this
end, some effective algorithms were developed in Bright and Taylor
\cite{Bri:1995, Bri:1997}, and also Liu et al. \cite{Liu:2020} provided some
practical examples of such a computation.
\end{Rem}

By using the sequence of numbers $\left\{  R_{k}:k\geq2\right\}  $, it follows
from Section 1.3 in Chapter one of Li \cite{Li:2010} that%
\[
\psi_{k+2}=\psi_{3}R_{k}R_{k-1}R_{k-2}\cdots R_{2},\text{ \ }k\geq1,
\]
and the seven positive numbers $\psi_{\left(  0,0\right)  }$, $\psi_{-2}$,
$\psi_{-1}$, $\psi_{0}$, $\psi_{1}$, $\psi_{2}$ and $\psi_{3}$ uniquely
satisfy the following system of linear equations:%
\begin{equation}
\left\{
\begin{array}
[c]{l}%
-\left(  a+b\right)  \psi_{\left(  0,0\right)  }+\mu\psi_{-2}+\mu p_{2}%
\psi_{2}+\mu\left(  p_{3}+\sum\limits_{k=2}^{\infty}p_{k+2}R_{k}R_{k-1}%
R_{k-2}\cdots R_{2}\right)  \psi_{3}=0,\\
-\left(  a+b\right)  \psi_{-2}+b\psi_{-1}=0,\\
b\psi_{\left(  0,0\right)  }+a\psi_{-2}-\left(  a+b\right)  \psi_{-1}%
+b\psi_{0}=0,\\
a\psi_{-1}-\left(  a+b\right)  \psi_{0}+b\psi_{1}=0,\\
a\psi_{\left(  0,0\right)  }+a\psi_{0}-\left(  a+b\right)  \psi_{1}+b\psi
_{2}=0,\\
a\psi_{1}-\xi_{2}\psi_{2}+b\psi_{3}=0,\\
\psi_{\left(  0,0\right)  }+\psi_{-2}+\psi_{-1}+\psi_{0}+\psi_{1}+\psi
_{2}+\left(  1+\sum\limits_{k=2}^{\infty}p_{k+2}R_{k}R_{k-1}R_{k-2}\cdots
R_{2}\right)  \psi_{3}=1.
\end{array}
\right.  \label{Equ-8}%
\end{equation}
Once the sequence of numbers $\left\{  R_{k}:k\geq2\right\}  $ is obtained
numerically, we further solve the system of linear equations (\ref{Equ-8}).
Based on this, we obtain the stationary probability vector $\Psi=\left(
\psi_{0,0},\psi_{-2},\psi_{-1},\psi_{0},\psi_{1},\psi_{2},\psi_{3}%
,\ldots\right)  $.

\begin{Rem}
Since the Markov process $Q_{\mathbf{E}}$ is not a birth-death process due to
the key state $\left(  0,0\right)  $ with infinitely many rates of entry,
there does not exist any explicit expression for the stationary probability
vector. Despite of this, it still has the structure of a birth-death process
just beyond the six boundary states: $\left(  0,0\right)  ,-2,-1,0,1,$ and
$2$. Based on this, we can apply the matrix-product solution to express and
further numerically compute the stationary probability vector of the Markov
process $Q_{\mathbf{E}}$.
\end{Rem}

In what follows we show how to use the stationary probability vector $\Psi$ to
study the expected profits of the blockchain selfish mining.

For the Markov process $Q_{\mathbf{E}}$, we denote by $\mathbf{R}_{\text{H}}$
and $\mathbf{R}_{\text{D}}$ the long-run average profits of the honest and
dishonest mining pools, respectively. Then%
\begin{align*}
\mathbf{R}_{\text{H}}  &  =\psi_{-2}2\mu\left(  r_{B}+r_{F}\right)
\Psi_{\text{H}}-\left(  \psi_{0,0}+\sum\limits_{i=-2}^{\infty}\psi_{i}\right)
\left(  c_{E}+c_{A}\right)  \left(  \beta-\gamma\right) \\
&  =\psi_{-2}2\mu\left(  r_{B}+r_{F}\right)  \Psi_{\text{H}}-\left(
c_{E}+c_{A}\right)  \left(  \beta-\gamma\right)  ,
\end{align*}
and%
\begin{align*}
\mathbf{R}_{\text{D}}  &  =\sum\limits_{k=2}^{\infty}\psi_{k}p_{k}k\mu\left(
r_{B}+r_{F}\right)  \Psi_{\text{D}}-\left(  \psi_{0,0}+\sum\limits_{i=-2}%
^{\infty}\psi_{i}\right)  \psi_{i}\left(  \widetilde{\alpha}+\gamma\right)
\left[  c_{E}+c_{A}\left(  1+\Re\right)  \right] \\
&  =\sum\limits_{k=2}^{\infty}\psi_{k}p_{k}k\mu\left(  r_{B}+r_{F}\right)
\Psi_{\text{D}}-\left(  \widetilde{\alpha}+\gamma\right)  \left[  c_{E}%
+c_{A}\left(  1+\Re\right)  \right]  ,
\end{align*}
where $\Psi_{\text{H}}$ and $\Psi_{\text{D}}$ are given in (\ref{BlockN-1})
and (\ref{BlockN-2}), respectively. Further, the total relative long-run
average profit of the blockchain is given by
\begin{align}
\mathbf{R}=  &  \rho_{1}\mathbf{R}_{\text{H}}+\rho_{2}\mathbf{R}_{\text{D}%
}\nonumber\\
=  &  \rho_{1}\psi_{-2}2\mu\left(  r_{B}+r_{F}\right)  \Psi_{\text{H}}%
+\rho_{2}\sum\limits_{k=2}^{\infty}\psi_{k}kp_{k}\mu\left(  r_{B}%
+r_{F}\right)  \Psi_{\text{D}}\nonumber\\
&  -\rho_{1}\left(  c_{E}+c_{A}\right)  \left(  \beta-\gamma\right)  -\rho
_{2}\left(  \widetilde{\alpha}+\gamma\right)  \left[  c_{E}+c_{A}\left(
1+\Re\right)  \right]  , \label{Equ-9}%
\end{align}
where%
\[
\rho_{1}=\frac{\psi_{-2}}{\psi_{-2}+\sum\limits_{k=2}^{\infty}\psi
_{k}^{\left(  2\right)  }},\text{ \ }\rho_{2}=\frac{\sum\limits_{k=2}^{\infty
}\psi_{k}}{\psi_{-2}+\sum\limits_{k=2}^{\infty}\psi_{k}},
\]


Finally, we provide a long-run economic ratio of the dishonest mining pool
over the honest mining pool by using per unit net mining rate as follows:%
\begin{equation}
\Im=\frac{\frac{1}{\widetilde{\alpha}+\gamma}\mathbf{R}_{\text{D}}}%
{\frac{1}{\beta-\gamma}\mathbf{R}_{\text{H}}}=\frac{\beta-\gamma}%
{\widetilde{\alpha}+\gamma}\frac{\mathbf{R}_{\text{D}}}{\mathbf{R}_{\text{H}}%
}. \label{Equ-10}%
\end{equation}
We can numerically indicate that the long-run ratio $\Im$ increases as the
jumping's mining rate $\gamma$ (or the efficiency-increased ratio $\Re$) increases.

\section{No Network Latency}

In this section, we further discuss a special case without network latency of
the one-dimensional Markov model discussed in Section 8.

When the network latency is very short so that it can be ignored, we assume
that $\mu=+\infty$ or $1/\mu=0$. In this case, Figure 9 provides a simple
state transition relation of the Markov process $\left\{  \mathcal{N}\left(
t\right)  :t\geq0\right\}  $.

\begin{figure}[h]
\centering                   \includegraphics[width=11cm]{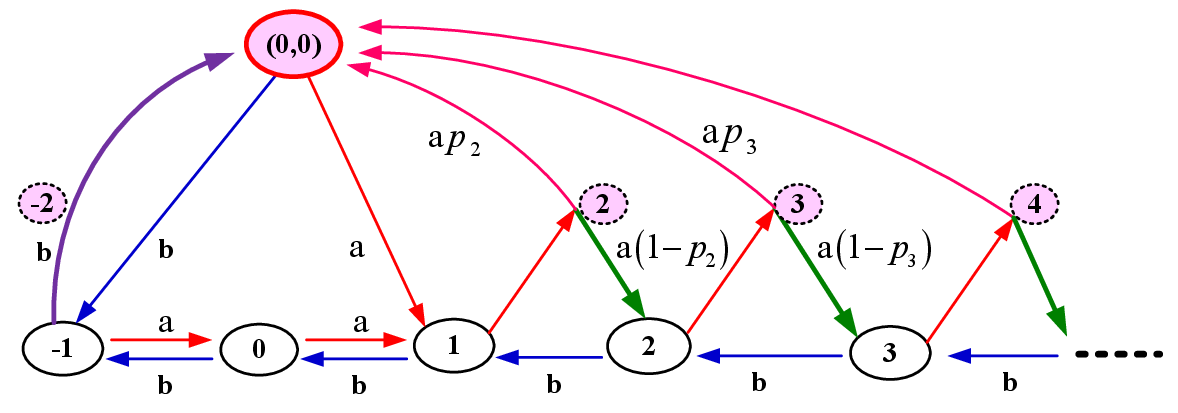}
\caption{The state transition relations of the Markov process for $\mu
=+\infty$}%
\label{figure:Fig-8}%
\end{figure}

From Figure 9, it is easy to check that the infinitesimal generator of the
Markov process $\left\{  \mathcal{N}\left(  t\right)  :t\geq0\right\}  $ is
given by%
\[
Q_{\mu=+\infty}=\left(
\begin{array}
[c]{ccccccc}%
-\left(  a+b\right)  & b & 0 & a &  &  & \\
b & -\left(  a+b\right)  & a & 0 &  &  & \\
0 & b & -\left(  a+b\right)  & a &  &  & \\
ap_{2} &  & b & -\left(  a+b\right)  & a\left(  1-p_{2}\right)  &  & \\
ap_{3} &  &  & b & -\left(  a+b\right)  & a\left(  1-p_{3}\right)  & \\
\vdots &  &  &  & \ddots & \ddots & \ddots
\end{array}
\right)  .
\]
Obviously, the Markov process $Q_{\mu=+\infty}$ is irreducible and positive
recurrent. Let $\mathbb{\Psi}^{\left(  \infty\right)  }=\left(  \psi
_{0,0}^{\left(  \infty\right)  },\psi_{-1}^{\left(  \infty\right)  },\psi
_{0}^{\left(  \infty\right)  },\psi_{1}^{\left(  \infty\right)  },\psi
_{2}^{\left(  \infty\right)  },\ldots\right)  $ be the stationary probability
vector of the Markov process $Q_{\mu=+\infty}$.

Let the sequence of numbers $\left\{  R_{k}^{\left(  \infty\right)
}:k=1,2,3,\ldots\right\}  $ be the minimal positive solution to the system of
nonlinear equations
\begin{equation}
a\left(  1-p_{k+1}\right)  -R_{k}^{\left(  \infty\right)  }\left(  a+b\right)
+R_{k}^{\left(  \infty\right)  }R_{k+1}^{\left(  \infty\right)  }b=0,\text{
\ }k=1,2,3,\ldots. \label{Equ-11}%
\end{equation}
Then from the system of linear equations $\Psi^{\left(  \infty\right)  }%
Q_{\mu=+\infty}=0$ and $\Psi^{\left(  \infty\right)  }\mathbf{e}=1$, by using
Section 1.3 in Chapter one of Li \cite{Li:2010}, we obtain that for
$k=2,3,4,\ldots$,%
\[
\psi_{k}^{\left(  \infty\right)  }=\psi_{1}^{\left(  \infty\right)  }%
R_{k-1}^{\left(  \infty\right)  }R_{k-2}^{\left(  \infty\right)  }\cdots
R_{2}^{\left(  \infty\right)  }R_{1}^{\left(  \infty\right)  },
\]
and $\psi_{0,0}^{\left(  \infty\right)  },\psi_{-1}^{\left(  \infty\right)
},\psi_{0}^{\left(  \infty\right)  }$ and $\psi_{1}^{\left(  \infty\right)  }$
uniquely satisfy the following system of linear equations%
\[
\left\{
\begin{array}
[c]{l}%
-\psi_{0,0}^{\left(  \infty\right)  }\left(  a+b\right)  +\psi_{-1}^{\left(
\infty\right)  }b+\psi_{1}^{\left(  \infty\right)  }\left(  ap_{2}+\sum
_{k=2}^{\infty}ap_{k+1}R_{k-1}^{\left(  \infty\right)  }R_{k-2}^{\left(
\infty\right)  }\cdots R_{2}^{\left(  \infty\right)  }R_{1}^{\left(
\infty\right)  }\right)  =0,\\
\psi_{0,0}^{\left(  \infty\right)  }b-\psi_{-1}^{\left(  \infty\right)
}\left(  a+b\right)  +\psi_{0}^{\left(  \infty\right)  }b=0,\\
\psi_{-1}^{\left(  \infty\right)  }a-\psi_{0}^{\left(  \infty\right)  }\left(
a+b\right)  +\psi_{1}^{\left(  \infty\right)  }b=0,\\
\psi_{0,0}^{\left(  \infty\right)  }a+\psi_{0}^{\left(  \infty\right)  }%
a-\psi_{1}^{\left(  \infty\right)  }\left(  a+b\right)  +\psi_{1}^{\left(
\infty\right)  }R_{1}^{\left(  \infty\right)  }b=0,\\
\psi_{0,0}^{\left(  \infty\right)  }+\psi_{-1}^{\left(  \infty\right)  }%
+\psi_{0}^{\left(  \infty\right)  }+\psi_{1}^{\left(  \infty\right)  }\left(
1+\sum_{k=2}^{\infty}R_{k-1}^{\left(  \infty\right)  }R_{k-2}^{\left(
\infty\right)  }\cdots R_{2}^{\left(  \infty\right)  }R_{1}^{\left(
\infty\right)  }\right)  =1.
\end{array}
\right.
\]
Similarly, we can solve the system of nonlinear equations (\ref{Equ-11}) by
means of the effective algorithms developed in Bright and Taylor
\cite{Bri:1995, Bri:1997}.

For the the Markov process $Q_{\mu=+\infty}$, we denote by $\mathbf{R}%
_{\text{H}}^{\left(  \infty\right)  }$ and $\mathbf{R}_{\text{D}}^{\left(
\infty\right)  }$ the long-run average profits of the honest and dishonest
mining pools, respectively. Then%
\[
\mathbf{R}_{\text{H}}^{\left(  \infty\right)  }=b\psi_{-1}^{\left(
\infty\right)  }\cdot2\left(  r_{B}+r_{F}\right)  \Psi_{\text{H}}-\left(
c_{E}+c_{A}\right)  \left(  \beta-\gamma\right)  ,
\]%
\[
\mathbf{R}_{\text{D}}^{\left(  \infty\right)  }=a\sum\limits_{k=1}^{\infty
}\psi_{k}^{\left(  \infty\right)  }\cdot p_{k+1}\left(  k+1\right)  \left(
r_{B}+r_{F}\right)  \Psi_{\text{D}}-\left(  \widetilde{\alpha}+\gamma\right)
\left[  c_{E}+c_{A}\left(  1+\Re\right)  \right]  ,
\]
where $\Psi_{\text{H}}$ and $\Psi_{\text{D}}$ are given in (\ref{BlockN-1})
and (\ref{BlockN-2}), respectively. Thus, the total long-run average profit of
the blockchain is given by%
\begin{align}
\mathbf{R}^{\left(  \infty\right)  }  &  =\widetilde{\rho}_{1}\mathbf{R}%
_{\text{H}}^{\left(  \infty\right)  }+\widetilde{\rho}_{2}\mathbf{R}%
_{\text{D}}^{\left(  \infty\right)  }\nonumber\\
&  =\widetilde{\rho}_{1}b\psi_{-1}^{\left(  \infty\right)  }\cdot2\left(
r_{B}+r_{F}\right)  \Psi_{\text{H}}+\widetilde{\rho}_{2}a\sum\limits_{k=1}%
^{\infty}\psi_{k}^{\left(  \infty\right)  }\cdot p_{k+1}\left(  k+1\right)
\left(  r_{B}+r_{F}\right)  \Psi_{\text{D}}\nonumber\\
&  -\widetilde{\rho}_{1}\left(  c_{E}+c_{A}\right)  \left(  \beta
-\gamma\right)  -\widetilde{\rho}_{2}\left(  \widetilde{\alpha}+\gamma\right)
\left[  c_{E}+c_{A}\left(  1+\Re\right)  \right]  , \label{Equ-14}%
\end{align}
where%
\[
\widetilde{\rho}_{1}=\frac{\psi_{-1}^{\left(  \infty\right)  }}{\psi
_{-1}^{\left(  \infty\right)  }+\sum\limits_{k=1}^{\infty}\psi_{k}^{\left(
\infty\right)  }},\text{ \ }\widetilde{\rho}_{2}=\frac{\sum\limits_{k=1}%
^{\infty}\psi_{k}^{\left(  \infty\right)  }}{\psi_{-1}^{\left(  \infty\right)
}+\sum\limits_{k=1}^{\infty}\psi_{k}^{\left(  \infty\right)  }}.
\]

Figure 15 shows how $\mathbf{R}_{\text{H}}$ and $\mathbf{R}_{\text{D}}$ depend
on the jumping's mining rate $\gamma$ and the efficiency-increased ratio $\Re$.

In what follows we first discuss a special case with $p_{2}=1$. Then we
provide a detailed comparison between our work and Eyal and
Sirer\cite{Eya:2014}. This is necessary and useful for understanding the
Markov chain method of Eyal and Sirer\cite{Eya:2014}.\begin{figure}[h]
\centering                     \includegraphics[width=11cm]{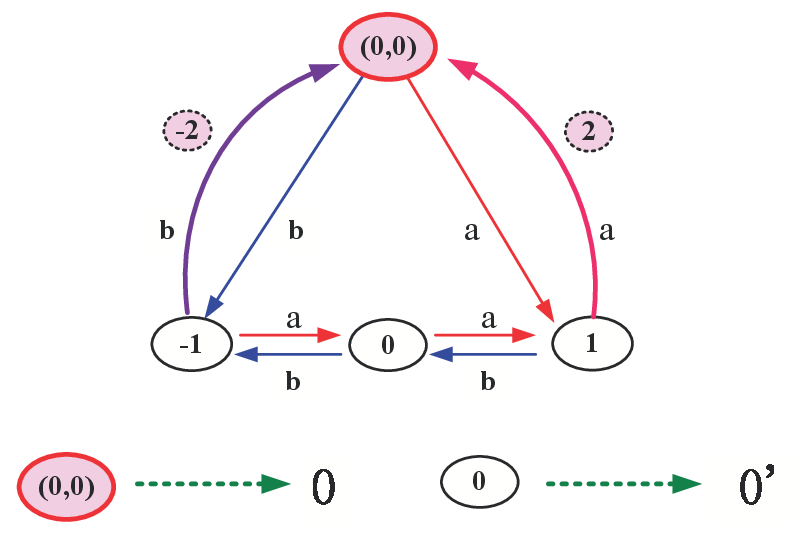}
\caption{A Markov process with $\mu=+\infty$ and $p_{2}=1$ is related to
Figure 1 of Eyal and Sirer\cite{Eya:2014}}%
\label{figure:Fig-9}%
\end{figure}

When $p_{2}=1$, Figure 10 depicts the state transition relations of the Markov
process $Q_{\mu=+\infty}$. It is easy to see that the Markov process
$Q_{\mu=+\infty}$ can not arrive at state $k$ for $k=2,3,4,\ldots$. By using
Figure 10, the infinitesimal generator of the Markov process $Q_{\mu=+\infty}$
is given by%
\[
Q_{\mu=+\infty}=\left(
\begin{array}
[c]{cccc}%
-\left(  a+b\right)  & b & 0 & a\\
b & -\left(  a+b\right)  & a & 0\\
0 & b & -\left(  a+b\right)  & a\\
a & 0 & b & -\left(  a+b\right)
\end{array}
\right)  .
\]
Since the Markov process $Q_{\mu=+\infty}$ is irreducible and positive
recurrent, its stationary probability vector is given by%
\begin{align*}
\Psi^{\left(  1\right)  }  &  =\left(  \psi_{0,0}^{\left(  1\right)  }%
,\psi_{-1}^{\left(  1\right)  },\psi_{0}^{\left(  1\right)  },\psi
_{1}^{\left(  1\right)  }\right) \\
&  =\left(  \frac{a^{2}+b^{2}}{2\left(  a+b\right)  ^{2}},\frac{b}{2\left(
a+b\right)  },\frac{ab}{\left(  a+b\right)  ^{2}},\frac{a}{2\left(
a+b\right)  }\right)  .
\end{align*}
Further, we obtain%
\begin{align}
\mathbf{R}^{\left(  1\right)  }=  &  \overline{\rho}_{1}\mathbf{R}_{\text{H}%
}^{\left(  1\right)  }+\overline{\rho}_{2}\mathbf{R}_{\text{D}}^{\left(
1\right)  }\nonumber\\
&  =\overline{\rho}_{1}b\psi_{-1}^{\left(  1\right)  }\cdot2\left(
r_{B}+r_{F}\right)  \Psi_{\text{H}}+\overline{\rho}_{2}a\psi_{1}^{\left(
1\right)  }\cdot2\left(  r_{B}+r_{F}\right)  \Psi_{\text{D}}\nonumber\\
&  -\overline{\rho}_{1}\left(  c_{E}+c_{A}\right)  \left(  \beta
-\gamma\right)  -\overline{\rho}_{2}\left(  \widetilde{\alpha}+\gamma\right)
\left[  c_{E}+c_{A}\left(  1+\Re\right)  \right]  , \label{Equ-15}%
\end{align}
where $\Psi_{\text{H}}$ and $\Psi_{\text{D}}$ are given in (\ref{BlockN-1})
and (\ref{BlockN-2}), respectively, and%
\[
\overline{\rho}_{1}=\frac{\psi_{-1}^{\left(  1\right)  }}{\psi_{-1}^{\left(
1\right)  }+\psi_{1}^{\left(  1\right)  }}=\frac{b}{a+b},\text{ }%
\overline{\rho}_{2}=\frac{\psi_{1}^{\left(  1\right)  }}{\psi_{-1}^{\left(
1\right)  }+\psi_{1}^{\left(  1\right)  }}=\frac{a}{a+b}.
\]

Now, we provide a long-run economic ratio of the dishonest mining pool over
the honest mining pool by using per unit net mining rate. We define%
\[
\Im^{\left(  1\right)  }=\frac{\frac{1}{\widetilde{\alpha}+\gamma}%
\mathbf{R}_{\text{D}}}{\frac{1}{\beta-\gamma}\mathbf{R}_{\text{H}}%
}=\frac{\beta-\gamma}{\widetilde{\alpha}+\gamma}\frac{\mathbf{R}_{\text{D}}%
}{\mathbf{R}_{\text{H}}}.
\]
We can numerically show that the long-run ratio $\Im^{\left(  1\right)  }$
increases as the efficiency-increased ratio $\Re$ or the jumping's mining rate
$\gamma$ increases.

\vskip                                         1.2cm

In the remainder of this section, by comparing our Figure 10 with Figure 1 of
Eyal and Sirer\cite{Eya:2014}, we provide some useful remarks for
understanding the Markov chain (Figure 1) and revenue analysis ((1) to (3)) of
Eyal and Sirer\cite{Eya:2014} as follows:

\textbf{(a) Do there exist the two boundary states }$0$\textbf{ and
}$0^{\prime}$\textbf{? }

In Figure 1 of Eyal and Sirer \cite{Eya:2014}, the two boundary states $0$ and
$0^{\prime}$ were introduced in a strange way which is not easy to understand.
From Figure 10 in this paper, we establish a one-dimensional Markov process
with state $(0,0)$, and it is observed that our state $(0,0)$ corresponds to
state $0$ of Eyal and Sirer \cite{Eya:2014}; while our state $0$ becomes state
$0^{\prime}$ of Eyal and Sirer \cite{Eya:2014}. However, we do not know in any
way how to set up states $0$ and $0^{\prime}$ of Eyal and Sirer
\cite{Eya:2014} by using the theory of Markov processes.

\textbf{(b) Can Eyal and Sirer's parameter }$\gamma$\textbf{ play a necessary
role in the revenue analysis? }

In Figure 1 of Eyal and Sirer \cite{Eya:2014}, it is seen that $1$ is the
state transition probability from state $0^{\prime}$ to state $0$. However,
the probability $1$ is decomposed into three parts (or separated
path-probabilities): $\alpha$, $\gamma\left(  1-\alpha\right)  $ and $\left(
1-\gamma\right)  \left(  1-\alpha\right)  $ by using $1=\alpha+\gamma\left(
1-\alpha\right)  +\left(  1-\gamma\right)  \left(  1-\alpha\right)  $.
Although Eyal and Sirer's parameter $\gamma$ is introduced, the Markov chain
(Figure 1 of Eyal and Sirer \cite{Eya:2014}) is independent of Eyal and
Sirer's parameter $\gamma$. For example, the stationary probability vector has
nothing to do with Eyal and Sirer's parameter $\gamma$. Furthermore, Eyal and
Sirer's parameter $\gamma$ is used in the revenue analysis, see (1), (2) and
(3) of Eyal and Sirer \cite{Eya:2014}. Clearly, their revenue computation is
based on the law of total probability, in which from Eyal and Sirer
\cite{Eya:2014}, its event probabilities are determined by using the Markov
chain (Figure 1), and its event revenue is derived in (1) and (2) by means of
discussing those reward cases from (a) to (h). Since the stationary
probability vector of the Markov chain (Figure 1 of Eyal and Sirer
\cite{Eya:2014}) is independent of Eyal and Sirer's parameter $\gamma$, the
selfish mining pool's revenue function ((1) to (3) of Eyal and Sirer
\cite{Eya:2014}) can linearly depend on Eyal and Sirer's parameter $\gamma$.
From Sections 8 and 9 of this paper, it is easy to see that the selfish mining
pool's revenue function ((3) of Eyal and Sirer \cite{Eya:2014}) is not based
on a rigorous mathematical calculation by using the Markov reward processes.
Therefore, it is easy to see that Eyal and Sirer's parameter $\gamma$ will not
play any role in the revenue analysis of blockchain selfish mining under a
rigorous mathematical setting.

\textbf{(c) States }$k$\textbf{ for }$k\geq2$\textbf{ can not exist in Figure
1 of Eyal and Sirer \cite{Eya:2014}.}

From Figure 10 of this paper, it is easy to see that there do not exist states
$2,3,4,\ldots$, unless introducting a block-detained probability sequence
$\left\{  p_{k}:k=2,3,4,\ldots\right\}  $. Therefore, there can not exist the
Markov chain (Figure 1 of Eyal and Sirer \cite{Eya:2014}) in the blockchain
selfish mining. Based on this, we explain the realistic background that the
Markov chain of Eyal and Sirer \cite{Eya:2014} can be related to.

In brief, one of our main findings demonstrates that the Markov chain and the
revenue analysis in Eyal and Sirer\cite{Eya:2014} should be confused. This
paper provides some new insights on improving the Markov chain method of Eyal
and Sirer \cite{Eya:2014}, and those works following Eyal and Sirer
\cite{Eya:2014} in the literature.

\section{Numerical Examples}

In this section, we use some numerical examples to verify our theoretical
results, and indicate how performance measures of our more general model of
blockchain selfish mining depend on some key parameters of blockchain.

Our numerical experiments are classified into three parts: (a) The orphan
blocks, (b) the long-run average profits, and (c) an approximate computational
model without network latency.

\textbf{Part one: The orphan blocks}

To study the influence of orphan blocks, we take some parameters:
$\widetilde{\alpha}=10$, $\beta=28$, the block-pegging rate $\mu=3$. Let the
jumping's mining rate $\gamma\in\left[  0.5,8.5\right]  $, and the
efficiency-increased ratio $\Re=0.5,$ $0.7,$ $0.9$.

For the average stationary lengths $L_{\text{M}}$ and $L_{\text{O}}$ of the
main chain and the chain of orphan blocks, Figure 11 shows that $L_{\text{M}}$
and $L_{\text{O}}$\textbf{ }are not monotonous for $\gamma$ or $\Re$, and they
start to decrease and then increase as $\gamma$ or $\Re$ increases. In fact,
such theoretical analysis is given at the end of Section 5.

\begin{figure}[h]
\centering         \includegraphics[width=7cm]{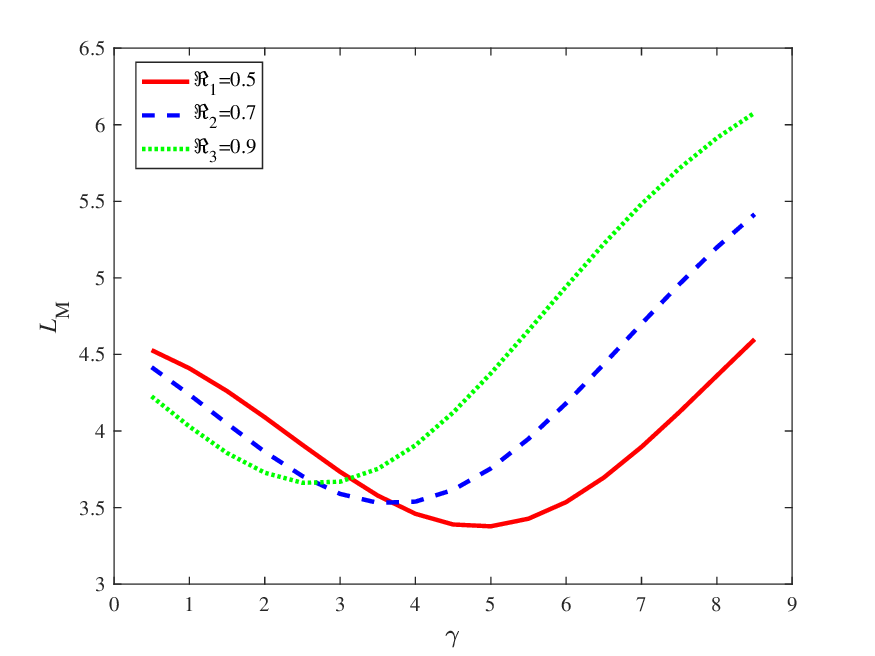}  \centering
\includegraphics[width=7cm]{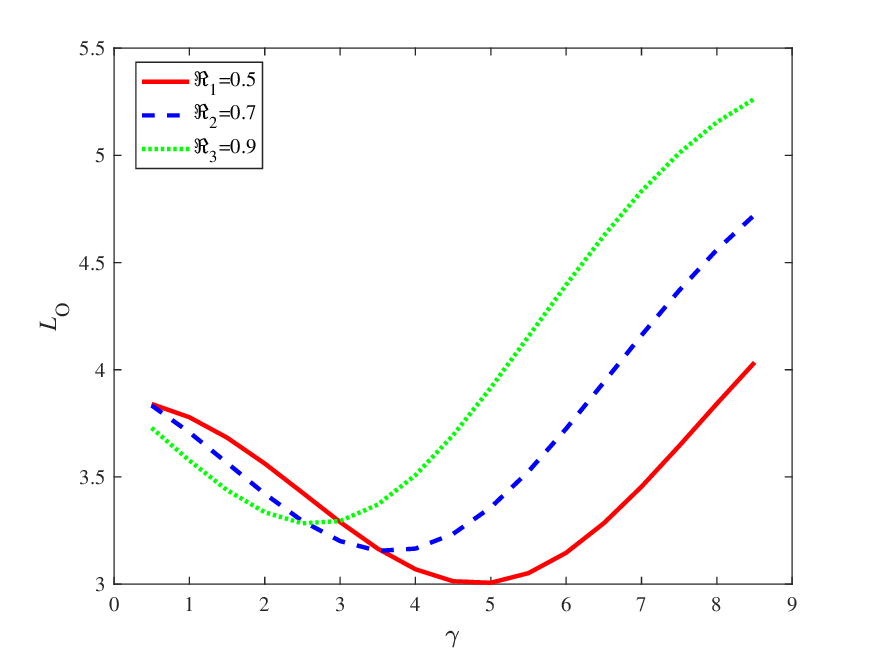}  \caption{$L_{\text{M}}$ and
$L_{\text{O}}$ vs. $\gamma$ for three different $\Re_{1}$, $\Re_{2}$, $\Re
_{3}$.}%
\label{Fig-10}%
\end{figure}

Now, we observe the ratio $\phi$ of the two average stationary lengths, and
the stationary ratio $\psi$ of the two block removing and pegging rates. From
Figure 12, it is seen that $0<\phi<\psi<1$. Also, $\phi$ and $\psi$ begin to
increase and then decrease, as $\gamma$ or $\Re$ increases. Thus $\phi$ and
$\psi$ are not monotonous in $\gamma$ or $\Re$.

\begin{figure}[h]
\centering         \includegraphics[width=7cm]{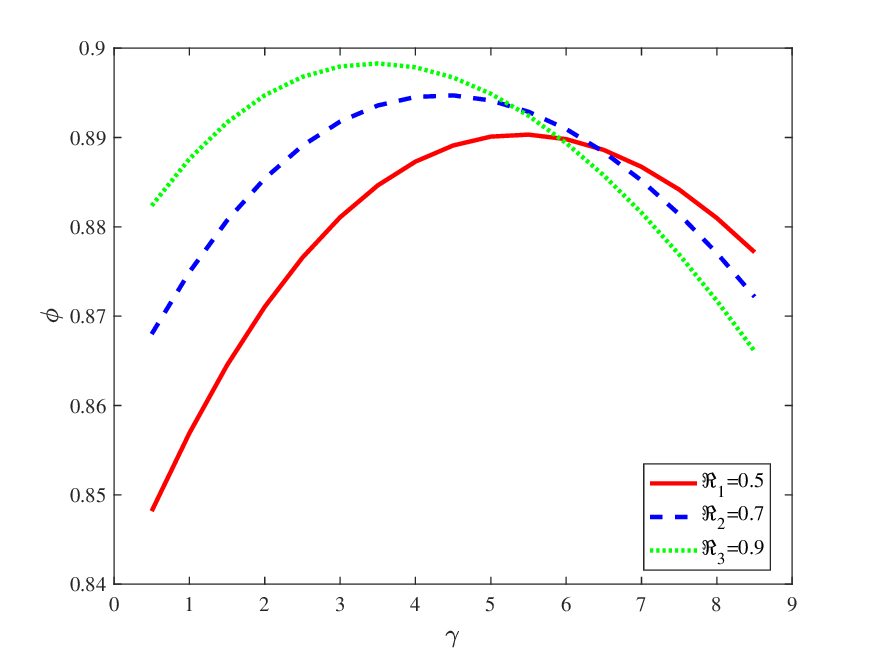}  \centering
\includegraphics[width=7cm]{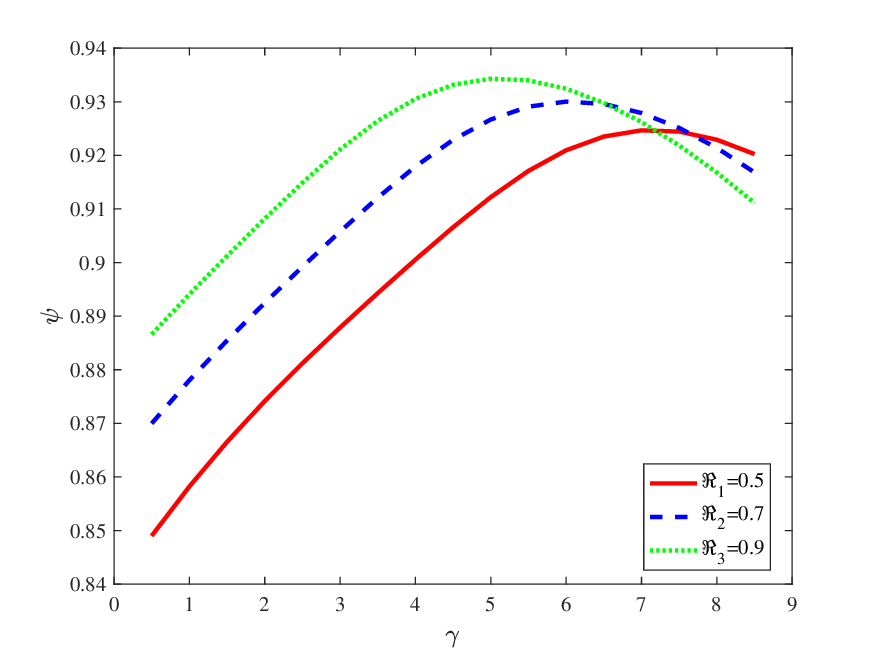}  \caption{$\phi$ and $\psi$ vs.
$\gamma$ for three different values of $\Re$}%
\label{Fig-11}%
\end{figure}

\textbf{Part two: The long-run average profits}

To discuss the two long-run average profits given in Section 6, we take some
parameters: $\widetilde{\alpha}=10$, $\beta=28$, the block-pegging rate
$\mu=3$, $r_{B}=0.5$, $r_{F}=0.5$, $c_{E}=0.5$, $c_{A}=0.5$. Let the jumping's
mining rate $\gamma\in\left[  5,8.5\right]  $, and the efficiency-increased
ratio $\Re=0.5,0.7,0.9$.

From the left half of Figure 13, it is seen that the long-run average profit
$\mathbf{R}_{\text{H}}$ of the honest mining pool decreases as the jumping's
mining rate $\gamma$ increases, and it also decreases as the
efficiency-increased ratio $\Re$ increases.

From the right half of Figure 13, it is observed that the long-run average
profit $\mathbf{R}_{\text{D}}$ of the dishonest mining pool increases as the
jumping's mining rate $\gamma$ increases, and it also increases as the
efficiency-increased ratio $\Re$ increases.

\begin{figure}[h]
\centering         \includegraphics[width=7cm]{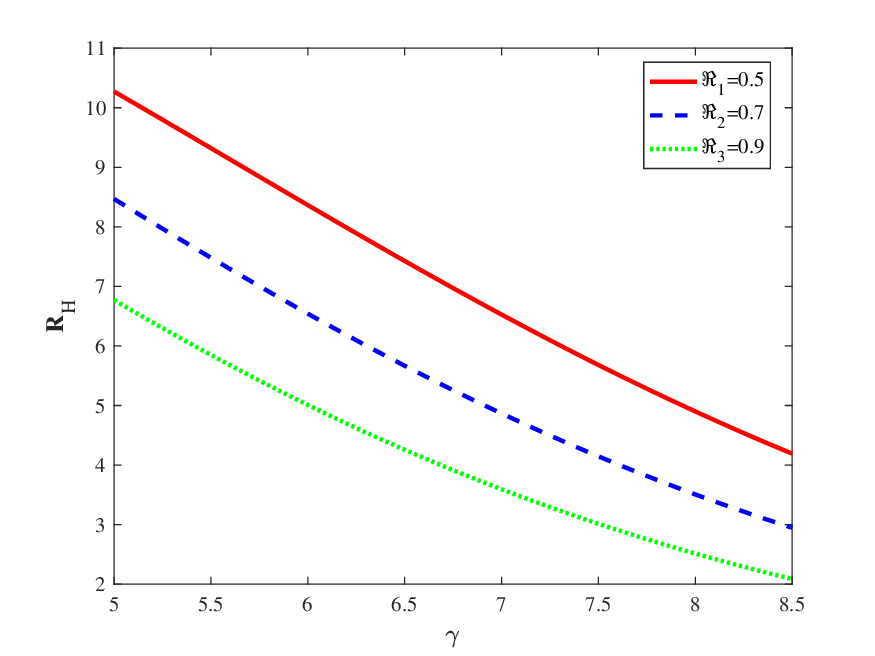}  \centering
\includegraphics[width=7cm]{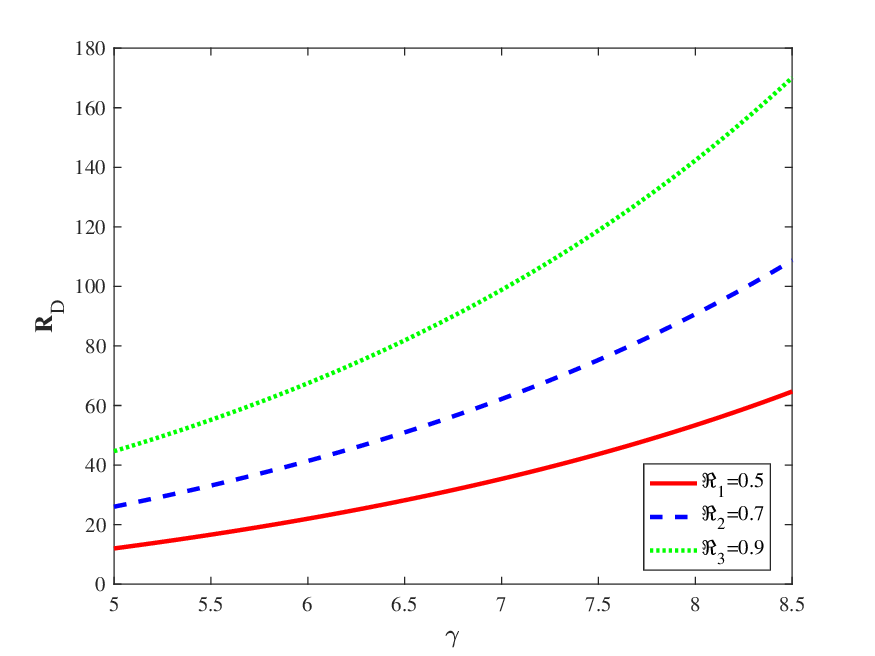}  \caption{$\mathbf{R}_{\text{H}}$
and $\mathbf{R}_{\text{D}}$ vs. $\gamma$ for three different $\Re_{1}$,
$\Re_{2}$, $\Re_{3}$.}%
\label{Fig-12}%
\end{figure}

Now, we discuss the long-run economic ratio $\Im$ of the dishonest mining pool
over the honest mining pool, and the long-run block-pegging rate ratio $\tau$
of the dishonest mining pool over the honest mining pool. The two long-run
ratios are necessary and useful in the study of blockchain selfish mining.

From the left half of Figure 14, it is seen that $\Im>1$. Also, $\Im$
increases as the jumping's mining rate $\gamma$ increases, and it also
increases as the efficiency-increased ratio $\Re$ increases.

From the right half of Figure 14, it is seen that $\tau>1$. Moreover, $\tau$
increases as the jumping's mining rate $\gamma$ increases, and it also
increases as the efficiency-increased ratio $\Re$ increases.

Corollary \ref{Cor:Mono} shows that each of the two long-run ratios $\Im$ and
$\tau$ increases, as the efficiency-increased ratio $\Re$ increases. This is
the same as our numerical experiment. However, we can not prove that the two
long-run ratios $\Im$ and $\tau$ are monotonically increasing in the jumping's
mining rate $\gamma$, because $\Im$ and $\tau$ have a complicated relation
with $\gamma$. While it is interesting that our numerical experiment indicates
such increasing monotonicity. That is, the bigger the selfish mining pool, the
more profit each selfish miner is obtained. This is the incentive for the
selfish mining pool to become bigger and bigger.

\begin{figure}[h]
\centering         \includegraphics[width=7cm]{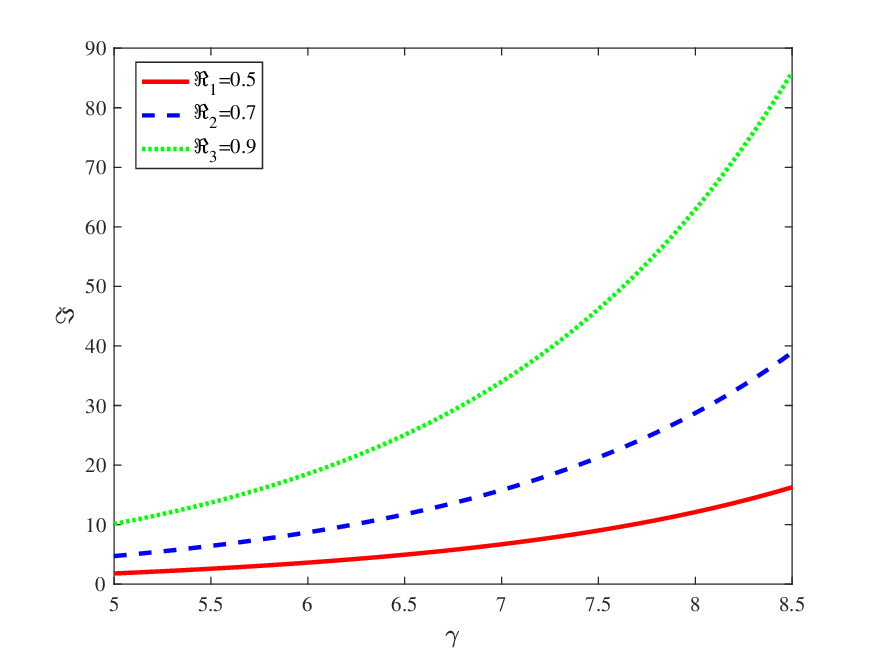}  \centering
\includegraphics[width=7cm]{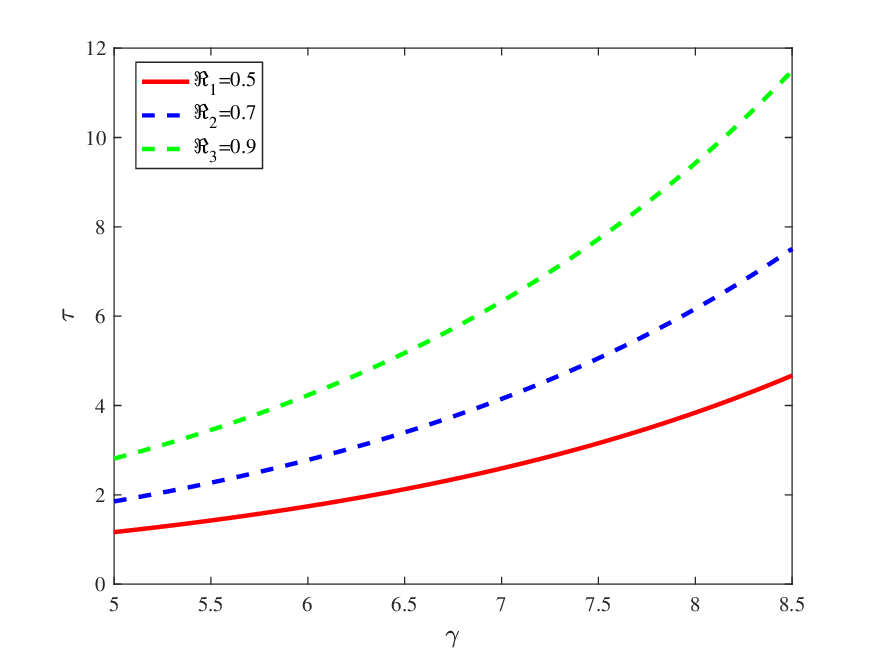}  \caption{$\Im$ and $\tau$ vs.
$\gamma$ for three different $\Re_{1}$, $\Re_{2}$, $\Re_{3}$.}%
\label{Fig-13}%
\end{figure}

\textbf{Part three: The approximate model without network latency}

In the approximate computational model without network latency given in
Section 9, we take some parameters: $\widetilde{\alpha}=10$, $\beta=28$, the
block-pegging rate $\mu=3$, $r_{B}=15$, $r_{F}=3$, $c_{E}=3$, $c_{A}=1$. Let
the jumping's mining rate $\gamma\in\left[  0.5,8\right]  $, and the
efficiency-increased ratio $\Re=0.5,0.7,0.9$.

We first analyze the relative long-run average profit $\mathbf{R}_{\text{H}%
}^{\left(  \infty\right)  }$ of the honest mining pool. From the left half of
Figure 15, it is seen that $\mathbf{R}_{\text{H}}^{\left(  \infty\right)  }$
decreases as the jumping's mining rate $\gamma$ increases, and it also
decreases as the efficiency-increased ratio $\Re$ increases.

Then we discuss the relative long-run average profit $\mathbf{R}_{\text{D}%
}^{\left(  \infty\right)  }$ of the dishonest mining pool. From the right half
of Figure 15, it is seen that $\mathbf{R}_{\text{D}}^{\left(  \infty\right)
}$ increases as the jumping's mining rate $\gamma$ increases, and it also
increases as the efficiency-increased ratio $\Re$ increases.

Note that the precise long-run average profit $\mathbf{R}_{\text{H}}$ and the
relative long-run average profit $\mathbf{R}_{\text{H}}^{\left(
\infty\right)  }$ have a similar monotonicity, but $\mathbf{R}_{\text{H}}$ is
smaller than $\mathbf{R}_{\text{H}}^{\left(  \infty\right)  }$. So are
$\mathbf{R}_{\text{D}}$ and $\mathbf{R}_{\text{D}}^{\left(  \infty\right)  }$.

\begin{figure}[h]
\centering         \includegraphics[width=7cm]{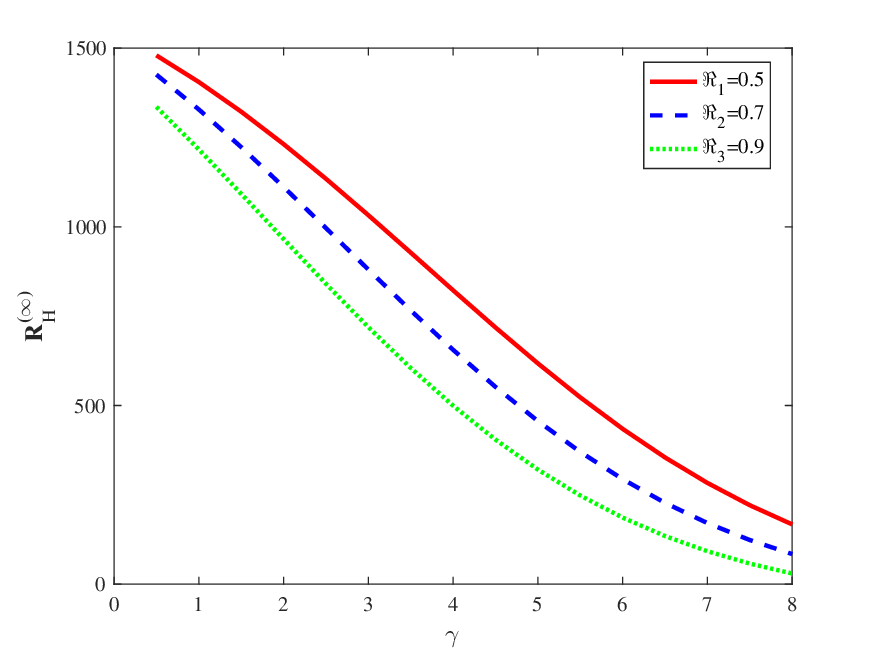}  \centering
\includegraphics[width=7cm]{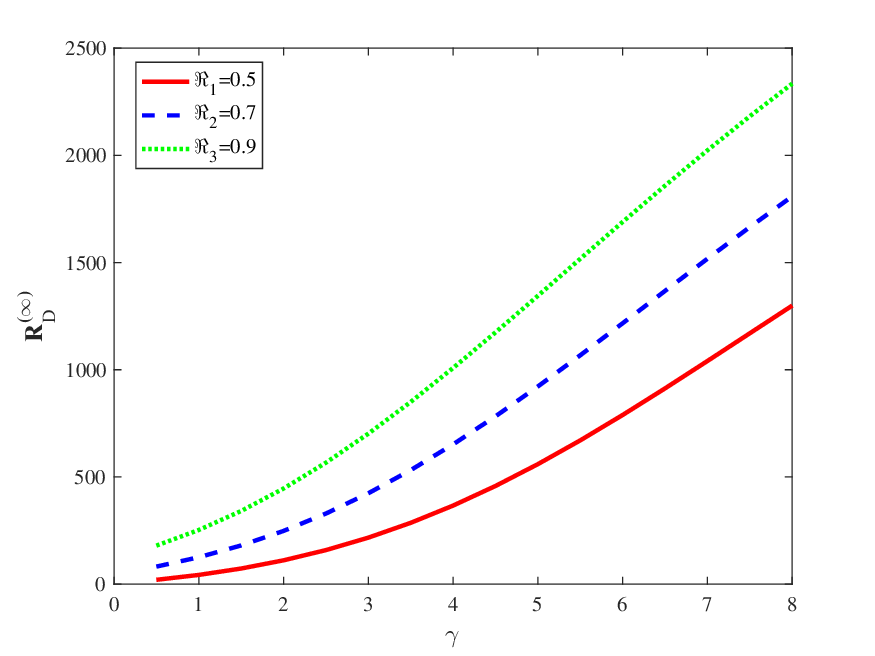}  \caption{$\mathbf{R}_{\text{H}%
}^{\left(  \infty\right)  }$ and $\mathbf{R}_{\text{D}}^{\left(
\infty\right)  }$ vs. $\gamma$ for three different $\Re_{1}$, $\Re_{2}$,
$\Re_{3}$.}%
\label{Fig-14}%
\end{figure}

\section{Concluding Remarks}

In this paper, we provide a new theoretical framework of pyramid Markov
processes in the study of blockchain selfish mining. We describe a more
general model of blockchain selfish mining with both a two-block leading
competitive criterion and a new economic incentive, both of which are
expressed by means of the block-detained probability sequence, the
efficiency-increased ratio and the jumping's mining rate. For such selfish
mining, we establish a pyramid Markov process, and show that the pyramid
Markov process is irreducible and positive recurrent, and the stationary
probability vector is matrix-geometric with an explicitly representable rate
matrix. Also, we use the stationary probability vector to analyze the
influence of orphan blocks on the waste of computing resource. Furthermore, we
set up a pyramid Markov reward process to investigate the long-run average
profits of the honest and dishonest mining pools, respectively. Based on this,
we can measure the mining efficiency of the dishonest mining pool through
comparing with the honest mining pool. As a by-product, we build three
approximative Markov processes when the system states are taken as the
difference of block numbers on the two forked branches at the common tree
root. For a special case without network latency, one of our main findings
demonstrates that the Markov chain in Eyal and Sirer \cite{Eya:2014} should be
incorrect, and thus the results by following the Markov chain method of Eyal
and Sirer \cite{Eya:2014} in the literature may not be true as well. Finally,
we use some numerical examples to verify our theoretical results.

Note that the pyramid Markov (reward) processes open a new avenue to the study
of blockchain selfish mining. We hope that the methodology and results
developed in this paper can shed light on the blockchain selfish mining and
lead to a series of potentially promising research. We will continue our
future research in the following directions:

-- Considering the blockchain selfish mining under a changing difficulty level
of PoW puzzle, which is described as a Markovian arrival process, or a
transient periodic point process.

-- Analyzing the case that a part of the main chain by the dishonest mining
pool is pegged on the blockchain, while the other part of the main chain is
left to support the next round of competition between two new block branches
forked at a common tree root. Note that the other part of the main chain by
the dishonest mining pool is possible to become the orphan blocks if its
subsequent new branch is no longer ahead of that by the honest mining pool.
This significant risk should be considered when the dishonest mining pool
decides whether to peg the whole main chain or peg a part of it.

-- Discussing the case with a $K$-block leading competitive criterion for
$K=1,2,3,4,5,\ldots$, and optimizing the positive integer $K$ to maximize the
mining efficiency and/or the long-run average profit.

-- Setting up a pyramid block-structure Markov process for Ethereum, and
developing an effective algorithm for computing the matrix-analytic solution.
Using the stationary probability vector to analyze the influence of orphan and
uncle blocks on the waste of computing resource, and further investigate the
long-run average profit of Ethereum.

-- Developing the fluid and diffusion approximation for analyzing the
blockchain selfish mining with multiple mining pools, providing the stable
conditions of the multi-dimensional blockchain systems, and establishing the
long-run average profits of the multiple mining pools.

-- Further developing stochastic optimization and dynamic control of the
blockchain selfish mining, for example, Markov decision processes, stochastic
game, and evolutionary game.

\section*{Acknowledgements}

Quan-Lin Li was supported by the National Natural Science Foundation of China
under grants No. 71671158 and 71932002 and by the Beijing Social Science
Foundation Research Base Project under grant No. 19JDGLA004.

\end{document}